\documentclass[structabstract]{aa}  
\usepackage{graphicx}
\usepackage[utf8]{inputenc}
\usepackage{caption}
\usepackage{textcomp}
\usepackage{txfonts}
\usepackage{float}
\usepackage{natbib}
\bibpunct{(}{)}{;}{a}{}{,} 
\DeclareUnicodeCharacter{207A}{^+}

\begin{document}

\newcommand{\tablenotea}[1]{\parbox{9.8cm}{ \indent \footnotesize{\textsc{Notes.--}~#1}}}
\newcommand{\tablenoteb}[1]{\parbox{8.7cm}{ \indent \footnotesize{\textsc{}~#1}}}
\newcommand{\tablenotec}[1]{\parbox{8.6cm}{ \indent \footnotesize{\textsc{}~#1}}}
\newcommand{\tablenoted}[1]{\parbox{8.6cm}{ \indent \footnotesize{\textsc{}~#1}}}
\newcommand{\tuc}{\rm $J$=1--0}            
\newcommand{\tdu}{\rm $J$=2--1}         
\newcommand{\ttd}{\rm $J$=3--2}         
\newcommand{\doce}{\rm $^{12}$CO}       
\newcommand{\trece}{\rm $^{13}$CO}      
\newcommand{\gsim}{\raisebox{-.4ex}{$\stackrel{>}{\scriptstyle \sim}$}}
\newcommand{\lsim}{\raisebox{-.4ex}{$\stackrel{<}{\scriptstyle \sim}$}}
\newcommand{\psim}{\raisebox{-.4ex}{$\stackrel{\propto}{\scriptstyle \sim}$}}
\newcommand{\kms}{\mbox{km~s$^{-1}$}}
\newcommand{\jyb}{\mbox{Jy~beam$^{-1}$}}
\newcommand{\s}{\mbox{$''$}}
\newcommand{\mloss}{\mbox{$\dot{M}$}}
\newcommand{\my}{\mbox{$M_{\odot}$~yr$^{-1}$}}
\newcommand{\ls}{\mbox{$L_{\odot}$}}
\newcommand{\ms}{\mbox{$M_{\odot}$}}
\newcommand{\mm}{\mbox{$\mu$m}}
\def\arcdeg{\hbox{$^\circ$}}
\newcommand{\secp}{\mbox{\rlap{.}$''$}}
\newcommand{\secs}{\mbox{\rlap{.}$^{\rm s}$}}
\newcommand{\um}{\mbox{$\mu$m}}
\newcommand{\h}{$^{\rm h}$}
\newcommand{\m}{$^{\rm m}$} 
\newcommand{\irc}{IRC\,+10420 }         
\newcommand{\afg}{AFGL\,2343}
\newcommand{\tas}{\mbox{T\rlap{$_A$}$^*$}}

   \title{A $\lambda$ 3\,mm and 1\,mm line survey toward the yellow hypergiant
   IRC\,+10420.  
   \thanks{Based on observations carried out with the IRAM 30 m Telescope. 
   IRAM is supported by INSU/CNRS (France), MPG (Germany) and IGN (Spain)}}

   \subtitle{N-rich chemistry and IR flux variations} 

   \author{G. Quintana-Lacaci
          \inst{1}
          \and
          M. Ag\'undez \inst{1}
          \and
          J. Cernicharo \inst{1}
          \and
          V. Bujarrabal \inst{2}
          \and
          C. S\'anchez Contreras \inst{3}
          \and
          A. Castro-Carrizo \inst{4}
          \and
          J. Alcolea \inst{5}
          }

   \institute{Instituto de Ciencia de Materiales de Madrid, 
   Sor Juana In\'es de la Cruz, 3, Cantoblanco, 28049 Madrid, Spain.\\
              \email{g.quintana@icmm.csic.es,jose.cenicharo@icmm.csic.es}
         \and
             Observatorio Astron\'omico Nacional (IGN), Ap 112, E–28803, Alcal\'a de Henares, Spain.\\
             \email{v.bujarrabal@oan.es}
         \and
             Department of Astrophysics, Astrobiology Center (CSIC-INTA), Postal address: ESAC campus, P.O. Box 78, E-28691 Villanueva
de la Ca\~nada, Madrid, Spain.\\
                \email{csanchez@cab.inta-csic.es}
         \and 
             Institut de RadioAstronomie Millim\'etrique, 300 rue de la Piscine, 38406 Saint Martin d'H\'eres, France.\\
                \email{ccarrizo@iram.fr}
         \and
             Observatorio Astron\'omico Nacional (IGN), Alfonso XII N$^{o}$3, 28014 Madrid, Spain.\\
                \email{j.alcolea@oan.es}
         }
         
   \date{Received September 15, 1996; accepted March 16, 1997}

 
  \abstract
    {}
   {Our knowledge of the chemical properties of the circumstellar ejecta of the most massive
   evolved stars is particularly poor. We aim to study the chemical characteristics of the prototypical yellow
   hypergiant star, IRC\,+10420. For this purpose, we obtained full line surveys at 1 and 3\,mm atmospheric
   windows.}
   {We have identified 106 molecular emission lines from 22 molecular species.
   Approximately half of the molecules detected are N-bearing species, in particular HCN, HNC, CN, NO, NS, PN, and N$_2$H$⁺$. 
   We used rotational diagrams to derive the density and rotational temperature of the different molecular
   species detected. We introduced an iterative method that allows us to take moderate line opacities into account.}
   {We have found that IRC\,+10420 presents high abundances of the N-bearing molecules compared with O-rich evolved stars.
   This result supports the presence of a N-rich chemistry, expected for massive stars. 
   Our analysis also suggests a decrease of the $^{12}$C/$^{13}$C ratio from $\gtrsim 7$ 
   to $\sim 3.7$ in the last 3800 years, which can
   be directly
   related to the nitrogen enrichment observed.
   In addition, we  found that
   SiO emission presents a significant intensity decrease for high-$J$ lines when compared with older observations. 
   Radiative transfer modeling shows that this variation can be explained by a decrease in the infrared (IR) flux of the dust.
   The origin of this decrease might be an expansion of the dust shell or a lower stellar temperature due {to the pulsation
   of the star.}   }
   {}
   

   \maketitle
%

\section{Introduction}

IRC\,+10420 is a yellow hypergiant (YHG) star. These objects are evolved massive stars, which
present extreme initial mases and very high luminosities ($\mathrm{log}(L/\ls)\sim$\,5.6, $M_{init}\,\gsim\, 20\ms$). In fact, 
this particular YHG has a luminosity of
$L\sim 5 \times 10^5 \ls$ and has an estimated initial mass of $M_{init} \sim 50 \ms$ \citep[][]{tiffany2010,nieuwen00}. 

Yellow hypergiants are post red-supergiant stars (RSGs) evolving toward higher temperatures
in the HR diagram. { In particular, the spectral type of \irc\ has changed from F8Ia to A5Ia in just 20 yr \citep{Klochkova97}.}
These objects are thought to lose as much as one half of their initial masses during the RSG phase \citep[e.g.,][]{MM88}.
In addition, during their post-RSG evolution the YHGs encounter an instability region, called
the yellow void \citep{dejager98}, which results in a new episode of effective mass ejection. 
These outburst were recently detected for $\rho$ Cas \citep{lobel2003}.
As a result of these outbursts, the ionized wind of these stars becomes optically thick and the resultant stellar 
wind spectrum mimics that of a lower $T_\mathrm{eff}$ star
\citep{CrossingYV}. 
As the ejected material dilutes into the interstellar medium (ISM), the apparent $T_\mathrm{eff}$ increases again. 
These apparent $T_\mathrm{eff}$ oscillations are frequently called bouncing against the yellow void.
Since the real $T_\mathrm{eff}$ continues increasing, at a certain moment, as the ejected material dilutes out, the YHG stars would 
eventually appear at the { just beyond the high-temperature edge} of the yellow void.
However, \citet{Smith04} found that the { high-temperature edge} of the yellow void is coincident with the S Doradus instability strip. 
These sources then form a pseudophotosphere that keeps the source in the { low-temperature edge} of the yellow void in the HR diagram.
Therefore, these authors suggested that the evolution of the YHGs remain hidden until they become slash stars and finally 
enter in the Wolf-Rayet phase.

While these mass ejections are predicted to be very important for these objects, 
only three YHGs, IRC+\,10420 and AFGL\,2343 \citep{cc07}, and 
recently  IRAS\,17163--3907 \citep{muller2015}, have shown molecular emission. 

The kinematics and structure of the molecular envelopes around IRC+\,10420 and AFGL\,2343 were studied in detail by \citet{cc07} 
thanks to high angular resolution interferometric maps of \doce. 
They found that, while these objects showed slight departures from the spherical symmetry, the data could
be reasonably modeled by adopting an isotropic mass loss with large variations with time. 
In particular, for the YHG IRC+\,10420 they found a detached circumstellar envelope (CSE) with an extent 
of $5 \times 10{^{17}}$cm expanding at velocities of $\sim$37 km/s.
Two strong mass ejection episodes, which occurred within a lapse of
1200 years and reached a mass loss rate of 3 $\times 10^{-4}$\my, are responsible for the formation of this CSE.
The total
mass derived for the CSE around this object is $\sim 1 \ms$.
\citet{gqtesis} showed that the ejection of this material could be explained in a similar way as the ejections
presented by the low mass AGB stars (i.e., the mass ejection is driven by radiation pressure on the dust grains). Also, 
\citet{gqtesis} argued that all the molecular material observed by \citet{cc07} was only ejected
during the YHG phase. Any gas ejected during the previous RGS phase should have been rapidly diluted in the ISM
and photodissociated by ISM ultraviolet (UV) radiation field.

The chemistry of IRC\,+10420 is particularly rich \citep{chemyhg}. 
The following species were detected in the CSE around this object: CO, $^{13}$CO, HCN, CN, H$^{13}$CN, SiO, $^{29}$SiO, SO, SiS, HCO$^{+}$, 
CN, HNC, HN$^{13}$C, and CS. Surprisingly, some species, such as HCN and HNC, showed particularly high abundances
compared with the O-rich AGB stars, which are the low mass counterpart of YHGs \citep[see, e.g.,][]{bujarrabal94}.
This is explained by an enrichment in nitrogen due to the hot bottom burning process \citep{boothroyd93}, 
which is present for stars with
masses above $\sim$3 \ms. This process transforms $^{12}$C into nitrogen, which changes the composition of the material that is
transported to the photosphere and later on expelled. The nitrogen enrichment is expected
to lead to a N-rich chemistry for the most massive
and evolved stars. This was recently confirmed by \citet{NO10420}. These authors found an abnormally
high abundance of NO in IRC\,+10420, directly related to an enhancement of the elemental abundance of nitrogen.  

On the other hand, SiO emission was found to come from regions
located at 10$^{17}$\,cm from the star, far from the radius where dust formation takes place 
and where the SiO is expected to be largely depleted in the gas \citep{ccsio}. These authors suggested 
that this SiO emission is related to a spherical shock front that heats up the grains,
releasing some amount of Si back to the gas phase.

Recently, \citet{teyssier2012} observed selected transitions of NH$_3$, OH, H$_2$O, CO, and $^{13}$CO toward this object with HIFI. 
They also showed that the model derived by \citet{cc07} for low-$J$ CO transitions
was applicable, with minor changes, to high excitation lines.

In this paper we present a full line survey of IRC\,+10420
obtained with the IRAM 30 m telescope in the atmospheric windows
centered at wavelengths of 1\,mm and 3\,mm. The survey
confirms that the chemistry of this object is particularly rich.
We detected 22 molecular species for which we determined the column 
density and rotational temperature.


\section{Observations}

We used the IRAM 30 m radiotelescope to obtain a complete line
survey at 3\,mm and 1\,mm for the YHG IRC\,+10420.
We observed IRC\,+10420 
at the position coordinates (J2000) 19\h26\m48$\secs$10, 
+11$\rm^o$21${'}$17\secp0 with an $\rm{v_{LSR}}$ = 76\,km/s.
The observations were obtained during February 2012. We used the EMIR receiver,  simultaneously 
using the receivers E090 and E230 with a bandwidth of 4\,GHz at two 
polarizations.
We used 16 different setups to cover the atmospheric windows at 
3\,mm and 1\,mm. Each setup was observed for one hour.
We used the wobbler switching mode to minimize the ripples in the baselines.
The system temperatures
during the observations were in the range 100--250\,K for the E090 receiver 
and between 200 and 425\,K
for the E230 receiver. The weather conditions during the observations 
were good with an amount of precipitable water vapor ranging between 
2 and 6 mm. 

The backends used were WILMA (spectral resolution 2\,MHz) and the 4\,MHz 
filter bank. Owing to the large width of the line profiles 
 observed in IRC\,+10420 ($\sim$60 km/s), this resolution is enough to 
resolve the line profiles. 

The pointing correction was checked frequently and, therefore, we expect 
pointing errors of $\sim$3''. The telescope beam size at 3\,mm is 21--29$''$ and 9--13$''$ at 1\,mm. 
The Atmospheric Transmission Model (ATM) is adopted
at the IRAM 30 m \citep{Cerni85,Pardo01}.
The data presented is calibrated in antenna 
temperature ($T^*_A$). The calibration error is expected to be 10\% at 3\,mm and 
30\% at 1\,mm.
The data were processed using the GILDAS package\footnote{See URL 
http://www.iram.fr/IRAMFR/GILDAS/}. The baselines were subtracted using 
only first grade polynomials.

\section{Line identification}

The full spectra obtained at 3 and 1\,mm are presented in Figs.\ref{Fig3mm}\&\ref{Fig1mm}.
The sideband rejection of the EMIR receiver is higher than 10\,db
and therefore only the strongest lines present their counterparts
in the image band. The spectral features whose origin is
the image band are labeled in red characters in Figs. \ref{Fig3mm}\&\ref{Fig1mm}.

The line identification was performed using the online catalogs of CDMS\footnote{http://www.astro.uni-koeln.de/cdms/catalog} \citep{cdms}, 
JPL\footnote{http://spec.jpl.nasa.gov/} \citep{jpl}, Splatalogue\footnote{http://www.splatalogue.net/}, 
and the catalog incorporated into the radiative
transfer code MADEX \citep{MADEX}.

In Tables A.1\&A.2, we list all the lines detected in the line survey in the 3\,mm and
1\,mm bands, respectively. In these tables we also present
the velocity-integrated intensity of the lines, the peak intensity and the {rms of the noise level} for a spectral 
resolution of 2\,MHz. As a result of the large width of the lines of \irc,
the hyperfine structure of the transitions of CN,
NO, and NS cannot be resolved. For these transitions we present
the global integrated intensity including all hyperfine components.

Thanks to the large frequency coverage of the spectral survey, we were able to confirm 
the presence of the species detected by inspecting their relative strengths expected for local thermodynamic equilibrium 
(LTE) conditions.

We detected a total of 106 spectral lines, which we identified as arising from 22 species
including different isotopologues.
The list of the species detected can be found in Table\,1.

 \begin{table}
\caption{Molecular species detected}             
\label{table:1}      
\centering                          
\begin{tabular}{l c | c c }        
\hline\hline                 
Molecule & Isotopologues  & Molecule  & isotopologues \\
\hline  
CO & $^{13}$CO &HCN &H$^{13}$CN\\
SiO & $^{29}$SiO, $^{30}$SiO,Si$^{18}$O&HNC&HN$^{13}$C\\
SiS&&CH$_3$OH&\\
SO&$^{34}$SO&CN&\\
SO$_2$&&PN&\\
CS&&NO&\\
HCO$^{+}$&&NS&\\
N$_2$H$^{+}$&&&\\
\hline                                   
\end{tabular}
\end{table}

\section{LTE modeling of the line emission.}

Once we successfully identified the emission lines observed, we aimed to estimate the 
physical conditions of the regions where these species are located.

In order to estimate the excitation temperatures and column densities
of the different species observed we used a LTE approach as that described by \citet{rtd}.
These authors showed that the rotational diagram of a certain molecule follows the 
next relationship.

\begin{equation}
 \mathrm{ln} \left(\frac{N_\mathrm{u}}{g_\mathrm{u}}\right) = 
 {\mathrm{ln}} \left(\frac{N}{Z(T_\mathrm{rot})}\right) - \mathrm{ln}\ C_{\tau} - \frac{E_\mathrm{u}}{kT_\mathrm{rot}}
,\end{equation}

\noindent where $N_\mathrm{u}$ and $g_\mathrm{u}$ are the total column density and statistical weight
of the upper level, respectively, $N$ is the total column density, $T_\mathrm{rot}$ is the rotational temperature,
$Z(T_\mathrm{rot})$ is the partition function, $E_\mathrm{u}$ is the energy of the upper level,
and $C_{\tau}$ is the term that takes the opacity effects into account,  
which corresponds to $C_{\tau} = \tau / (1-e^{-\tau})$. The value of $\mathrm{ln}({N_\mathrm{u}}/{g_\mathrm{u}})$ is
directly proportional to the integrated intensity of the molecular line observed \citep{rtd}.

Here we use the opacity-corrected column density $N'_\mathrm{u} = N_\mathrm{u}\ C_{\tau}$. Therefore, relation
(1) can be rewritten as

\begin{equation}
 \mathrm{ln} \left(\frac{N'_\mathrm{u}}{g_\mathrm{u}}\right) = 
 {\mathrm{ln}} \left(\frac{N}{Z(T_\mathrm{rot})}\right) - \frac{E_\mathrm{u}}{kT_\mathrm{rot}}
.\end{equation}

This relationship allows us to estimate $T_\mathrm{rot}$ and $N$ of a molecule by fitting a straight
line to the rotational diagram. 

As an approximation for the opacity, we use that presented by \citet{chemyhg}, 

\begin{equation}
 \tau = \mathrm{ln}\ \left( \frac{1}{1-\frac{T_\mathrm{mb}}{S_\nu}\frac{\Omega_\mathrm{A}}{\Omega_\mathrm{S}}} \right) 
,\end{equation}

\noindent where $T_\mathrm{mb}$ is the peak temperature in main beam temperature scale, $S_\nu$ is the source function, and
$\Omega_\mathrm{S}/\Omega_\mathrm{A}$ is the geometrical dilution factor between source
size and telescope main beam. 
{ We adopt the source size from \citet{chemyhg}, i.e., an angular radius of
11\secp0 for} \doce\ and \trece, and 3\secp3 { for the rest of the molecular species.
{ We also took this beam dilution correction into account in the calculation of ${N_\mathrm{u}}/{g_\mathrm{u}}$, 
in particular by correcting the integrated intensity of each molecular transition.}
The source function is calculated as follows:}

\begin{equation}
 S_\nu = \frac{h\,\nu}{k}\left( \frac{1}{e^{\frac{h\,\nu}{k\,T_\mathrm{ex}}}-1} - \frac{1}{e^{\frac{h\,\nu}{k\,T_\mathrm{bg}}}-1} \right) 
,\end{equation}

\noindent where $T_\mathrm{ex}$ is the excitation temperature and $T_\mathrm{bg}$ is the temperature of the cosmic background.

\begin{figure}[h!]
   \centering
   \includegraphics[angle=0,width=9cm]{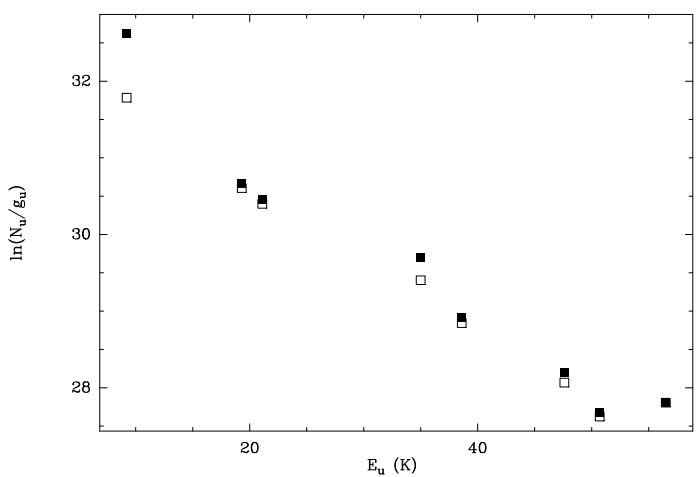}
   \caption{Rotational diagram for SO. Empty squares represent the values for the SO       transitions
   assuming a negligible opacity. Filled squares show the values for $N_\mathrm{u} / g_\mathrm{u}$ 
   corrected for opacity effects.}
              \label{tau}%
    \end{figure}

The term $C_\tau$ is different for each rotational transition and
depends on $T_\mathrm{rot}$ via the source function. As a result of this, $T_\mathrm{rot}$ cannot
be directly obtained by fitting a straight line to equation (1). In order to solve
this problem, we used an iterative procedure. We introduce an initial rotational temperature in the opacity term (1000\,K) and
fit to the expression in Eq.\,(1). The value of $T_\mathrm{rot}$ obtained is then introduced as the new temperature to compute
the opacity corrections. This process was repeated until the difference between the 
temperature introduced in the opacity term and that derived was below a certain tolerance (0.1\,K).
At $T_\mathrm{rot} = 1000$\,K, the opacity term $\mathrm{ln}(C_\tau)$ is in general negligible at the 
frequencies covered by the surveys presented and 
$N_\mathrm{u} = N'_\mathrm{u}$. In that sense, assuming an initial rotational temperature of 1000\,K for the calculation of the 
opacity is equal to assume $\tau << 1$. 
An example of the effect of the opacity correction on the rotational diagram is shown in Fig.\,\ref{tau}.

The opacity correction applied can only account for moderate values of the optical 
depth. For high opacities the line emission we detect
comes only from the outer layers of the circumstellar envelope. Because of this, we impose a higher limit to the opacity of 
$\tau = 2$. 

Furthermore, in case the opacity is remarkably high, and if there is a large opacity variation for the different
transitions, the method fails to obtain estimates of  $T_\mathrm{rot}$. As shown by \citet{rtd}, 
when the opacity is high, the rotational diagram given by eq.\,(1) deviates from the straight line expected for low opacities.
The optical depth of the transition does not vary proportionally to the energy of the transition. 
These authors showed that
for linear molecules in LTE conditions the $J_\mathrm{up, \tau max}$ of the transition with the maximum opacity is $J_\mathrm{up, \tau max} = 4.6 
\sqrt{T_\mathrm{rot}(K)/B_\mathrm{o}(\mathrm{GHz})}$ at a frequency of $\nu(\mathrm{GHz})_{\tau max} = 9.13 \sqrt{T_\mathrm{rot}(K) B_\mathrm{o}(\mathrm{GHz})}$.
Therefore, depending on the $T_\mathrm{rot}$ of the gas and of the transitions observed, the opacity effects 
could lead to both higher or lower slopes for the fitting than that expected for low opacities. 
Only in those cases in which we can estimate  the $T_\mathrm{k}$ of the
gas where a particular emission 
arises by any other method, as deriving it from fitting the rotational diagrams of low-abundance isotopologues, can 
we obtain a limit to the column density of a given molecule. 
This can be accomplished for instance by ignoring those lines available
with frequencies closer to $\nu(\mathrm{GHz})_{\tau max}$. 

In case we only observe two transitions,  we can obtain a realistic derivation of 
the value of $T_\mathrm{rot}$ if the opacity from both lines 
is relatively low. If the optical depth significantly 
affects the intensity of the lines, however, we could only obtain a first 
approximation to the value of $T_\mathrm{rot}$ assuming optically thin emission.

We use the molecular distribution of the two shells  described by \citet{cc07}. 
{ As mentioned}, in a similar approach as  \citet{chemyhg}, we assume that the molecular emission from all molecules but
those of CO and $^{13}$CO come from the inner shell. In particular, the diameter of the emitting region would be
11$''$ for CO and $^{13}$CO and 3\secp3 for the rest of the molecules. 

\subsection{Column density and rotational temperature determination}

As expected, the method described above is not well suited for certain molecules because of the high opacity of some
of their transitions. These molecules are HCN, H$^{13}$CN, SiO, $^{29}$SiO, and $^{30}$SiO.

In the particular cases of SiO and their isotopologues, 
the origin of the emission of these molecules was claimed to be related to the presence of a shock front \citep{ccsio}. 
However, recently \citet{teyssier2012} were not able to detect high-$J$ lines with HIFI. In addition, no vibrationally excited lines
were detected in the current survey. This shows that the excitation temperature of the SiO emission is not be particularly
high. Since we cannot estimate the kinetic temperature of the gas where this emission arises, we cannot rely on the method proposed
to derive the abundance and excitation temperature of this molecule, but just to constrain these values. 
We assumed that the emission is optically thin to obtain a
lower limit to the abundance.
Also, since only two lines were detected for the rest of the molecules, which presented opacity effects that avoided the convergence of the
procedure described above (HCN, H$^{13}$CN), 
we  also assumed that their emission is optically thin.

For those molecules for which only a single line was observed, we had to impose a certain value of
$T_\mathrm{rot}$ to estimate the column density.
In case the molecule has an isotopologue for which we could determine the $T_\mathrm{rot}$, 
we assumed the same $T_\mathrm{rot}$ value for both isotopologues. In the rest of the cases,
the value of $T_\mathrm{rot}$ was imposed to be the mean $T_\mathrm{rot}$ value obtained for all the
molecules located in the same region of the envelope, assuming the molecule distribution proposed by \citet{chemyhg}. As mentioned above, we
assume that the emission from all molecules but $^{12}$CO and $^{13}$CO 
arises from the inner shell observed by \citet{cc07}.  We found 
that $\langle T_\mathrm{rot} \rangle = 11.7 \pm 5.7\,K$ for this inner inner
shell.
The column density for these molecules was obtained by modeling the observed spectra assuming LTE 
conditions with the radiative transfer code MADEX.

In the case  lines with hyperfine structures (CN, NO, and NS), we used the LTE approach instead of the rotational diagram analysis in MADEX to reproduce the
observed profiles. Since MADEX computes the line opacity, we were able to estimate whether the column densities obtained
are a lower limit to the value of $N$.

We also confirmed that the results obtained for the column density and $T_\mathrm{rot}$ using the rotational diagrams were
compatible with the results derived adopting the LTE approximation within the MADEX code.
In the particular case of HCO$^{+}$, although only one line was
detected, we could estimate a lower limit to the column density
and an upper limit to the rotational temperature thanks to an upper
limit to the intensity of the $J= 3-2$ transition, lying at 267.557\,GHz.

The rotational diagrams are presented in Fig.\,\ref{rtd} for those species for which we estimated the opacity and in
Fig.\,\ref{rtd2} for those molecules for which we assumed an optically thin regime.
The $N$ and $T_\mathrm{rot}$ results obtained are presented in Table 2. 
{ In this table, the error of the column density 
is only presented
for those species with more that two transitions detected and for which the opacity has not been imposed by a certain value.
For the species with only two transitions detected, the error derived is unrealistically low, as the only source of error in the fitting
of the two points by a line is the error in the intensity of the line, which in general was very low. The species, as SiS or SO, for which
a larger number of species were observed, present more realistic errors and therefore they are prompted in the table. }
{  We used MADEX to generate
synthetic spectra with the parameters presented in table 2 to produce a global estimate of the error in the abundance determination
for the rest of the species. The uncertainty in the determination of the abundance was 
found to be on average $\sim$15\%.}

 \begin{table*}
\caption{Column densities and rotational temperatures. $^*$: tau fixed to a value of 10$^{-3}$. }             
\centering                          
\begin{tabular}{l c c c c c}        
\hline\hline                 
Molecule & $N$ (cm$^{-2}$) & $T_{rot}$ (K) &  $\langle X \rangle$&N. Obs. Line & Comment\\
\hline 
CO              & $2.6 \times 10^{17}$                  &$24.9\pm 4.3$  & $5.3 \times 10^{-4}$    &2      &       \\   
$^{13}$CO       & $3.7 \times 10^{16}$                  &$56.4\pm8.9$   & $7.5 \times 10^{-5}$    &2      &       \\   
SiO             & $>4.8 \times 10^{15}$                 &$10.0\pm2.2$   & $>1.3 \times 10^{-6}$   &3      & $^*$  \\   
$^{29}$SiO      & $>8.3 \times 10^{14}$                 &$9.7\pm2.9$    & $>2.8 \times 10^{-7}$   &3      & $^*$  \\   
$^{30}$SiO      & $>8.1 \times 10^{15}$                 &$9.1\pm0.3$    & $>2.7 \times 10^{-7}$   &3      & $^*$  \\   
Si$^{18}$O      & $1.7  \times 10^{14}$                 &$9.1\pm0.3$    & $5.7 \times 10^{-8}$    &1      &       \\   
HCN             & $>3.4 \times 10^{15}$                 &$>2.4\pm0.1$   & $>1.1 \times 10^{-6}$&2 & $^*$  \\   
H$^{13}$CN      & $>4.2 \times 10^{14}$                 &$>6.0\pm0.3$   & $>1.4 \times 10^{-7}$   &2      & $^*$  \\   
HNC             & $2.9 \times 10^{14}$                  &$11.3\pm0.9$   & $9.7  \times 10^{-8}$   &2      &       \\   
HN$^{13}$C      & $8.0  \times 10^{13}$                 &$11.3\pm0.9$   & $2.7  \times 10^{-8}$&1 &       \\   
CN              & $4.0  \times 10^{15}$                 &$5.0\pm0.5$    & $1.3  \times 10^{-6}$   &2      &       \\   
CS              & $3.7 \times 10^{14}$                  &$7.9\pm0.4$    & $1.2  \times 10^{-7}$   &2      &       \\   
PN              & $1.1  \times 10^{14}$                 &$7.9\pm0.4$    & $3.7   \times 10^{-8}$  &2      &       \\   
NO              & $3.0  \times 10^{16}$                 &$9.0\pm0.1$    & $6.7  \times 10^{-6}$   &1      &       \\
NS              & $5.0  \pm 0.5 \times 10^{15}$         &$8.0\pm0.5$    & $1.7  \pm 0.2 \times 10^{-6}$   &4      &       \\   
SiS             & $2.1  \pm 0.1 \times 10^{14}$         &$20.5\pm2.0$   & $7.0  \pm 0.5 \times 10^{-8}$   &5      &       \\   
SO              & $3.3  \pm 1.2 \times 10^{16}$         &$11.4\pm1.2$   & $1.1  \pm 0.4 \times 10^{-6}$   &9      &       \\  
$^{34}$SO       & $9.0  \times 10^{14}$                 &$11.4\pm1.2$   & $2.4 \times 10^{-7}$&1  &       \\   
SO$_2$          & $1.0  \pm 0.4 \times 10^{15}$         &$22.1\pm5.2$   & $3.4  \pm 1.0 \times 10^{-7}$   &14     &       \\   
HCO$^{+}$         & $>2 \times 10^{13}$                   &$<9.2\pm0.5$   & $>6.7  \times 10^{-9}$&2        &       \\   
N$_2$H$^{+}$      & $4.4  \times 10^{13}$                 &$11.7\pm5.7$   & $1.5  \times 10^{-8}$&1 &       \\   
CH$_3$OH        & $2.1  \times 10^{14}$                 &$8.6\pm0.5$    & $7.0  \times 10^{-8}$&2 &       \\   

\hline                                   
\end{tabular}
\end{table*}
  
\section{Results}

The results obtained for the column density 
in the previous section allow us to understand the characteristics 
of the chemical processes occurring in the CSE around the massive evolved star IRC\,+10420 in greater detail.
To obtain the fractional abundance of each molecule relative to H$_2$, $X$, we
adopted the density profile presented by \citet{cc07}. This density profile allows us to calculate
the column density of H$_2$ at each region of the shell and, therefore, to determine the fractional
abundance of each molecule as $X_\mathrm{mol} = N_\mathrm{mol}/N_\mathrm{H_2}$. 
{ These fractional abundances are also shown in Table\,2.}

We discuss the results for the species detected separately.

\subsection{$^{12}$CO and $^{13}$CO}

In their work, \citet{cc07} 
assumed that to derive the H$_2$ density profile of the envelope around \irc, the fractional abundance of $^{12}$CO in IRC\,+10420 was
the standard abundance found for { AGB stars}, i.e., $3 \times 10^{-4}$.
From this assumption, these authors derived the amount of mass ejected by the star.
Since we used the density profiles obtained by these authors, 
we expected to find similar values for the fractional abundance of \doce\ to those assumed
by \citet{cc07}. 
In fact, the ratio between both $^{12}$CO abundances is $\sim 1.8$, which is higher than that 
derived in this work.
The LVG modeling approach carried out by \citet{cc07} estimates the excitation more accurately than 
the LTE approach. This factor provides an estimate of the error expected by the method used in this work,
compared with more accurate radiative transfer models.

{ The $^{12}$CO  abundance derived here is similar to that obtained toward the RSG VY\,CMa by \citet{ziurys09}
for the red flow, while those abundances obtained by these authors for the spherical flow and blue flow are lower.}

Even though the \doce\ abundance is somehow fixed, and since the opacity effects do not critically affect the density fitting of both $^{12}$CO and $^{13}$CO, we can 
estimate the $^{12}$C/$^{13}$C ratio.
It is found to be $\sim$7, which is slightly below the range found by \citet{cisotopic} for O-rich evolved stars ($^{12}$C/$^{13}$C $= 10-35$).
This was expected in any case, as the hot bottom burning process transforms $^{12}$C in $^{14}$N in massive stars as \irc.

\subsection{CN, HCN, and HNC}

The fractional abundance obtained for CN in this object is $1.3 \times 10^{-6}$.
This abundance is high compared with
O-rich evolved stars ($6.6 \times 10^{-8}$), as already noticed by \citet{cnbachiller}.
This value is 3.4 times higher than that derived in \citet{chemyhg}. In that work we used an approximation to
simplify the hyperfine structure of CN to a simple structure {of two Hund rotational levels 
without hyperfine structure}. This approximation is only reasonable 
for optically thin emissions, however, the approach presented in the current work 
reveals that the hyperfine transitions present moderate opacities. 
Therefore, in this particular case, the most accurate approach to estimate the abundance and rotational temperature of CN is that presented
in this work, {i.e., generating a synthetic spectra to fit the different hyperfine lines.}

Since the formation of CN is mainly produced by the photodissociation of HCN and HNC
due to UV radiation, a high CN value is expected for those objects presenting a hot central
star surrounded by very diluted material, as in the proto-PN phase. While this could be the case for the 
yellow hypergiant IRC\,+10420, the ratio $X_\mathrm{CN}/X_\mathrm{HCN}$, which traces the
UV field, is remarkably low ($\sim$ 0.12). In addition, \citet{Alcolea2013} found a weak emission
of HCN $J$= 13--12, which reveals that HCN is not highly excited in this CSE.

The photodissociation of HCN in IRC\,+10420 might not be very effective because of the high density of the ejected material 
\citep[$n \sim 2 \times 10^4$ cm$^{-3}$,][]{cc07}. 
The reason for the high abundance of CN, as well as for that of HCN, is most probable an enhanced nitrogen abundance 
in the photosphere of the star (see below).

In contrast, the fractional abundance obtained for HNC is $9.7 \times 10^{-8}$.
The opacities obtained for the HNC lines are relatively low (0.2 for HNC $J=1-0$ and 0.5 for HNC $J=3-2$) 
and, therefore, the abundance obtained is probably a good approximation to reality. 
This value is slightly higher than
the mean values obtained for the O-rich AGB stars \citep[][$\langle X_\mathrm{HNC} \rangle = 8.2 \times 10^{-8}$]{bujarrabal94}.

The $^{12}$C/$^{13}$C ratio derived from the abundances obtained for HNC and its $^{13}$C isotopologue 
is $\sim$ 3.6. As mentioned in Sect.\,5.1., this ratio is remarkably low compared with the 
standard value found for the AGB stars, which is consistent with the low value deduced from CO (Sect.\,5.1). 

{ It is also particularly relevant to compare these abundances
with those of the RSG VY\,CMa, as IRC\,+10420 is expected to be a post-RSG object.
The lower limit for the HCN abundance found for \irc\ is similar to the lowest value found
for VY\,CMa \citep{ziurys09}. Also, the HNC abundances found are similar for both objects. 
On the contrary, the abundance found for CN by these authors is $\sim$72 times lower in VY\,CMa than in IRC\,+10420.
This is most probably a sign of the nitrogen enrichment of the photosphere of the star along its
RSG -- post-RSG evolution.}

\subsection{Si-bearing species}

The presence of 
{ SiO in most of the evolved stars (AGBs, RSGs, and YHGs) is believed to be restricted to the inner layers}
of the circumstellar envelopes 
where the dust is
being formed \citep[see, e.g.,][]{lucas92}. The refractory materials, such as silicon, are rapidly attached to the grains that
are formed and removed from the gas. 

\citet{cc07} showed that the amount of molecular gas in the inner regions of the circumstellar envelope
around \irc, where the SiO emission would be expected to arise, is negligible. Because of this, in principle, 
the anticipated intensity of the Si-rich lines would be low. Despite this, theses species are particularly abundant in 
\irc .

In \irc the SiO emission was found to come from a shell located
at $\sim 10^{17}$cm \citep{ccsio}. Because of the large distance from the star where SiO emission is located,
these authors suggested that SiO emission is actually tracing a shocked region, where the dust grains have
been heated, evaporating part of the silicon attached to them.

{ In their study of the molecular emission of IK\,Tau \citet{Gobrecht16} found that while both SiS and SiO are
destroyed by the shock, SiS molecules are reformed in the post-shocked regions with the same abundance as before it is destroyed.
This might indicate that both Si-bearing species can be located in different regions of the ejecta of the evolved stars.}

The lines observed of SiO and its isotopologues are optically thick, as shown by \citet{chemyhg}. Therefore, the values obtained for the 
column densities are lower limits. 
The LTE approximation is not an accurate option for SiO to determine the 
physical conditions of the region where this molecule is located. Solving the radiative transfer equations
using an LVG approximation would be a better option (see Sect.\,6.1).
On the contrary, the SiS emission is only affected by small opacity effects. 

While, as already mentioned, the origin of both SiO and SiS emission is different in the AGB stars and in \irc a comparison of the abundance ratio between both Si-bearing molecules from an O-rich evolved star and \irc\ gives a 
similar result of { $X_\mathrm{SiO}/X_\mathrm{SiS} \gsim 10$}.
{ At the distances from the star at which \citet{ccsio} found the SiO shell, the reaccretion of the SiO in the grains
would be extremely slow because of the dust dilution lasting hundreds of years. Therefore, once these molecules evaporated from the grains, the
SiO/SiS abundance ratio would be expected to be similar to that found in the innermost layers of the star.

\subsection{Cations: N$_2$H$^{+}$ and HCO$^{+}$.}

The formation of these two cations is strongly coupled to the cosmic-ray 
ionization rate of H$_2$ ($\zeta$) via the cation H$_3^{+}$. Chemical modeling 
calculations similar to those carried out by \citet{NO10420} and \citet{sanchezcontreras15} indicate that at 
the low edge of the range of values of $\zeta$ in the Galaxy (10$^{-17}$--10$^{-15}$ s$^{-1}$; 
\citealt{Ionizationrate}), the formation of HCO$^{+}$ in O-rich envelopes is dominated by the reactions
\begin{equation}
\rm CO^{+} + H_2 \rightarrow HCO^{+} + H, \label{reac:co+_h2}
\end{equation}
\begin{equation}
\rm C^{+} + H_2O \rightarrow HCO^{+} + H, \label{reac:c+_h2o}
\end{equation}
where CO$^{+}$ is formed by the reaction between C$^{+}$ and OH. At high values 
of $\zeta$, HCO$^{+}$ is mainly formed by the reaction
\begin{equation}
\rm H_3^{+} + CO \rightarrow HCO^{+} + H_2,
\end{equation}
where H$_3^{+}$ is formed in the reaction between H$_2$ and H$_2^{+}$, where the latter 
is produced by the cosmic-ray ionization of H$_2$. In contrast, the main 
formation route to N$_2$H$^{+}$, independent of the value of $\zeta$, is the reaction

\begin{equation}
\rm H_3^{+} + N_2 \rightarrow N_2H^{+} + H_2.
\end{equation}
These chemical calculations indicate that in IRC~+10420, where there is an 
important nitrogen enhancement \citep{NO10420}, HCO$^{+}$ should be much more 
abundant than N$_2$H$^{+}$ if $\zeta$ is on the order of 10$^{-17}$ s$^{-1}$, 
while if it is approximately 10$^{-15}$ s$^{-1}$ both cations could reach comparable 
abundances. A high ionization rate could be driven by cosmic rays but also 
by soft X-rays emitted by the central stars. In fact, \citet{zhang7027} found 
that the abundance of N$_2$H$^{+}$ in the planetary nebulae NGC\,7027 was abnormally 
high compared with the rest of evolved stars, and this abundance was related to its particularly 
strong X-ray field.
In addition to this, HCO$^{+}$ can also be formed in the presence of an ionizing 
shock front by reactions~(\ref{reac:co+_h2}) and (\ref{reac:c+_h2o}), as found 
by \citet{Rawlings2004}.

In the case of \irc\,we find a lower limit for the abundance of HCO$^{+}$ of 
$6.7 \times 10^{-9}$. This lower limit for the abundance corresponds to the upper limit obtained for 
the HCO$^{+}$ $J= 3-2$ transition. In contrast, the abundance derived for N$_2$H$^{+}$ is 
$1.5 \times 10^{-8}$. The abundance ratio $X_\mathrm{N_2H^{+}}/X_\mathrm{HCO^{+}}$ 
is $\lsim\, 2.2$. This value is very high compared with NGC~7027, where 
$X_\mathrm{N_2H^{+}}/X_\mathrm{HCO^{+}} = 0.07$ \citep{zhang7027}. In fact, the 
N$_2$H$^{+}$ abundance derived in \irc\ is the second highest value found in an evolved 
star. The highest fractional abundance has been found for the extreme massive 
star Eta Carinae \citep[$X_\mathrm{N_2H^{+}} = 2\times 10^{-7}$;][]{Loinard12}.

The high N$_2$H$^{+}$/HCO$^{+}$ abundance ratio found in \irc\ points to a high 
ionization rate driven by cosmic rays or X-rays, although shocks could also 
be at the origin of these two molecules. In the first hypothesis, the formation 
of both cations would be chemically coupled, as H$_3^{+}$ would participate in both 
cases, and we could expect a similar emission distribution for HCO$^{+}$ and N$_2$H$^{+}$. 
Whether the main ionization mechanism is cosmic rays or X-rays is difficult to say. 
Recently, \citet{xrays} did not detect X-rays in IRC\,+10420 using the XMM-Newton 
Space telescope. This does not necessarily imply that X-rays are not an important 
ionizing source in this object as they could be obscured to a large extent by the 
massive envelope. In the second hypothesis, the emission of these two cations would 
tend to follow that of other molecules such as SiO, whose emission in \irc\ 
probably related to a shock front \citep{ccsio}. Additional single-dish observations 
of N$_2$H$^{+}$, in particular $J = 2-1$ and 
$J=3-2$, would allow us to determine the temperature of gas traced by N$_2$H$^{+}$ 
and, therefore,,accurately disentangle its origin. If the N$_2$H$^{+}$ emission is a 
consequence of the interaction of X-rays on the innermost regions of the ejected 
material the kinetic temperature of the gas traced by this molecule should be high, 
while the temperature should be low if its origin were the interaction of the cosmic rays with gas at the outer 
regions of the envelope .

In any case, the particularly high abundance of N$_2$H$^{+}$ found for \irc\ corresponds directly to a high abundance of N$_2$.

\subsection{CH$_3$OH}

We have identified two lines of CH$_3$OH. The number of molecular transitions of this molecule
that fall within the
observed ranges is relatively large (35 lines with excitation temperatures above 500\,K in the 3\,mm
and 47 in the 1\,mm band). 

In order to confirm that the two lines actually correspond to CH$_3$OH, we 
used the column density and rotational temperature derived from the rotational diagrams 
to produce
a synthetic spectra
with the MADEX code, assuming LTE conditions, to reproduce the intensity of the rest of the nondetected CH$_3$OH lines 
from the column density and temperature obtained from the rotational diagram.

{ We found that all the remaining lines either present an intensity that prevented their detection or 
are blended with other more intense lines}. Furthermore, the model fits a feature at 241.8\,GHz, which was not identified 
and was found to be related to the presence of several CH$_3$OH lines in these region of the spectra (see Fig.\,\ref{ch3oh241}).
The identification of these lines confirm the presence of CH$_3$OH in this object and the abundance and temperature derived for this molecule.

{ It has been suggested that the formation of CH$_3$OH might take place in warm 
environments \citep{Hartquist95}, such as  the environment found in the CSE around IRC\,+10420.
}

\begin{figure}[h!]
   \centering
   \includegraphics[angle=0,width=9cm]{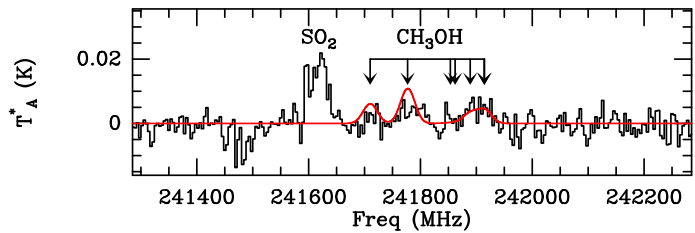}
   \caption[]{Spectral feature that corresponds to CH$_3$OH emission lines in the range  241.8\,GHz.}
              \label{ch3oh241}%
\end{figure}

\subsection{N-rich chemistry}

The most remarkable result of this survey is the detection of a wide number of 
N-bearing molecules. These molecules are HCN, HNC, PN, NS, NO, CN, and N$_2$H$^{+}$ plus some isotopologues.
In addition to these molecules, \citet{Menten1995} detected NH$_3$. 

Our abundance estimates revealed high abundances found for CN, HCN, HNC, and N$_2$H$^{+}$ when compared with
O-rich evolved stars \citep[see, e.g.,][]{bujarrabal94}. In particular, \irc\ presents{ one of the highest} N$_2$H$^{+}$
abundance found for an evolved star.

These facts could be explained by an enrichment in nitrogen due to the hot bottom burning process \citep[HBB;][]{boothroyd93} 
which, as mentioned, is present for stars with
masses above $\sim$3 \ms. 
For the most massive and evolved stars, the nitrogen enrichment is expected
to lead to a N-rich chemistry. In addition, the $^{12}$C/$^{13}$C isotopic ratio is therefore expected to be
low, as has been found for \irc\ (see Sect.\,5.1\&5.2).

Recently, \citet{NO10420} has shown that to reproduce the NO
profiles, the initial abundance of nitrogen in the photosphere of the star has to be
significantly enhanced with respect to the standard values expected for the O-rich evolved stars,
i.e., the enhancement expected as a result of the HBB is confirmed.

{ We obtained a lower limit to the column density of P$^{15}$N to compare the $^{14}$N/$^{15}$N isotopic ratio with that found for
N-rich Nova CK\,Vul by \citet{kaminsky15}. The lower limit of the integrated intensity was calculated
following \citet{chemyhg}. We choose this molecule as it does not present hyperfine transitions and since the PN emissions
detected are optically thin. We found an upper limit for the integrated intensity $W < 4 \times 10^{-3}$ K\,\kms, which results
in a column density of $N < 4 \times 10^{13}$ cm$^{-2}$. The upper limit of the isotopic ratio so derived is $^{14}$N/$^{15}$N $>$ 3.
This low value is similar to the range found by \citet{kaminsky15} for CK\,Vul ($4-26$). Deeper integrations toward selected 
$^{15}$N-bearing molecules would allow us to confirm this ratio.}

A paper devoted to a detailed study of the N-rich chemistry in \irc\ is under preparation.

\begin{figure*}[h!]
   \centering
   \includegraphics[angle=0,width=14cm]{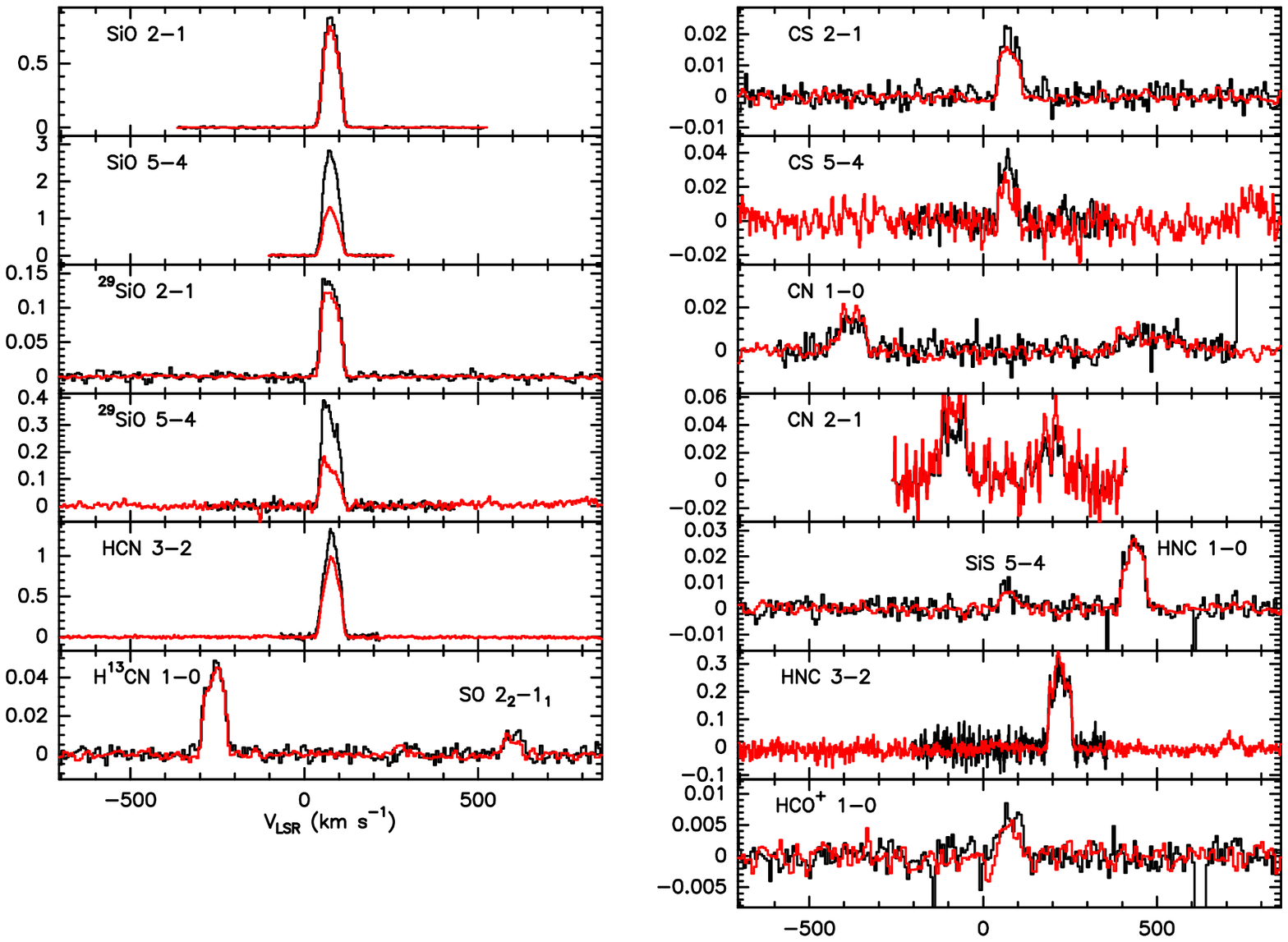}
   \caption[]{Comparison between the spectra observed by \citet{chemyhg} (black lines) and
   that are presented in this work (red lines). The temperatures are in mean beam temperature scale. 
}
              \label{Comp}%
\end{figure*}

\begin{figure*}[h!]
   \centering
   \includegraphics[angle=0,width=14cm]{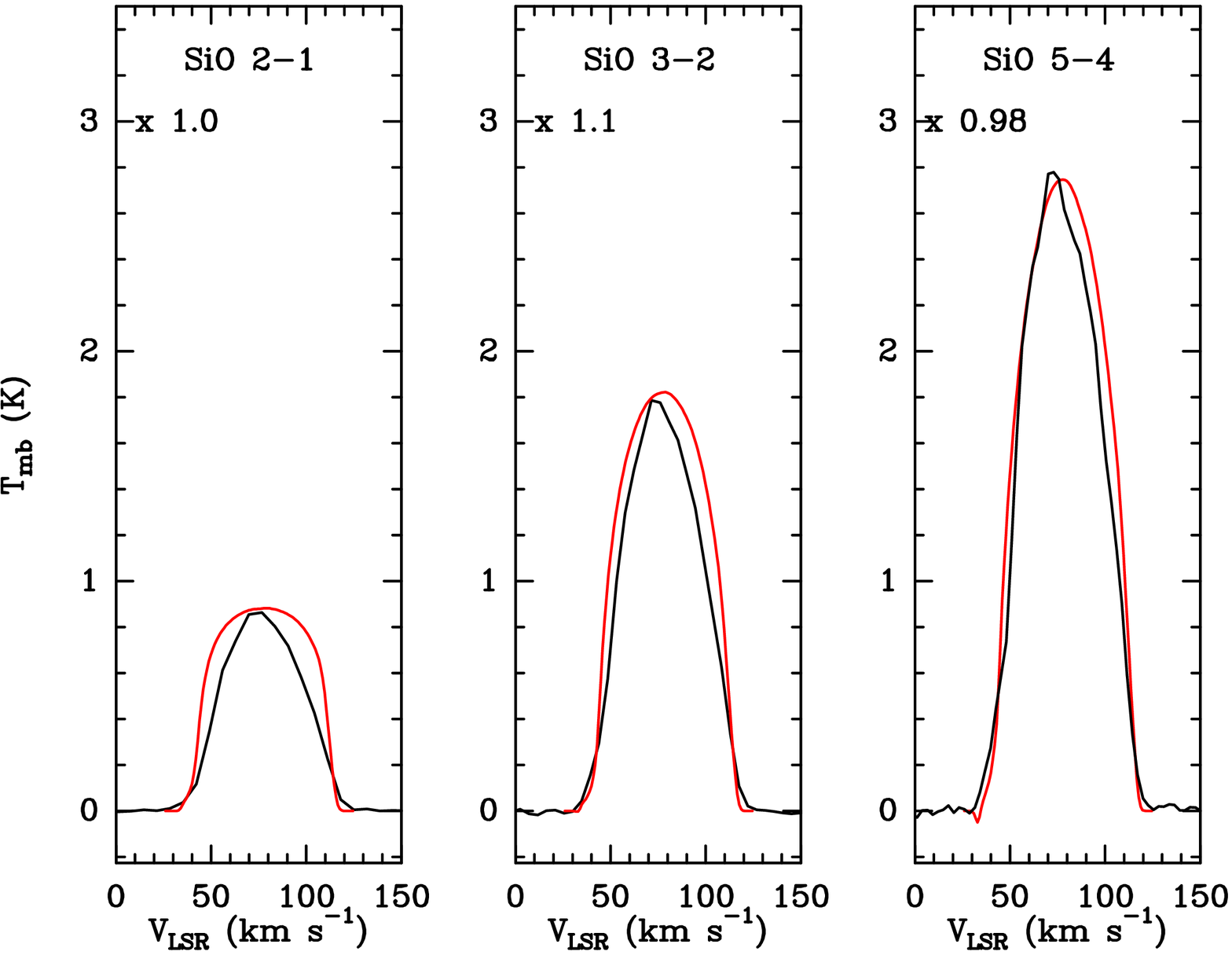}
   \caption[]{SiO line fitting for the observations presented in \citet{chemyhg}. 
   The temperature derived for the dust shell is 750\,K.
   The line fitting is depicted by a red line. In the upper left corner
   we present the calibration factor applied to the observed spectra.}
              \label{Comp2000}%
\end{figure*}

\begin{figure*}[h!]
   \centering
   \includegraphics[angle=0,width=14cm]{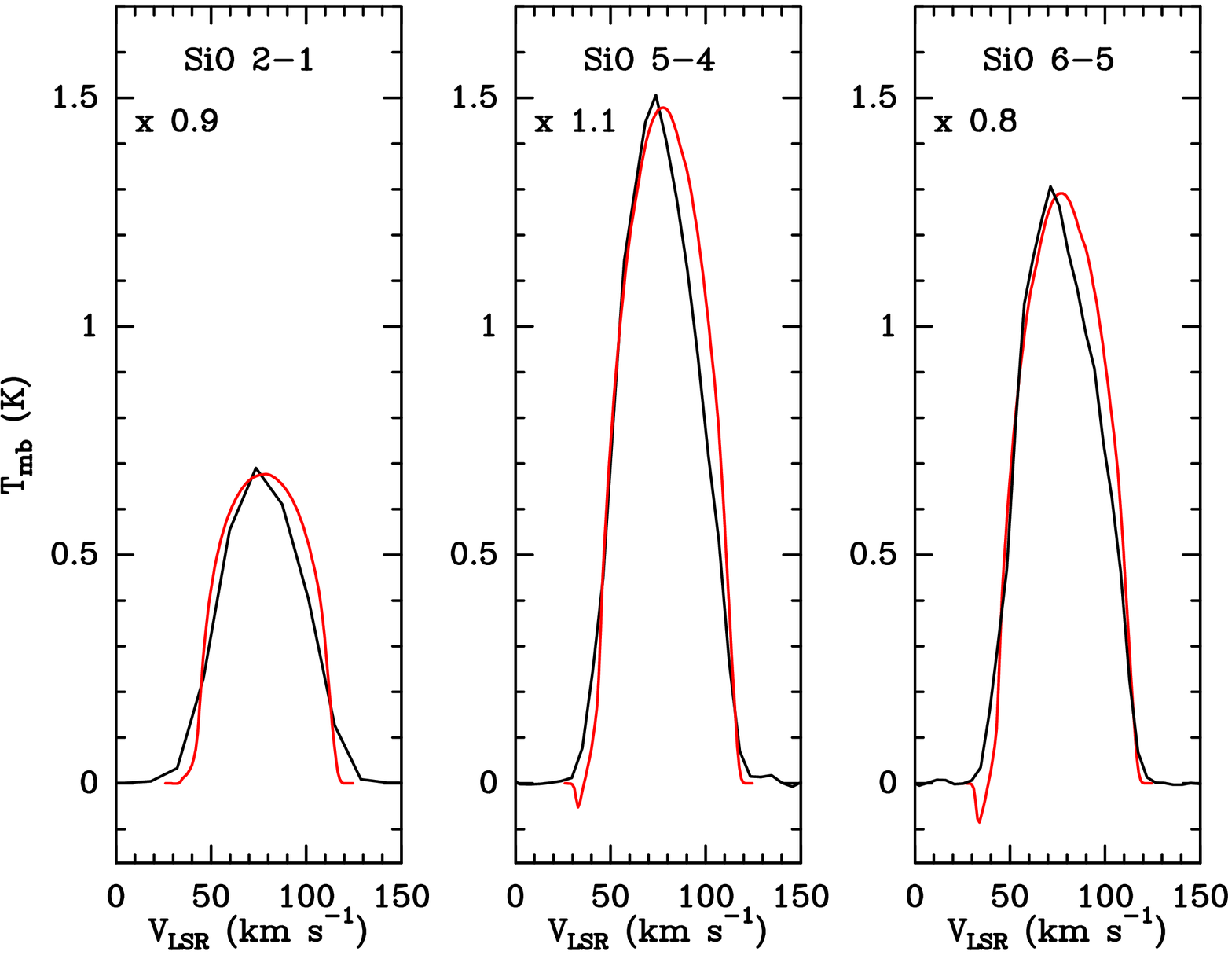}
   \caption[]{SiO line fitting for the observations presented in this work. 
   The temperature derived for the dust shell is 550\,K.
   The line fitting is depicted by a red line. In the upper left corner
   we present the calibration factor applied to the observed spectra.}
              \label{Compnow}%
\end{figure*}

\section{Time variations of the SiO lines}

The abundances we found for SiO and $^{29}$SiO are $\sim$9 and $\sim$4 times lower that those found by \citet{chemyhg},
respectively. 

A careful comparison between the line profiles obtained by \citet{chemyhg} and those presented here reveal 
that while the intensities of most of the molecular lines do not present significant changes, the $J$=5--4 lines of SiO and $^{29}$SiO are
much weaker (see Fig.\ref{Comp}).

The yellow hypergiants are known to present large variations in their effective temperatures \citep{nieuwen00}.
These changes in their effective temperatures would result in a remarkable variations in the emission of certain 
lines due to the IR pumping. SiO lines are particularly affected by this effect. A decrease in the $T_{\mathrm{eff}}$
of IRC\,+10420 would result in a decrease of the SiO emission, especially for the high-$J$ lines.

A second possible explanation for the variation in the SiO line emission is that the silicon evaporated from the 
dust grains during the shock is reattached to the grains. However, { as mentioned above,} at the large distances where \citet{ccsio} located
the SiO emission, the re-absorption of SiO would last hundreds of years.

\citet{nieuwen00} found that the $T_{\mathrm{eff}}$ of IRC\,+10420 showed an increase of $\sim$ 3000\,K in the 
last 20 years, reaching a value of 8500 in 1997.
These authors predicted the effective temperature of \irc\ to grow continuously in its evolution toward blue regions
of the HR diagram. The decrease in the intensity of the high-$J$ lines can confirm this scenario, as the dust ejected 
during the last encounter with the yellow void dilutes into the ISM.

In order to obtain an accurate view of the properties of the SiO emission, we used an LVG approximation to 
model the lines observed. We adopted the extent of the SiO shell from \citet{ccsio}, the density
profile from \citet{cc07}, and the SED model from \citet{Dinh09}.
We found that we are unable to reproduce the SiO emission from IRC\,+10420 assuming the 
mass distribution derived by \citet{cc07}. 

\citet{WongThesis} found a similar difficulty with the observed SiO $J$ = 1 -- 0 VLA maps. This author suggested that 
the emission of SiO is located in clumps, which were ejected from localized regions of the photosphere of the star. 
\citet{WongThesis} argue therefore that the properties of these clumps are not tied with the global mass ejection traced by CO. 
This author suggested a two-shell model to reproduce the SiO $J=1-0 $ and $J=2-1$ emission maps. These maps consisted of a radial 
density law, $n_\mathrm{H_2} = 2.3 \times 10^{4}$ cm$^{-3} \times (r/10^{17}$cm)$^{-0.7}$, and an abundance profile, which present two different 
regions with a constant SiO abundance: $X_\mathrm{SiO} = 1.25 \times 10^{-6}$ between $5 \times 10^{16}$ cm and $2.35 \times 10^{17}$ cm, and 
$X_\mathrm{SiO} = 6 \times 10^{-8}$ between $2.35 \times 10^{17}$ cm and $4.2 \times 10^{17}$ cm.
As already mentioned, the IR emission from the star has an important impact on the intensities of the SiO lines via the
infrared pumping. 
The emission from the star is obscured by the optically thick dust layers. To reproduce the infrared emission from the star,
\citet{WongThesis} assumed that 
the continuum source consisted in an optically thick shell of dust at 400\,K and at a radius of 8 10$^{15}$cm. This assumption was based on
the flux observed at 8$\mu{m}$ by \citet{Jones1993} in March 1992. 
However, the IR flux from this object is known to vary and the low-$J$ SiO lines are less sensitive to 
this variation. Therefore, while \citet{WongThesis} fitted the observations with a dust temperature of 400\,K, a higher temperature could
have also fitted the observations.

While the flux observed by \citet{Jones1993} at 8 $\mu{m}$ is $\sim$1000\,Jy, the flux observed with ISO in 2001 at this same wavelength
is only $\sim 350$\,Jy\footnote{Data obtained from the ISO data archive http://iso.esac.esa.int/ida/}. In fact, \citet{egan2003} showed that
in 1997 the flux at 8.28 $\mu{m}$ was 148.7\,Jy. This confirms the large flux variations that this object has undergone. 
Therefore, \citet{WongThesis} probably overestimated the 8 $\mu{m}$ flux in their calculations of the abundance and temperature.

Taking into account these variations, we used the ISO observations from 
year 2001 to fit the SiO lines presented by \citet{chemyhg}
observed in year 2000. The flux of 350\,Jy can be simulated, following 
\citet{WongThesis}, by a blackbody with a radius of 8 10$^{15}$cm, 
and a temperature of 331\,K. We also assumed the gas distribution of 
\citet{WongThesis}.
Simultaneously we fitted the observations presented in the current work. 
We left as free parameter the IR flux from the star for the 2012 observations, and 
allowed minor modifications to the
SiO fractional abundance of both regions suggested by \citet{WongThesis}.

We found that in order to fit the two sets of data, we needed to decrease the SiO fractional 
abundance in the inner region to 10$^{-5}$. The temperature of the dust shell 
obtained for the new set of data is 200\,K, which corresponds to a 8 $\mu{m}$ flux
of $\sim$8\,Jy. 

This reveals that an accurate knowledge on the 
variations of the IR flux of these objects is crucial to obtain 
realistic abundance estimates of species, such as
SiO, whose emission is strongly correlated with the IR emission.

The YHGs are known to present two main pulsation periods \citep{dejager98}. These consist of a
short period of several hundred of days corresponding to 
a quiescent pulsation \citep{PercyKim14,lecoroller03}, and a larger period of 
tens of years in which these objects present a quasi-explosive mass ejection \citep{dejager98}

A detailed study on the flux variation of this object, as well as high angular resolution observations of the SiO 
distribution, are essential to understand the SiO emission observed in IRC\,+10420.
Also, future SiO observations are fundamental to confirm whether the source of the IR variation 
is related to the expansion of the dust shell, which would result in a constant 
decrease of this flux, or to the quiescent pulsation or a new explosive mass ejection 
period, in case future observations present an increase of the SiO lines.
The first scenario would support the idea that \irc\ is finally crossing the yellow void, 
while the latter would suggest that this object is undergoing a new
encounter with the yellow void.

Future observations of the SiO emission lines{, in addition to bolometric observations at 8 $\mu{m}$} 
would allow us track this decrease in the IR flux from the dust{ and to better constrain the properties of the SiO emitting region.}
A new increase in the intensity of high-$J$ SiO emission lines could also be related to new mass ejections.

\section{Conclusions}

We have performed a full line survey in the atmospheric windows at 1 and 3\,mm for the yellow
hypergiant IRC\,+10420. We detected a total of 22 species. We have obtained LTE abundances for these species.
These abundances showed a particularly high value of all N-bearing species compared with standard O-rich evolved
stars. In particular, the abundance of N$_2$H$^{+}$ is found to be the second highest found for an evolved star.
These results seem to confirm the nitrogen enrichment found by \citet{NO10420}.

In addition to this, the isotopic $^{12}$C/$^{13}$C ratio is particularly low ($\sim$7--3.6, see Sects.\,5.1\&5.2). 
{ This difference cannot be explained by the errors in the fitting of the abundance (see Sect.4.1), therefore this
variation in the isotopic ration seems to be genuine.} 
Both effects are expected
for massive stars in which hot bottom burning process takes place, transforming $^{12}$C into $^{14}$N.

Also, the nitrogen enrichment and the $^{13}$C/$^{12}$C ratio increase with time as the hot bottom burning
is active \citep{Marigo2007}. The outer regions of the envelope, where only CO was detected, trace gas ejected between 1200 and 3000 
years before the ejection 
of the innermost shell \citep{cc07}. Therefore we might expect a highest $^{12}$C/$^{13}$C. The differences in the ratios derived from CO and HNC might
be caused by opacity effects. However, for high opacities we might expect higher abundances of $^{12}$CO, and therefore a lower
$^{12}$C/$^{13}$C ratio. Most probably, the difference in the ratios found for both regions are due to the continuous effect of the HBB along 
the last 3800 years. High angular resolution maps of optically thin emission lines of $^{12}$C and $^{13}$C species could confirm this result.

High-$J$ SiO line intensities observed in this paper are remarkably low compared with the same transitions observed by
\citet{chemyhg}, while the variation of the low-$J$ SiO lines observed is low. This variation was also observed in HCN and CS, while
HNC, CN, SiS, and HCO$^{+}$ line intensity have remained constant.
The effective temperature of the star has a direct effect on the SiO emission via infrared pumping 
\citep[see, e.g.,][]{cerni2014c}.
The SiO line intensity variation probably corresponds to a recent decrease in the IR flux emitted from the 
dust, which has a direct effect on the intensity of the SiO lines via infrared pumping.
\irc\ has been claimed to be evolving blueward in the HR diagram crossing the yellow void, i.e., 
to present a constant increase in its $T_\mathrm{eff}$.
The changes in the intensity of the SiO lines revealed a decrease in IR flux from the dust around the star.
Tracking the evolution of the IR flux of this object is important to determine its evolutionary status, i.e., if the source
is actually crossing the yellow void or if it is undergoing a new mass ejection.
{ Future SiO observations in combination with IR 8 $\mu{m}$ observations would allow us to trace the temperature of the dust
around evolved stars in particular around
massive evolved stars, which will allow us to disentangle their evolutionary status and to constrain the 
properties of the SiO emitting region. }

\begin{acknowledgements}
The research leading to these results has received funding from the European Research Council
under the European Union's Seventh Framework Programme (FP/2007-2013) / ERC Grant
Agreement n. 610256 (NANOCOSMOS). 
We would also like to thank the Spanish MINECO for funding support from grants CSD2009-00038,
AYA2009-07304 \& AYA2012-32032.
\end{acknowledgements}

\bibliographystyle{aa} 
\bibliography{new} 

\appendix

\Online 
\section{Line properties}

\onecolumn
\longtab{2}{
\begin{longtable}{lccccc}
\caption{Line transitions detected at 3\,mm. H: Lines presenting hyperfile structure.}\\
\hline\hline
$\nu_{rest}$ (GHz)      &Molecule       &Transition     &\tas (mK) &Area (K km/s) &Notes\\    
\hline
\endfirsthead
\caption{continued.}\\
\hline\hline
\hline
\endhead
\hline
\endfoot
83.688093      &SO$_2$          &$8_{1,7}$--8$_{0,8}$   &12.5 $\pm$1.2 &0.55  &       \\
84.746170      &$^{30}$SiO      &2--1                   &84.0 $\pm$1.3 &4.89  &       \\
85.759199      &$^{29}$SiO      &2--1                   &103.0 $\pm$0.8 &6.13  &       \\
86.093950      &SO              &2$_2$--1$_1$           &8.9 $\pm$1.27 &0.4  &       \\
86.339922      &H$^{13}$CN      &1--0                   &37.4 $\pm$1.2 &2.4  &       \\
86.846960      &SiO             &2--1                   &666.2 $\pm$0.9 &34.32  &       \\
88.631602      &HCN             &1--0                   &234.2 $\pm$0.9 &14.6  &       \\
89.188525      &HCO$^{+}$ &1--0                   &3.9 $\pm$1.3 &0.24  &       \\
90.663568      &HNC             &1--0                   &21.9 $\pm$0.9 &1.29  &       \\
90.771564      &SiS             &5--4                   &4.4 $\pm$ 0.8  & 0.287           &              \\
93.173392      &N$_2$H$^{+}$      &1--0                   &5.9 $\pm$0.3 &0.42  &       \\
93.979770      &PN              &2--1                   &7.0 $\pm$1.0 &0.46  &       \\
96.741375      &CH$_3$OH        &2--1                   &3.2 $\pm$0.6 &0.15  &       \\
97.715317      &$^{34}$SO       &2$_3$--1$_2$           &5.1 $\pm$0.5 &0.27  &       \\
97.980953      &CS              &2--1                   &13.9 $\pm$0.3 &0.825  &       \\ 
99.299870      &SO              &2$_3$--1$_2$           &89.2 $\pm$0.6 &3.84  &       \\
104.029418     &SO$_2$          &3$_{1,3}$--2$_{0,2}$   &39.1 $\pm$0.6 &1.71  &       \\
104.239295     &SO$_2$          &10$_{1,9}$--10$_{0,10}$&8.6 $\pm$0.3 &0.43  &       \\
108.924301     &SiS             &6--5                   &9.2 $\pm$ 0.7  &  0.392         &               \\
109.252220     &SO              &3$_2$--$2_1$           &12.7 $\pm$1.3 &0.65  &       \\
110.201354     &$^{13}$CO       &$1-0$                  &57.1 $\pm$ 1.4 &3.42           &            \\
113.123370     &CN              &1$_{1/2,1/2}$--0$_{1/2,1/2}$   &16.5 $\pm$1.9 &3.48  &  H     \\
113.144157     &CN              &1$_{1/2,1/2}$--0$_{1/2,3/2}$   &--     &--           &  H         \\
113.170492     &CN              &1$_{1/2,3/2}$--0$_{1/2,1/2}$   &--     &--           &  H         \\
113.191279     &CN              &1$_{1/2,3/2}$--0$_{1/2,3/2}$   &--     &--           &  H         \\
113.488120     &CN              &1$_{3/2,3/2}$--0$_{1/2,1/2}$   &--     &--           &  H         \\
113.490970     &CN              &1$_{3/2,5/2}$--0$_{1/2,3/2}$   &--     &--           &  H         \\
113.499644     &CN              &1$_{3/2,1/2}$--0$_{1/2,1/2}$   &--     &--           &  H         \\
113.508907     &CN              &1$_{3/2,3/2}$--0$_{1/2,3/2}$   &--     &--           &  H         \\
113.520432     &CN              &1$_{3/2,1/2}$--0$_{1/2,3/2}$   &--     &--           &  H         \\
115.153935     &NS              &$\Pi^{+}$ 5/2$_{7/2}$--3/2$_{5/2}$       &30.6 $\pm$2.7 &2.09  &   H    \\
115.156812     &NS              &$\Pi^{+}$ 5/2$_{5/2}$--3/2$_{3/2}$       &--     &--           &  H         \\
115.162982     &NS              &$\Pi^{+}$ 5/2$_{3/2}$--3/2$_{1/2}$       &--     &--           &  H         \\
115.185336     &NS              &$\Pi^{+}$ 5/2$_{3/2}$--3/2$_{3/2}$       &--     &--           &  H         \\
115.191456     &NS              &$\Pi^{+}$ 5/2$_{5/2}$--3/2$_{5/2}$       &--     &--           &  H         \\
115.271202      &CO             &1--0                           & 657.5$\pm$ 0.9 & 42.8 &       \\
115.556253     &NS              &$\Pi^-$ 5/2$_{7/2}$--3/2$_{5/2}$       &21.8 $\pm$3.0 &1.52  &   H    \\
115.570763     &NS              &$\Pi^-$ 5/2$_{5/2}$--3/2$_{3/2}$       &--     &--           &  H         \\
115.571954     &NS              &$\Pi^-$ 5/2$_{3/2}$--3/2$_{1/2}$       &--     &--           &  H         \\
\end{longtable}
}

\longtab{4}{
\begin{longtable}{l c c c c c}
\caption{Line transitions detected at 1\,mm. H: Lines presenting hyperfile structure. 8 MHz: Spectral
resolution degraded to 8\,Mhz to increase the S/N ratio.}\\
\hline\hline
$\nu_{rest}$ (GHz)      &Molecule       &Transition     &\tas (mK) &Area (K km/s) &Notes\\    
\hline
\endfirsthead
\caption{continued.}\\
\hline\hline
$\nu_{rest}$ (GHz)      &Molecule       &Transition     &\tas (mK) &Area (K km/s) &Notes\\    
\hline
\endhead
\hline
\endfoot
199.672229      &SiS            &11--10         &24 $\pm$ 2     &  0.463         &              \\
201.751489      &Si$^{18}$O     &5--4           &10.6 $\pm$2.5 &0.40  &       \\ 
203.391550      &SO$_2$         &12$_{0,12}$--11$_{1,11}$       &17.2 $\pm$2.0 &0.82  &       \\
206.176005      &SO             &$5_4$--$4_3$           &40.3 $\pm$2.17 &1.8  &       \\
207.436051      &NS             &$\Pi^{+}$ 9/2$_{11/2}$--7/2$_{9/2}$      &48.8 $\pm$3.4 &1.68  &   H    \\
207.436636      &NS             &$\Pi^{+}$ 9/2$_{7/2}$--9/2$_{7/2}$       &--     &--           &  H         \\
207.438692      &NS             &$\Pi^{+}$ 9/2$_{7/2}$--7/2$_{5/2}$       &--     &--           &  H         \\
207.566407      &U              &--                     &12.0 $\pm$3.7 &0.50  &      \\ 
207.777535      &NS             &$\Pi^-$ 9/2$_{9/2}$--7/2$_{9/2}$       &53.2 $\pm$3.2 &1.65  &   H    \\
207.792951      &NS             &$\Pi^-$ 9/2$_{7/2}$--7/2$_{7/2}$       &--     &--           &  H         \\
207.834866      &NS             &$\Pi^-$ 9/2$_{11/2}$--7/2$_{9/2}$      &--     &--           &  H         \\
207.838365      &NS             &$\Pi^-$ 9/2$_{9/2}$--7/2$_{7/2}$       &--     &--           &  H         \\
208.700336      &SO$_2$         &3$_{2,2}$--2$_{2,1}$   &50.1 $\pm$3.8 &2.40  &       \\
211.853474      &$^{30}$SiO     &5--4                   &169.7 $\pm$4.1 &8.71  &       \\
214.385758      &$^{29}$SiO     &5--4                   &118.7 $\pm$4.7 &5.67  &       \\
215.220653      &SO             &$5_5$--$4_4$           &21.9 $\pm$4.9 &1.28  &       \\
217.104980      &SiO            &5--4                   &865.8 $\pm$4.2 &42.37  &       \\
217.817663      &SiS            &12--11                 &10 $\pm$ 2     &0.334          &         \\ 
218.440050      &CH$_3$OH       &4--3                   &9.2 $\pm$2.0 &0.20  &       \\
219.949442      &SO             &$5_6$--$4_5$           &138.6 $\pm$3.1 &5.70  &       \\
220.398684      &$^{13}$CO      &2--1                   &315 $\pm$3 &19.4      &         \\
221.965220      &SO$_2$         &11$_{1,11}$--10$_{0,10}$       &22.6 $\pm$2.8 &1.05  &       \\
226.616571      &CN             &2$_{3/2,1/2}$--1$_{1/2,3/2}$&37.2 $\pm$2.8 &3.29  &  H     \\
226.632190      &CN             &2$_{3/2,3/2}$--1$_{1/2,3/2}$&--        &--           &  H         \\
226.659558      &CN             &2$_{3/2,5/2}$--1$_{1/2,3/2}$&--        &--           &  H         \\
226.663693      &CN             &2$_{3/2,1/2}$--1$_{1/2,1/2}$&--        &--           &  H         \\
226.679311      &CN             &2$_{3/2,3/2}$--1$_{1/2,1/2}$&--        &--           &  H                 \\
226.874191      &CN             &2$_{5/2,5/2}$--1$_{3/2,3/2}$&--        &--           &  H         \\
226.874781      &CN             &2$_{5/2,7/2}$--1$_{3/2,5/2}$&--        &--           &  H         \\
226.887420      &CN             &2$_{5/2,3/2}$--1$_{3/2,3/2}$&--        &--           &  H         \\
226.892128      &CN             &2$_{5/2,5/2}$--1$_{3/2,5/2}$&--        &--           &  H         \\
226.905357      &CN             &2$_{5/2,3/2}$--1$_{3/2,5/2}$&--        &--           &  H         \\
230.538000      &CO             &2--1                   & 2.511 10$^{3}$ $\pm$ 3.2     & 159          &         \\
234.935695      &PN             &6--5                   &13.6 $\pm$2.9 &0.45  &       \\ 
235.151720      &SO$_2$         &4$_{2,2}$--3$_{1,3}$   &59.1 $\pm$2.1 &2.58  &       \\
235.961363      &SiS            &13--12                 &13.5 $\pm$3.5 &0.43  &       \\
241.615797      &SO$_2$         &5$_{2,4}$--4$_{1,3}$   &20.3 $\pm$1.9 &1.04  &       \\
244.365156      &U              &--                     &9.2 $\pm$2.9 &0.61  &  8MHz     \\ 
244.935557      &CS             &5--4                   &13.5 $\pm$3.5 &0.72  &       \\ 
250.436848      &NO             &$\Pi^{+}$ 5/2$_{7/2}$--3/2$_{5/2}$                       &22.8 $\pm$2.4 &1.49  &  H     \\
250.440659      &NO             &$\Pi^{+}$ 5/2$_{5/2}$--3/2$_{3/2}$                       &--     &--           &  H         \\
250.448530      &NO             &$\Pi^{+}$ 5/2$_{3/2}$--3/2$_{1/2}$                       &--     &--           &  H         \\
250.475414      &NO             &$\Pi^{+}$ 5/2$_{3/2}$--3/2$_{3/2}$                       &--     &--           &  H         \\
250.482939      &NO             &$\Pi^{+}$ 5/2$_{5/2}$--3/2$_{5/2}$                       &--     &--           &  H         \\
250.796436      &NO             &$\Pi^-$ 5/2$_{7/2}$--3/2$_{5/2}$                       &19.0 $\pm$2.3 &1.51  &  H    \\
250.815594      &NO             &$\Pi^-$ 5/2$_{5/2}$--3/2$_{3/2}$                       &--     &--           &  H         \\
250.816954      &NO             &$\Pi^-$ 5/2$_{3/2}$--3/2$_{1/2}$                       &--     &--           &  H         \\
251.199675      &SO$_2$         &13$_{1,13}$--12$_{0,12}$&24.0 $\pm$2.5 &1.12  &       \\
251.826156      &SO             &$6_{5}$--$5_{4}$               &35.9 $\pm$3.0 &1.06  &       \\
251.912005      &U              &3--2   &7.0 $\pm$2.2 &0.09  &       \\ 
253.968393      &NS             &$\Pi^-$ 11/2$_{13/2}$--9/2$_{11/2}$                    &38.3 $\pm$2.7 &1.39  &  H     \\
253.970581      &NS             &$\Pi^-$ 11/2$_{11/2}$--9/2$_{9/2}$                     &--     &--           &  H         \\
254.216656      &$^{30}$SiO     &6--5                   &115.3 $\pm$3.7 &4.70  &       \\
254.280536      &SO$_2$         &6$_{3,3}$--6$_{2,4}$   &14.4 $\pm$3.6 &0.31  &       \\ 
255.553302      &SO$_2$         &4$_{3,1}$--4$_{2,2}$           &14.1 $\pm$2.9 &0.51  &       \\
255.958044      &SO$_2$         &3$_{3,1}$--3$_{2,2}$           &10.3 $\pm$2.6 &0.74  &       \\
256.246945      &SO$_2$         &5$_{3,3}$--5$_{2,4}$           &26.0$\pm$3.7 &0.89  &       \\
257.099966      &SO$_2$         &7$_{3,5}$--7$_{2,6}$           &12.9 $\pm$2.4 &0.85  &       \\
257.255216      &$^{29}$SiO     &6--5                           &179.4 $\pm$2.6 &8.59  &       \\
258.255826      &SO             &6$_{6}$--5$_{5}$               &2.7 $\pm$3.3 &1.62  &       \\
259.011821      &H$^{13}$CN     &3--2                           &80.0 $\pm$3.3  &4.11   &       \\ 
260.518020      &SiO            &6--5                           &959 $\pm$2.2   &47.9   &       \\
261.259318      &HN$^{13}$C     &3--2                           &15.7   $\pm$4.8        &0.78   &       \\
261.843721      &SO             &6$_7$ -- 5$_6$                 &77.5   $\pm$6.0        &2.58         &       \\
265.886180      &HCN            &3--2                           &529.3  $\pm$4.1        &24.1   &       \\
271.529014      &SO$_2$         &7$_{2,6}$--6$_{1,5}$           &19.9   $\pm$3.8        &0.823  &       \\
271.981142      &HNC            &3--2                           &167.6  $\pm$4.6        &8.24   &       \\
\end{longtable}
}

\twocolumn

\begin{figure*}[h!] 
\centering 
\includegraphics[angle=0,width=12cm]{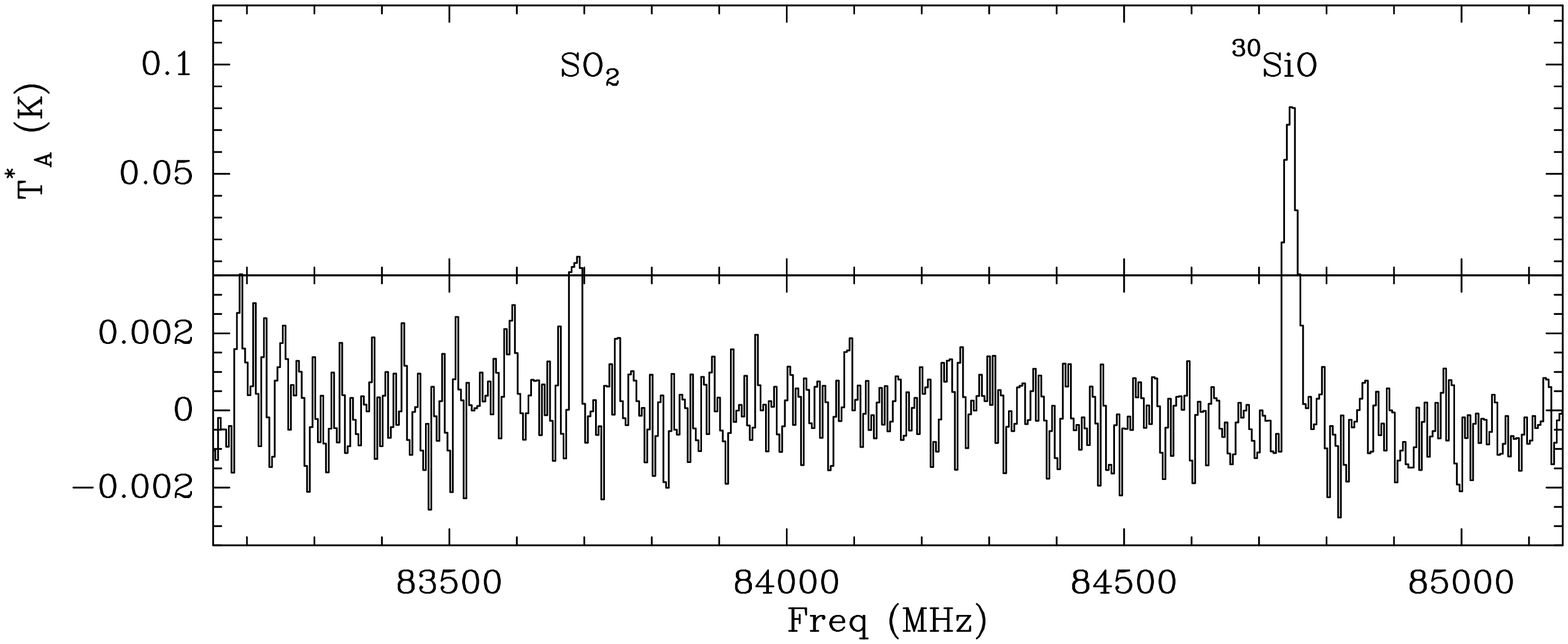} 
\includegraphics[angle=0,width=12cm]{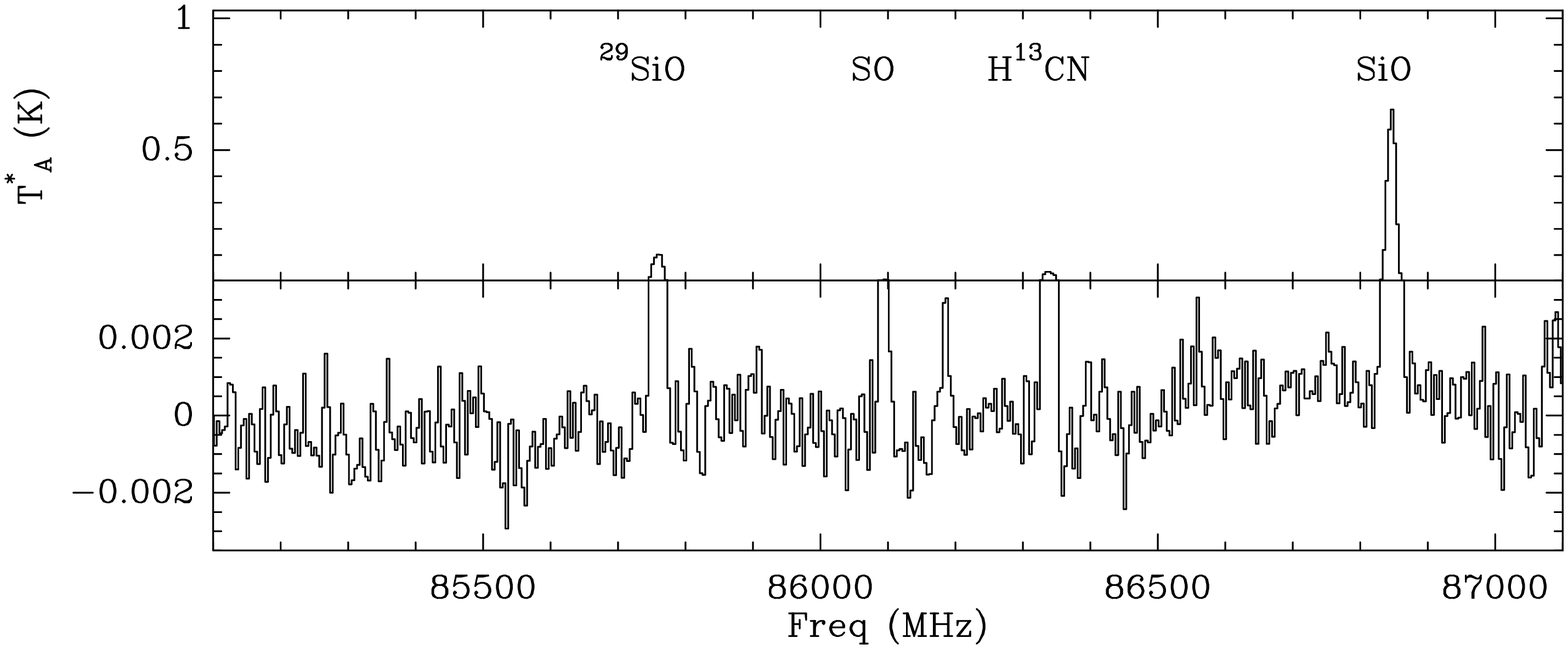} 
\includegraphics[angle=0,width=12cm]{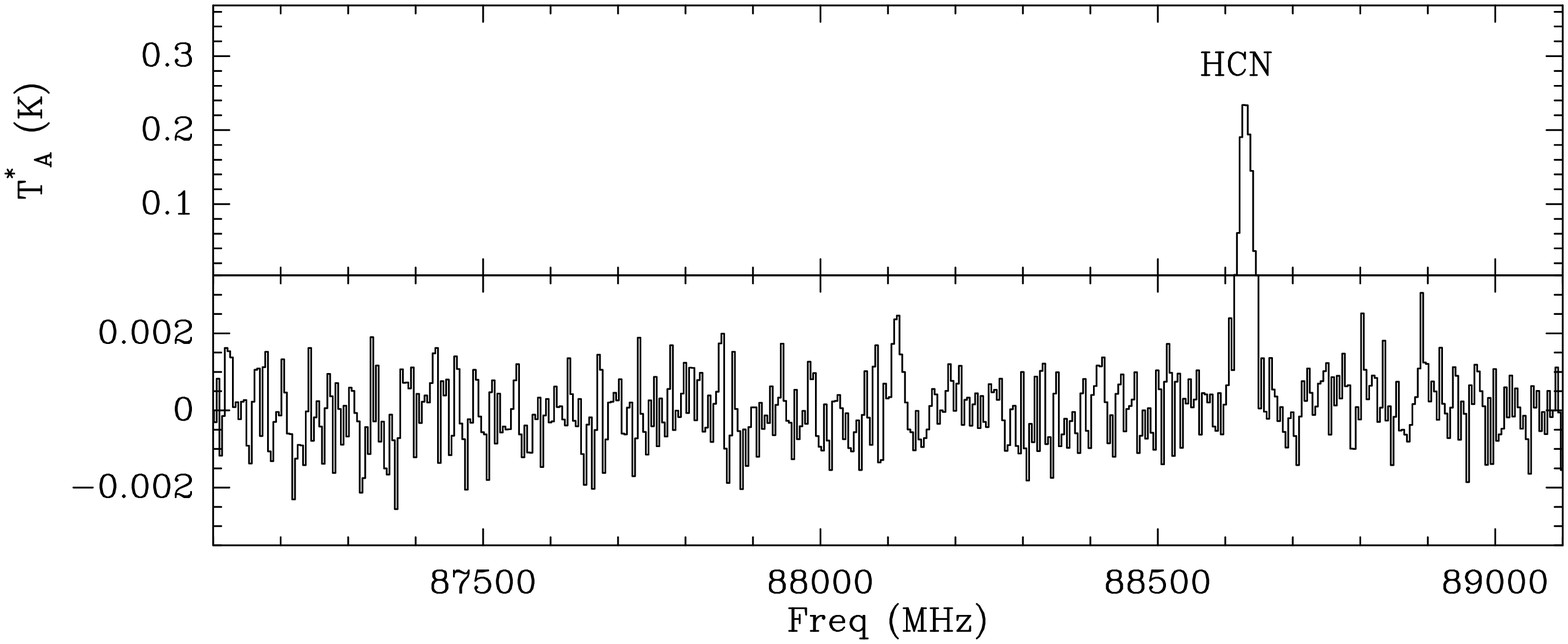} 
\includegraphics[angle=0,width=12cm]{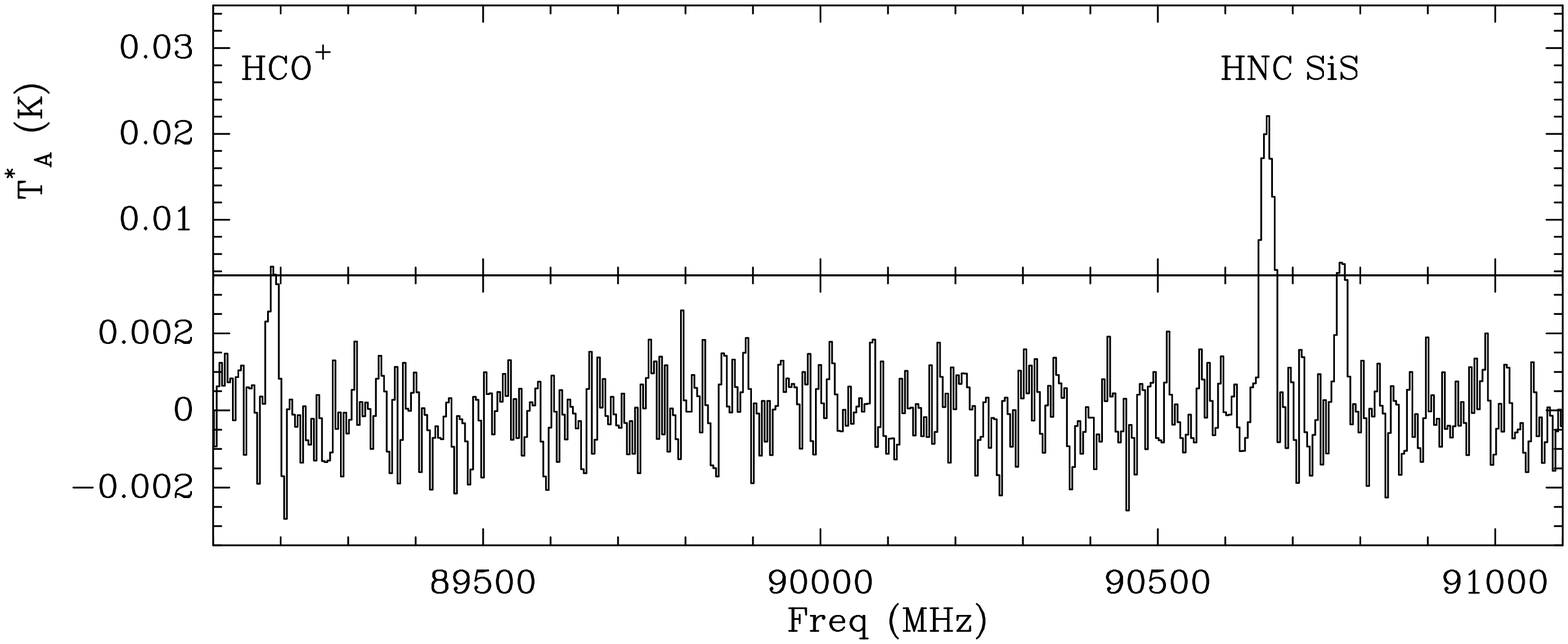} 
\caption{Spectral survey obtained with the IRAM 30 m telescope at the atmospheric window of 3\,mm. The $\rm{v_{LSR}}$ of of \irc used to calculate the frequencies
   of the spectra is 76\,km/s. } 
\label{Fig3mm}%
\end{figure*} 
\begin{figure*}[h!] 
\centering 
\ContinuedFloat 
\includegraphics[angle=0,width=12cm]{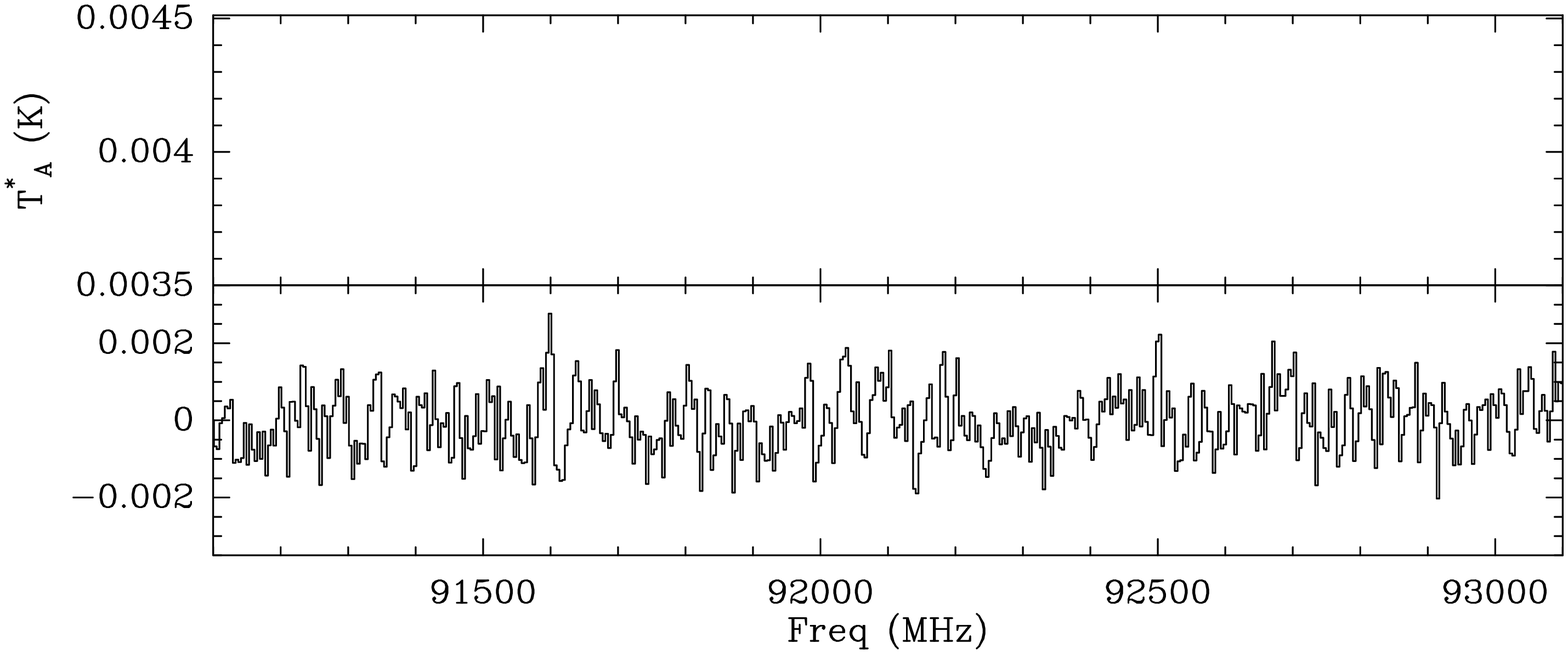} 
\includegraphics[angle=0,width=12cm]{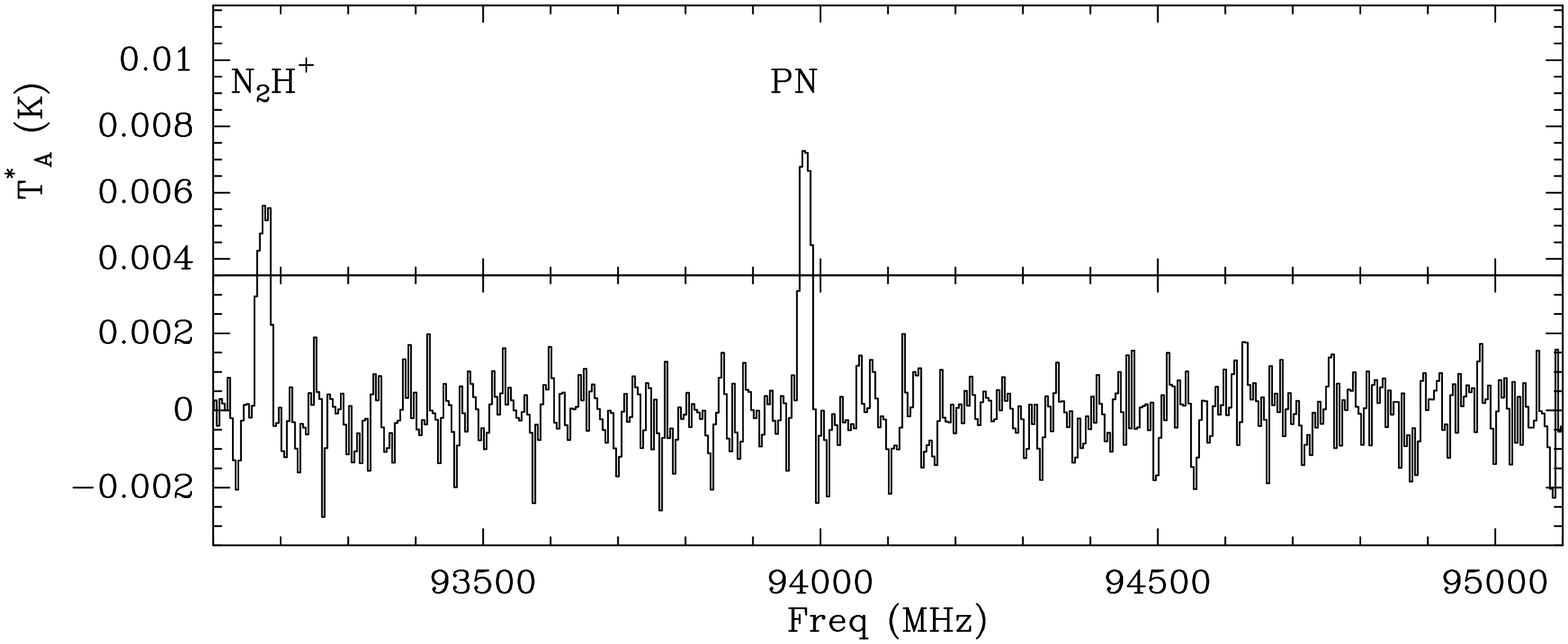} 
\includegraphics[angle=0,width=12cm]{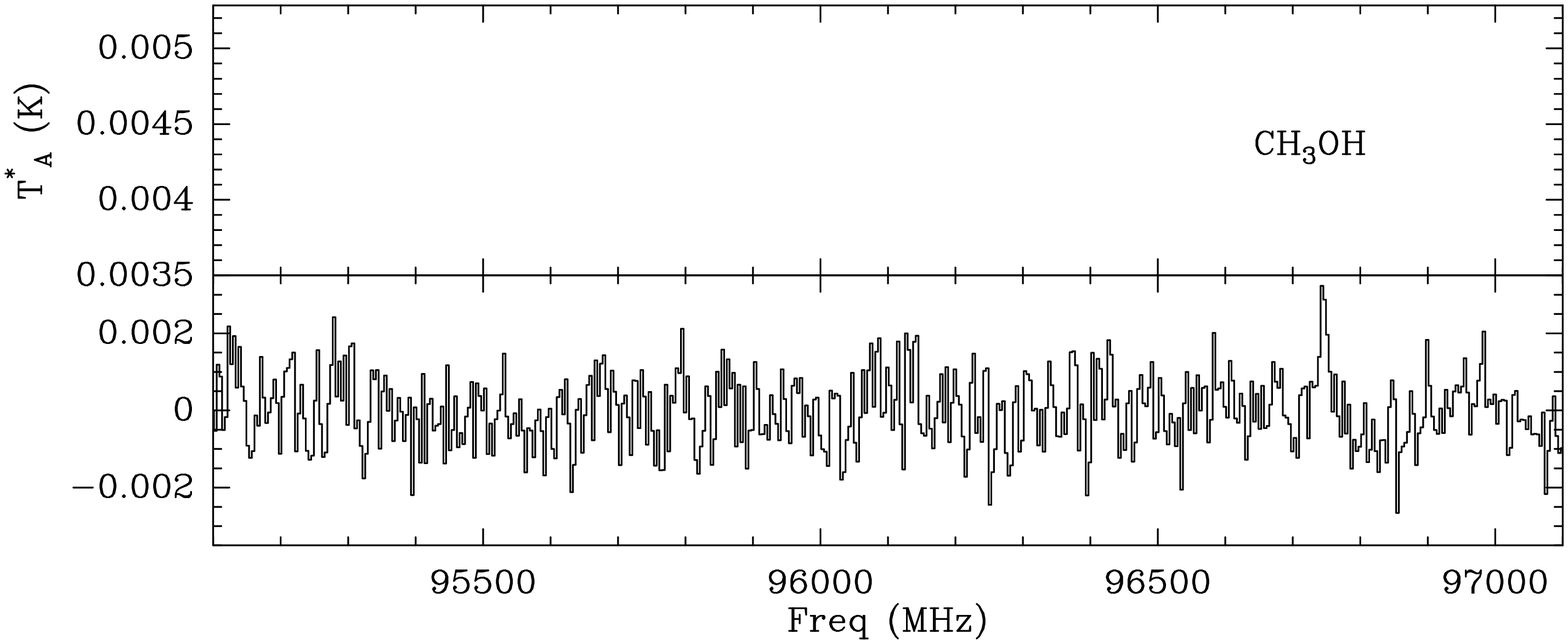} 
\includegraphics[angle=0,width=12cm]{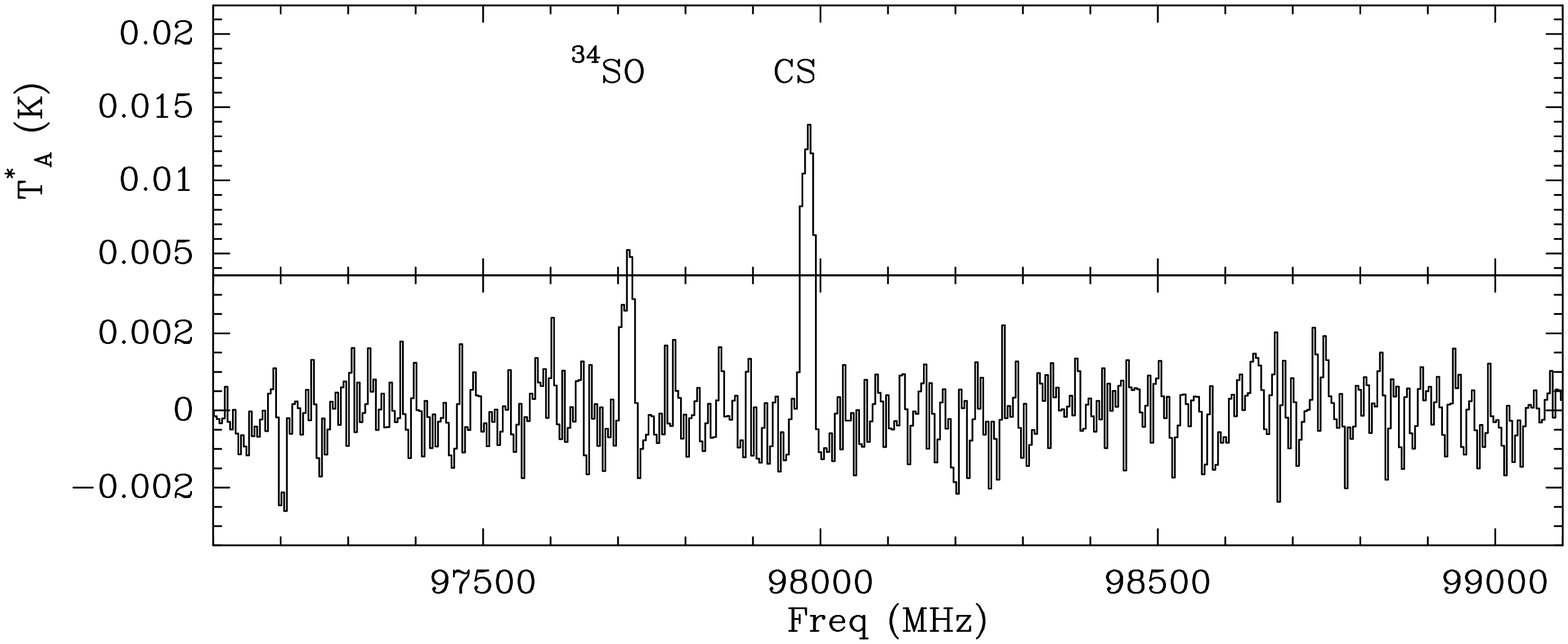} 
\caption{. (continued) } 
\label{Fig3mm}%
\end{figure*} 
\begin{figure*}[h!] 
\centering 
\ContinuedFloat 
\includegraphics[angle=0,width=12cm]{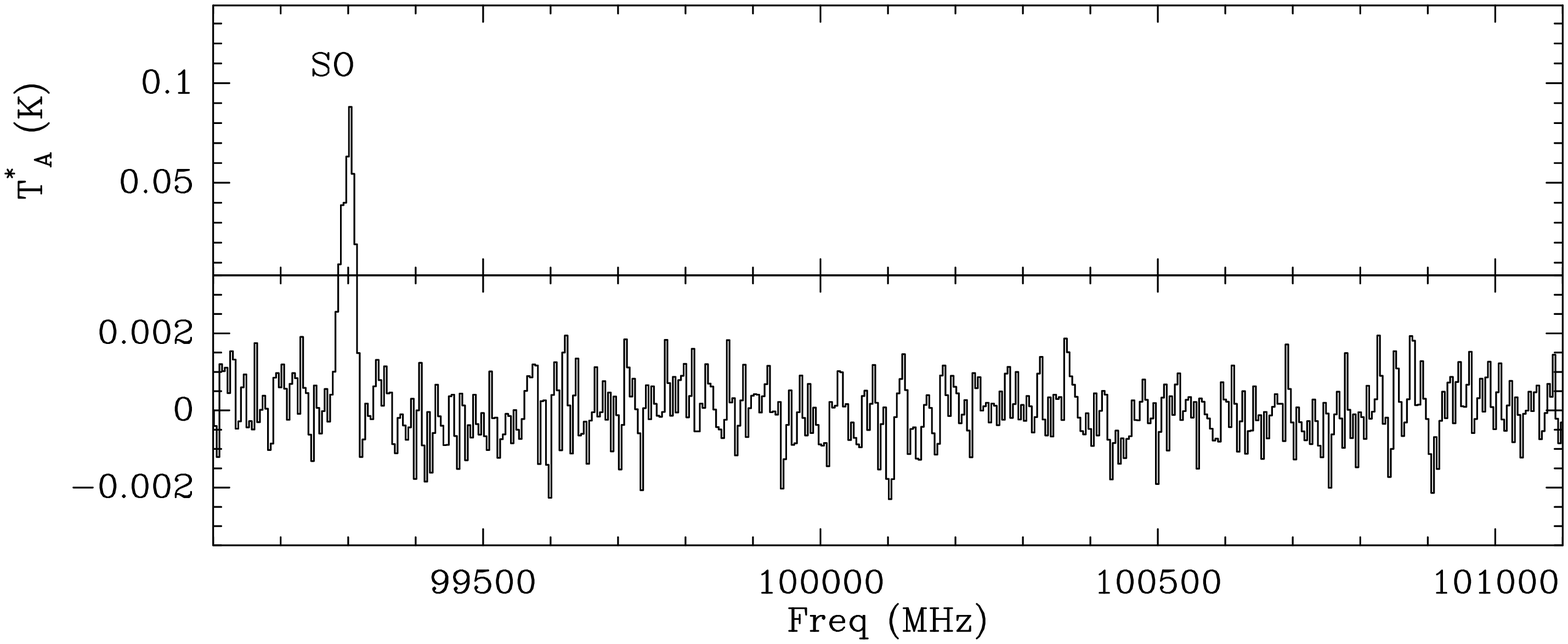} 
\includegraphics[angle=0,width=12cm]{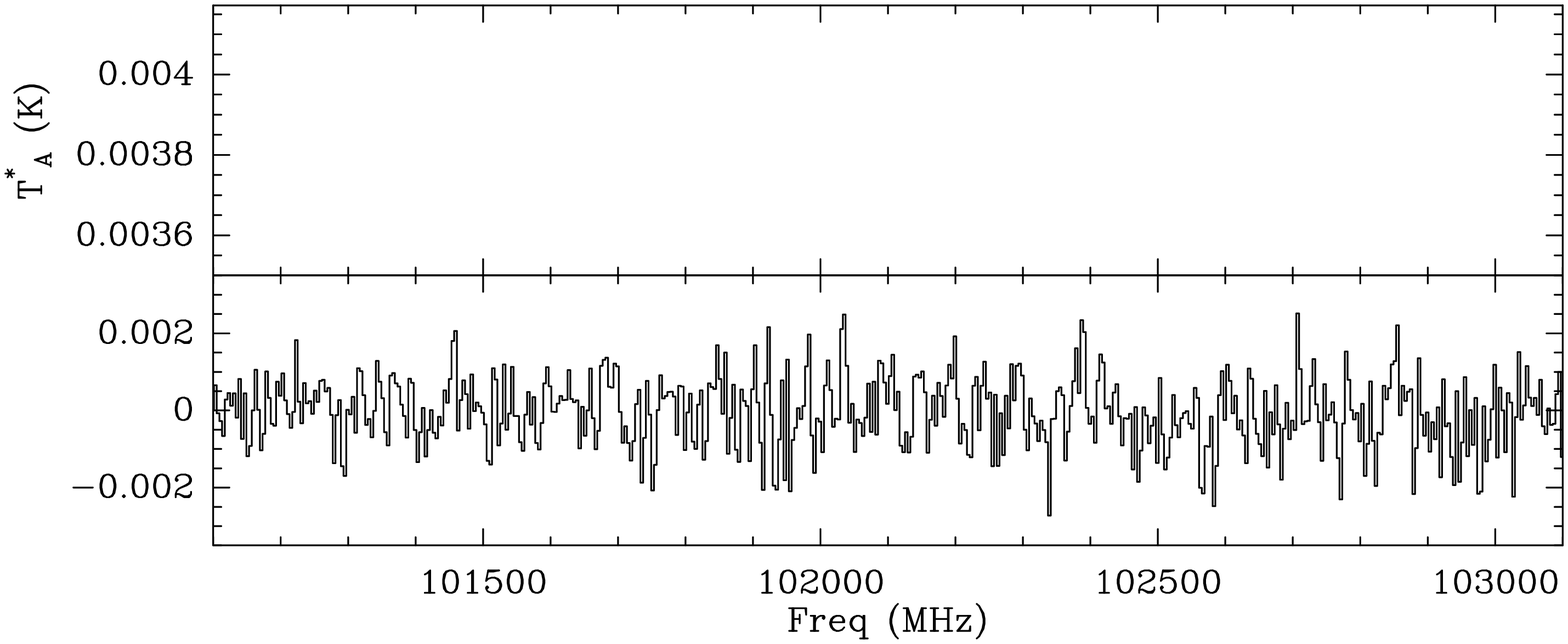} 
\includegraphics[angle=0,width=12cm]{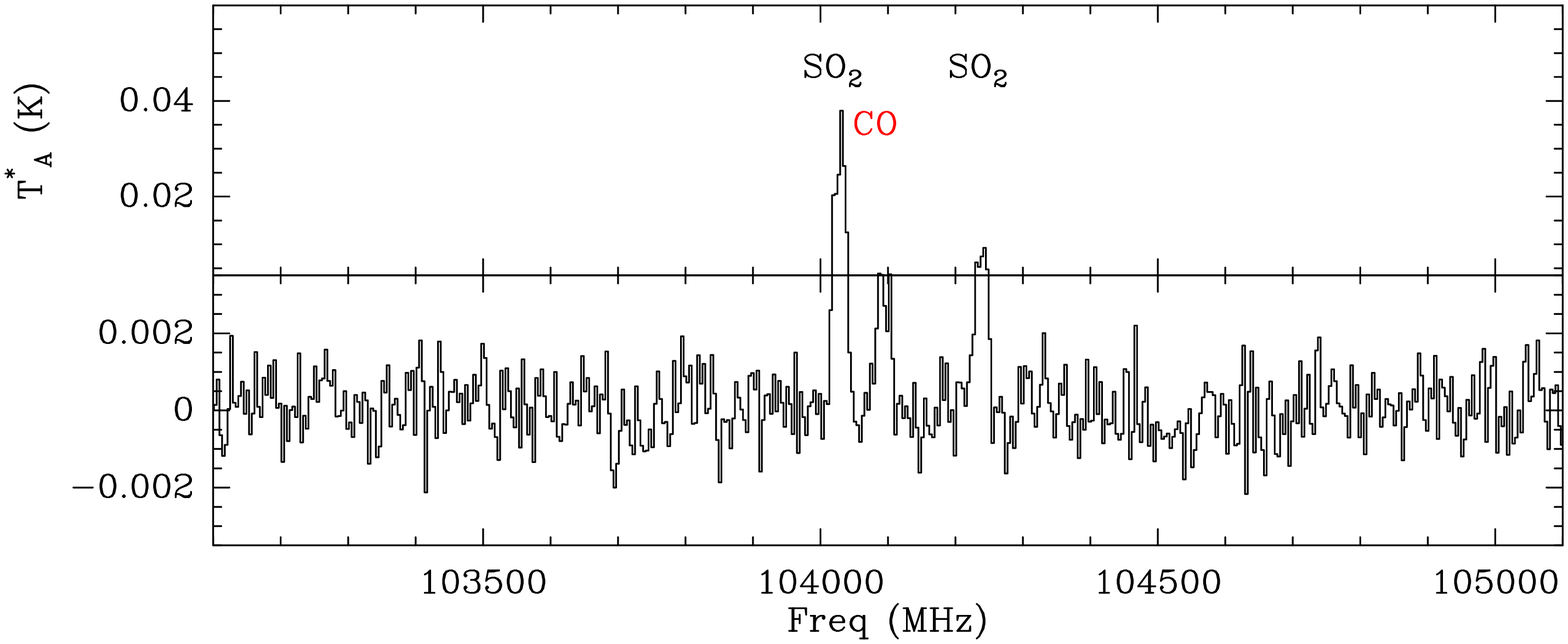} 
\includegraphics[angle=0,width=12cm]{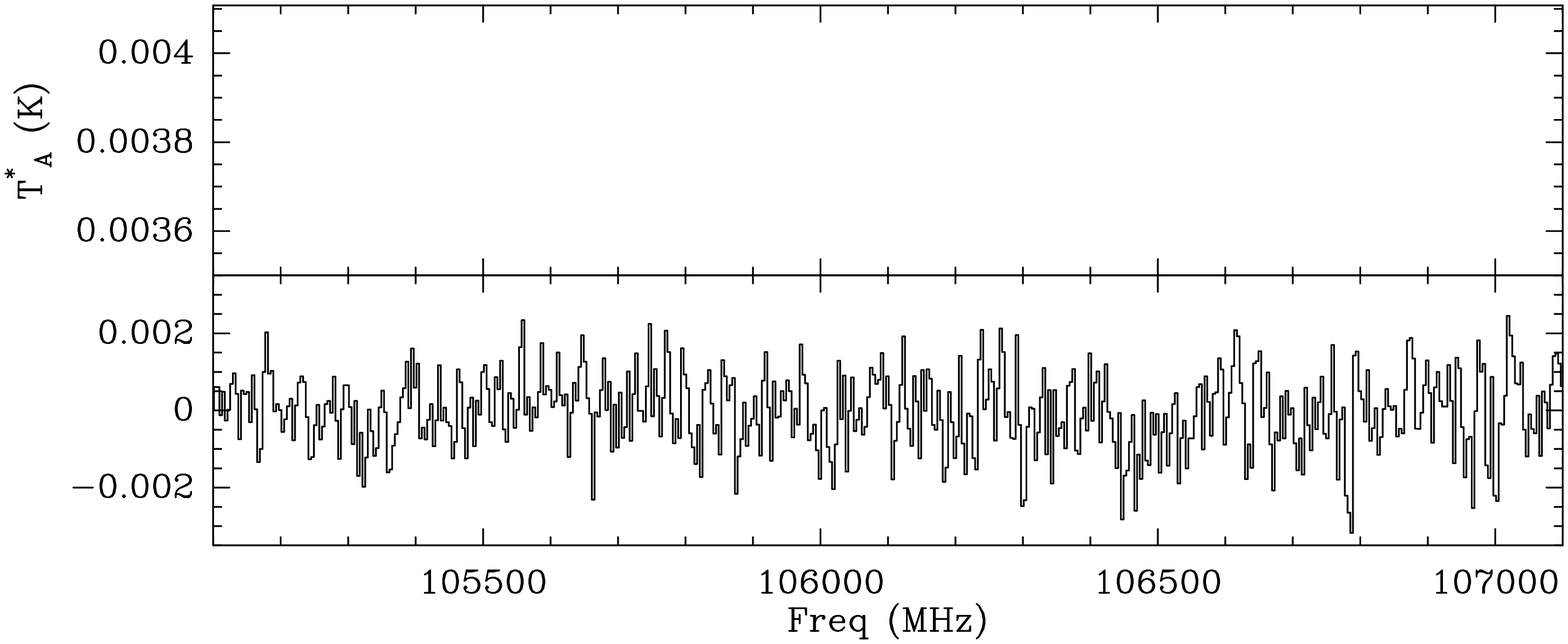} 
\caption{. (continued) } 
\label{Fig3mm}%
\end{figure*} 
\begin{figure*}[h!] 
\centering 
\ContinuedFloat 
\includegraphics[angle=0,width=12cm]{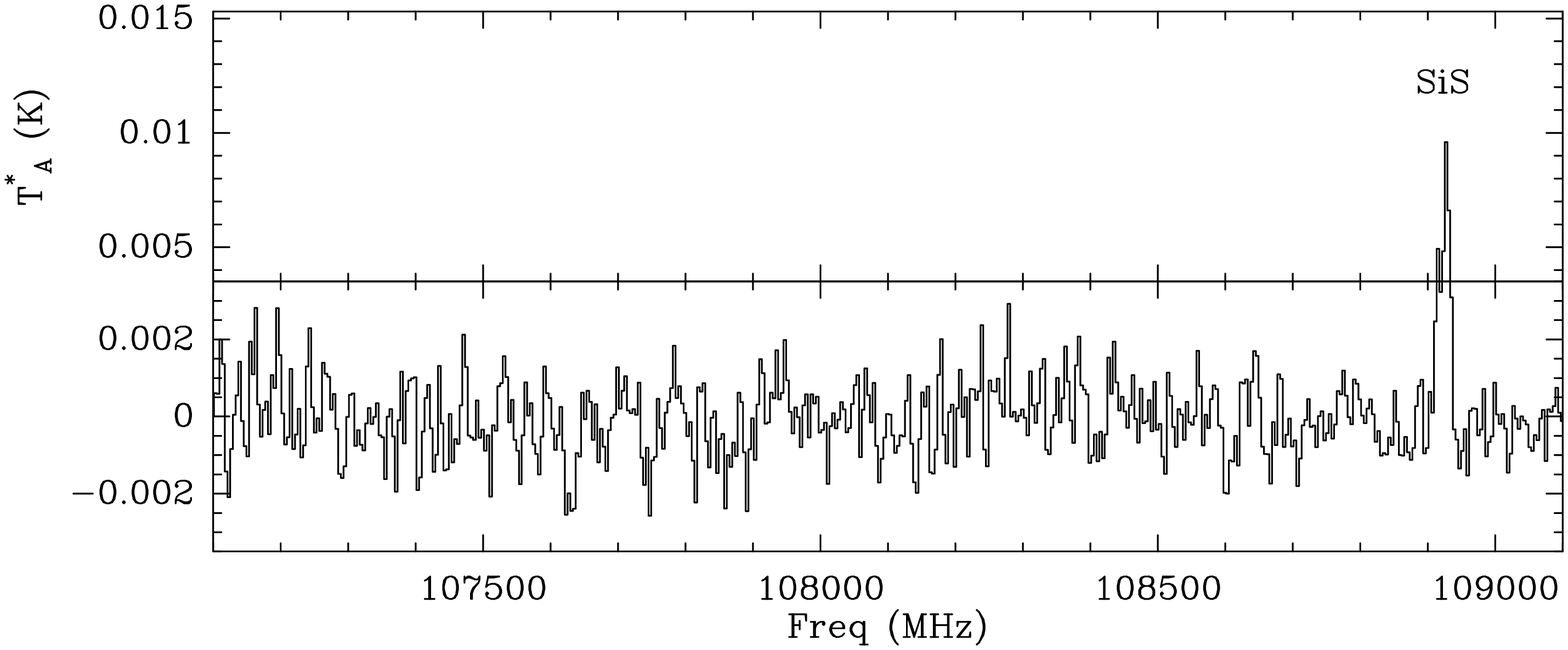} 
\includegraphics[angle=0,width=12cm]{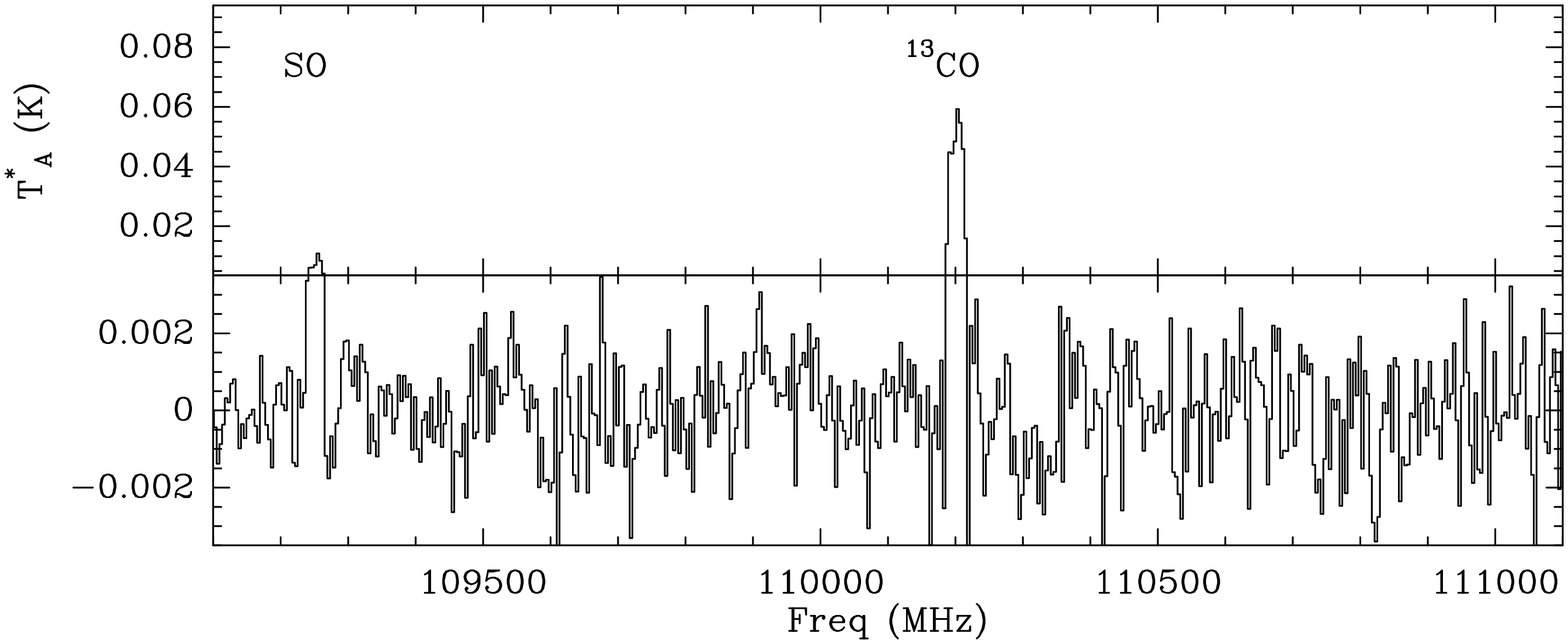} 
\includegraphics[angle=0,width=12cm]{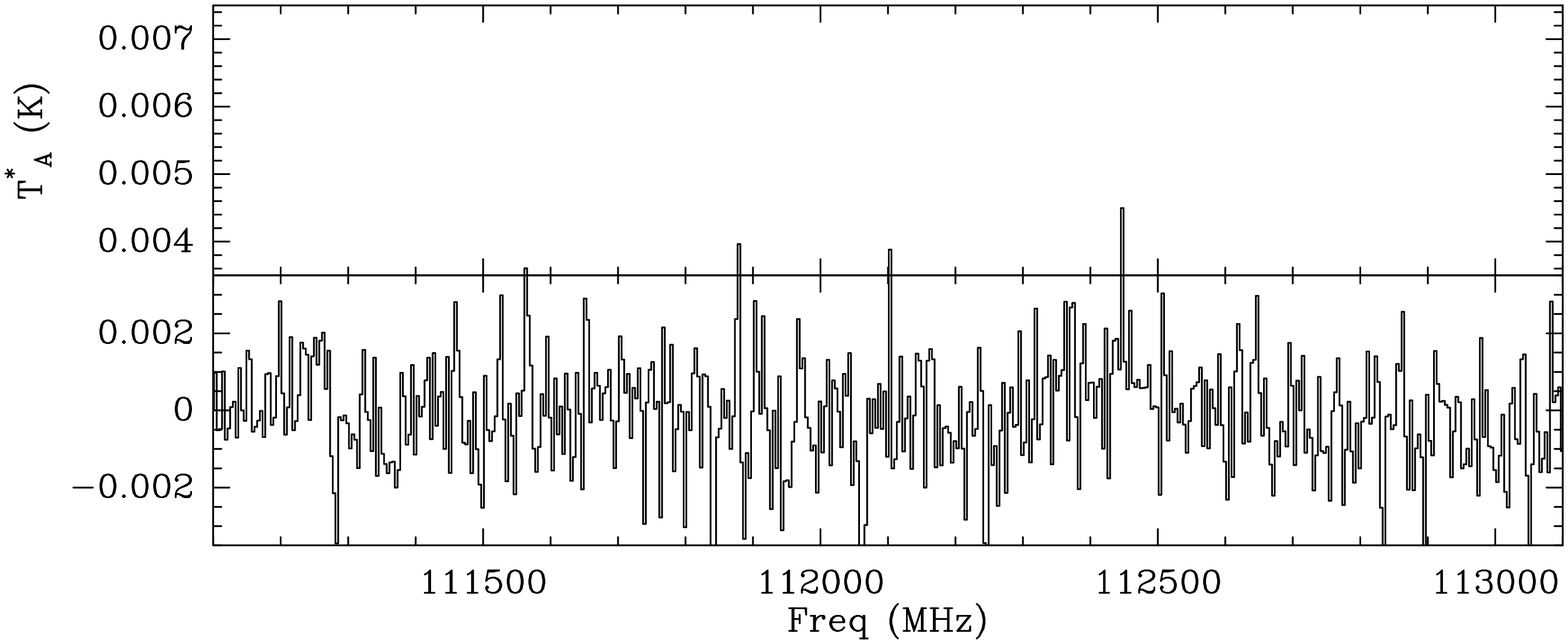} 
\includegraphics[angle=0,width=12cm]{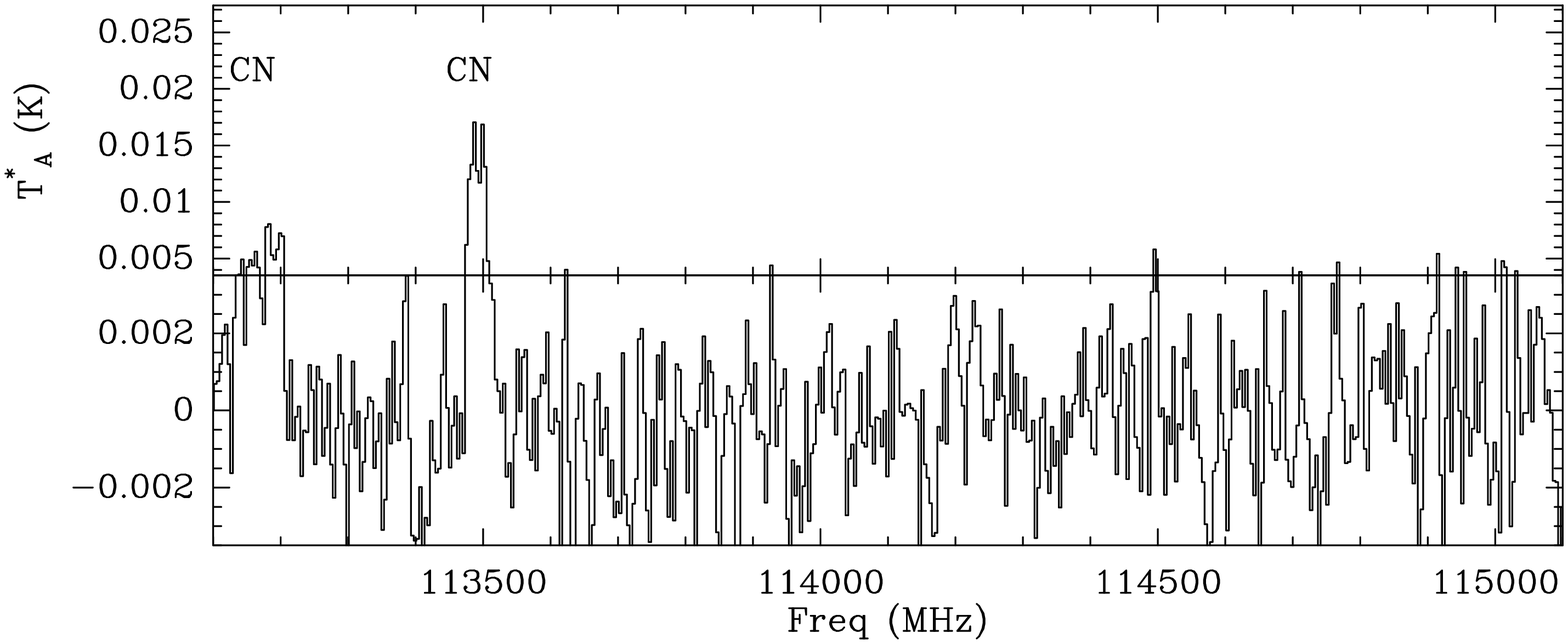} 
\caption{. (continued) } 
\label{Fig3mm}%
\end{figure*} 
\begin{figure*}[h!] 
\centering 
\ContinuedFloat 
\includegraphics[angle=0,width=12cm]{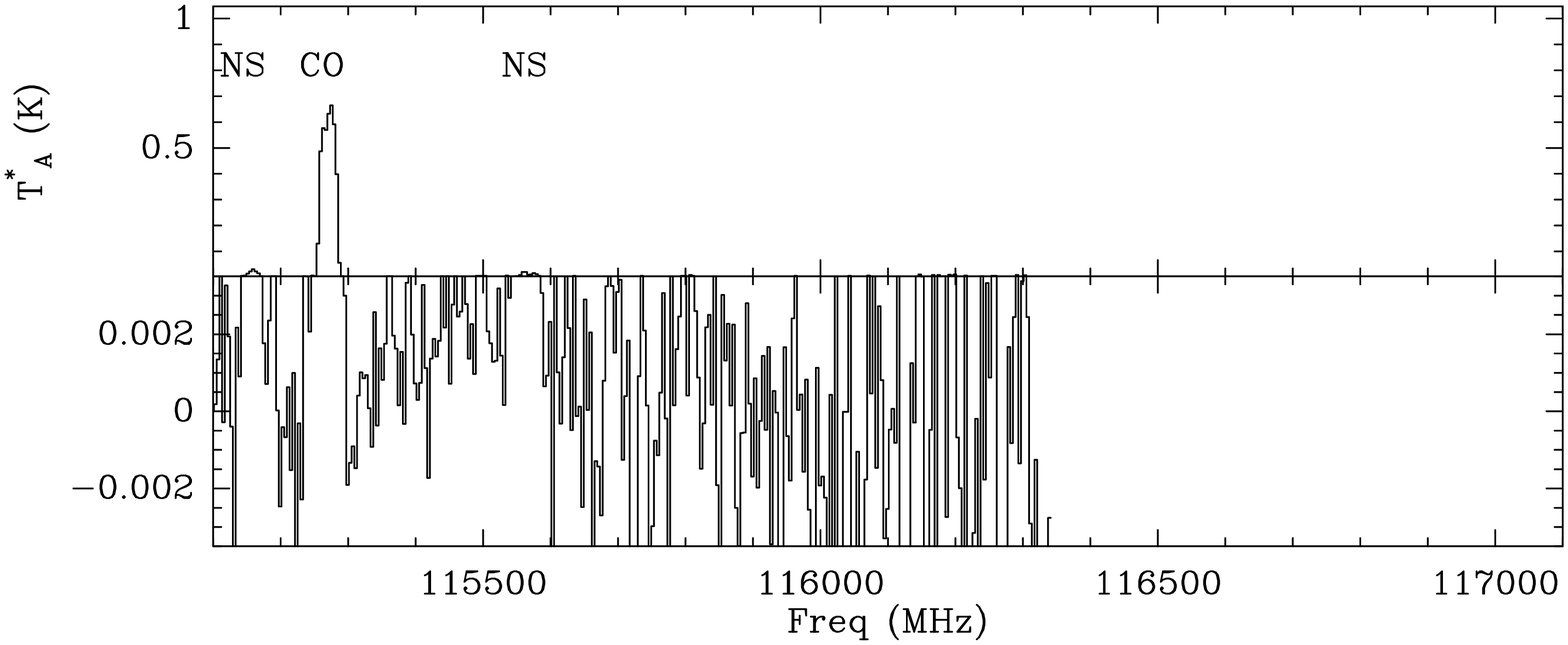} 
\caption{. (continued) } 
\label{Fig3mm}%
\end{figure*} 

\begin{figure*}[h!] 
\centering 
\includegraphics[angle=0,width=12cm]{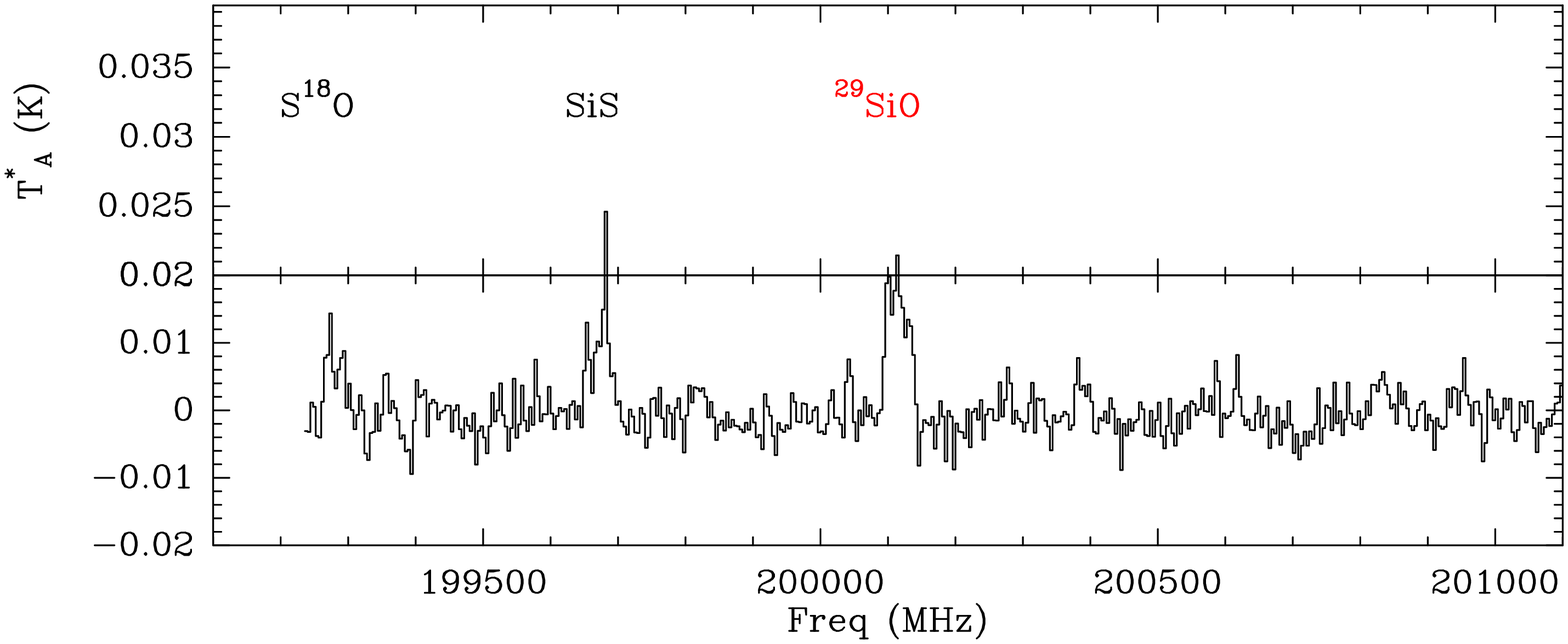} 
\includegraphics[angle=0,width=12cm]{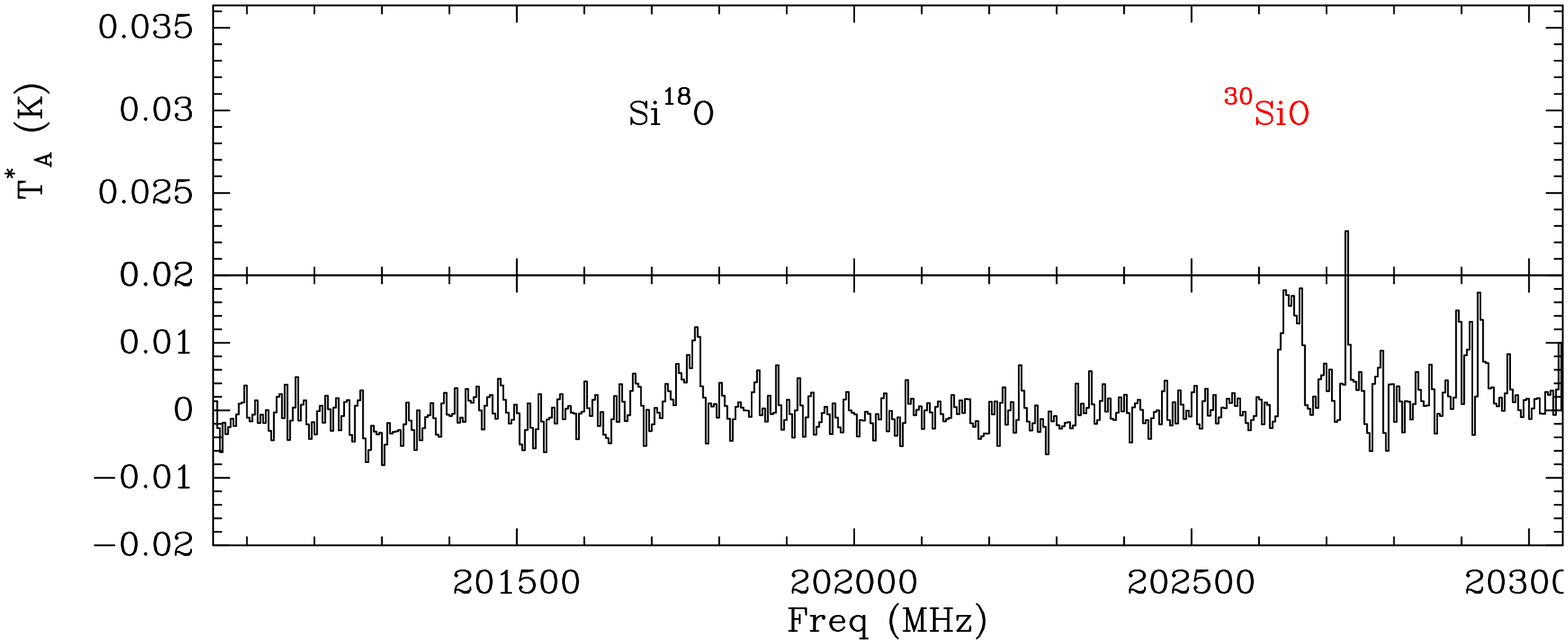} 
\includegraphics[angle=0,width=12cm]{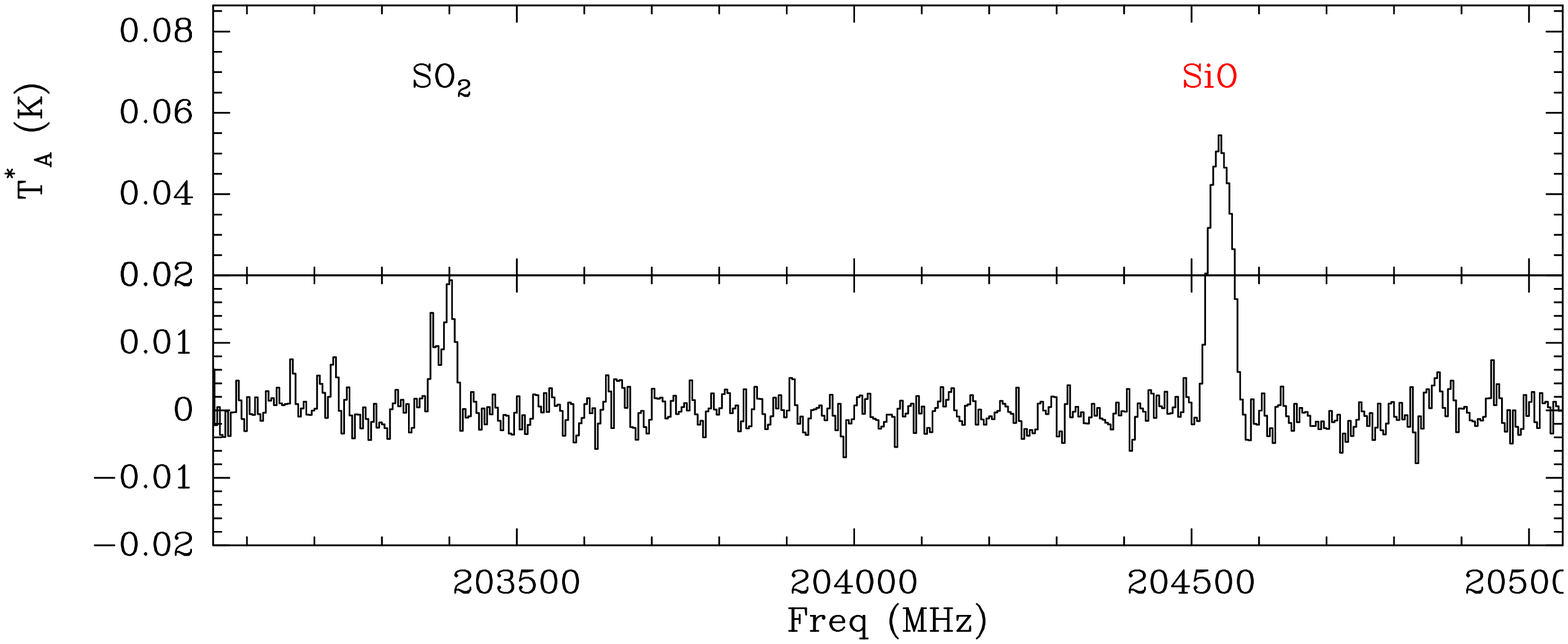} 
\includegraphics[angle=0,width=12cm]{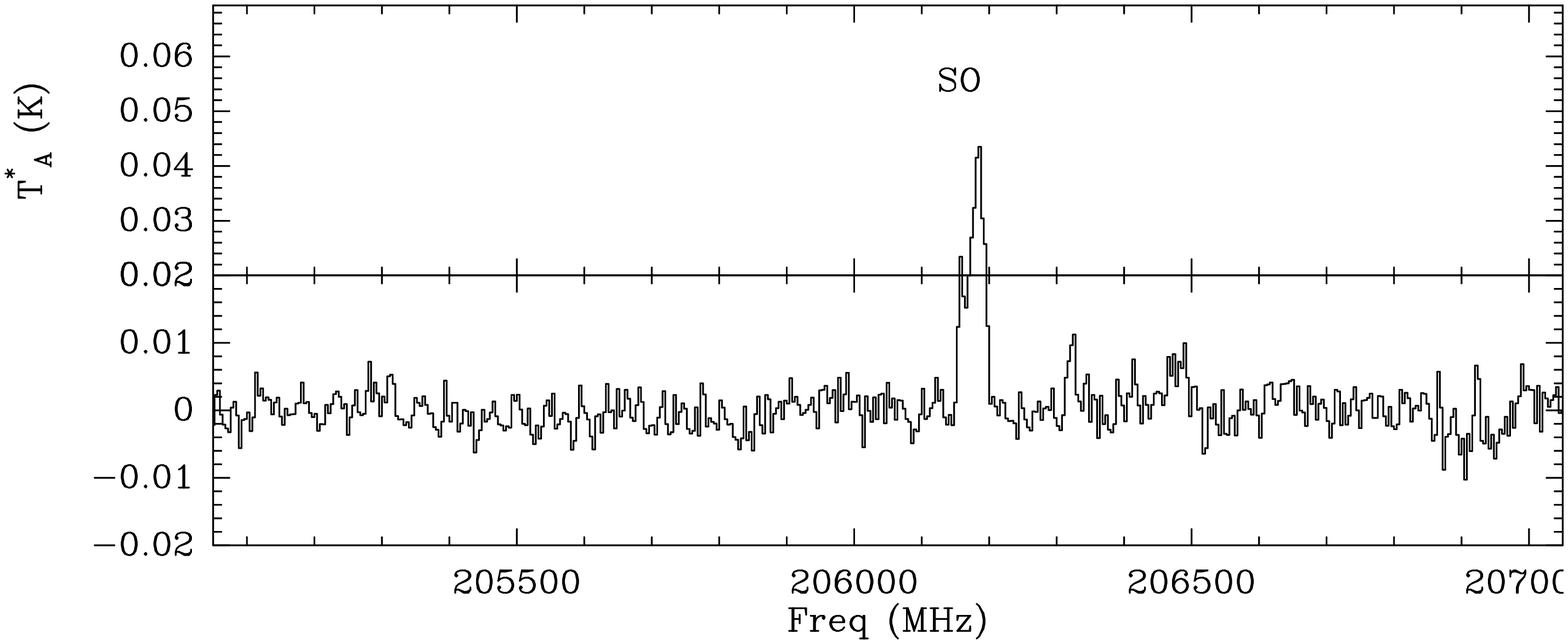} 
\caption{Spectral survey obtained with the IRAM 30 m telescope at the atmospheric window of 1\,mm. The $\rm{v_{LSR}}$ of of \irc used to calculate the frequencies
   of the spectra is 76\,km/s. } 
\label{Fig1mm}%
\end{figure*} 
\begin{figure*}[h!] 
\centering 
\ContinuedFloat 
\includegraphics[angle=0,width=12cm]{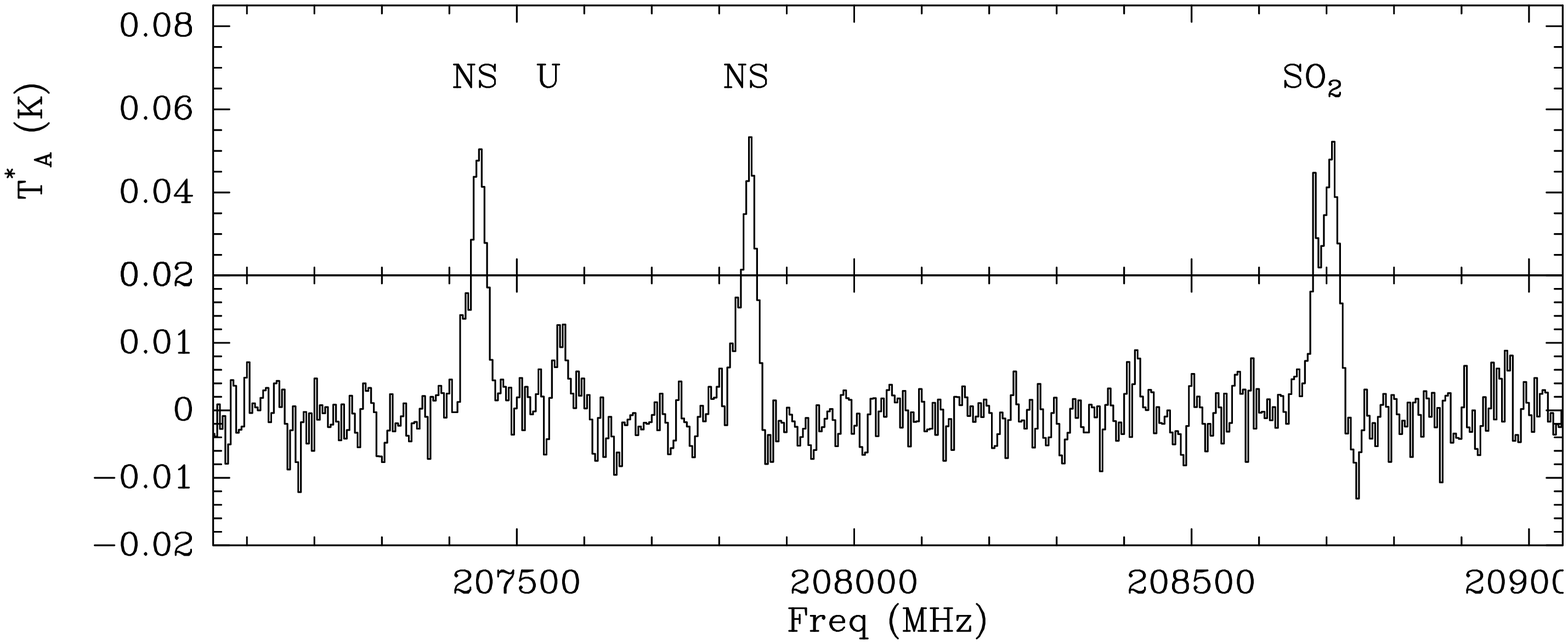} 
\includegraphics[angle=0,width=12cm]{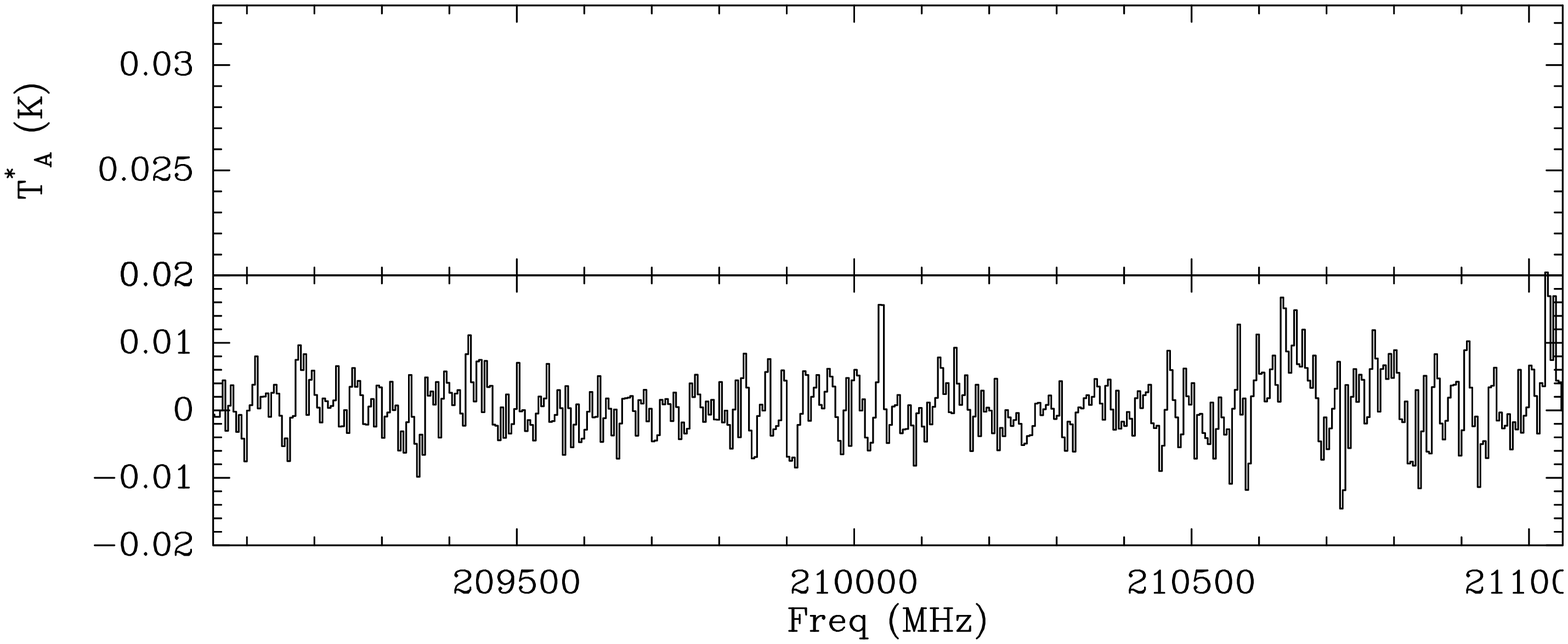} 
\includegraphics[angle=0,width=12cm]{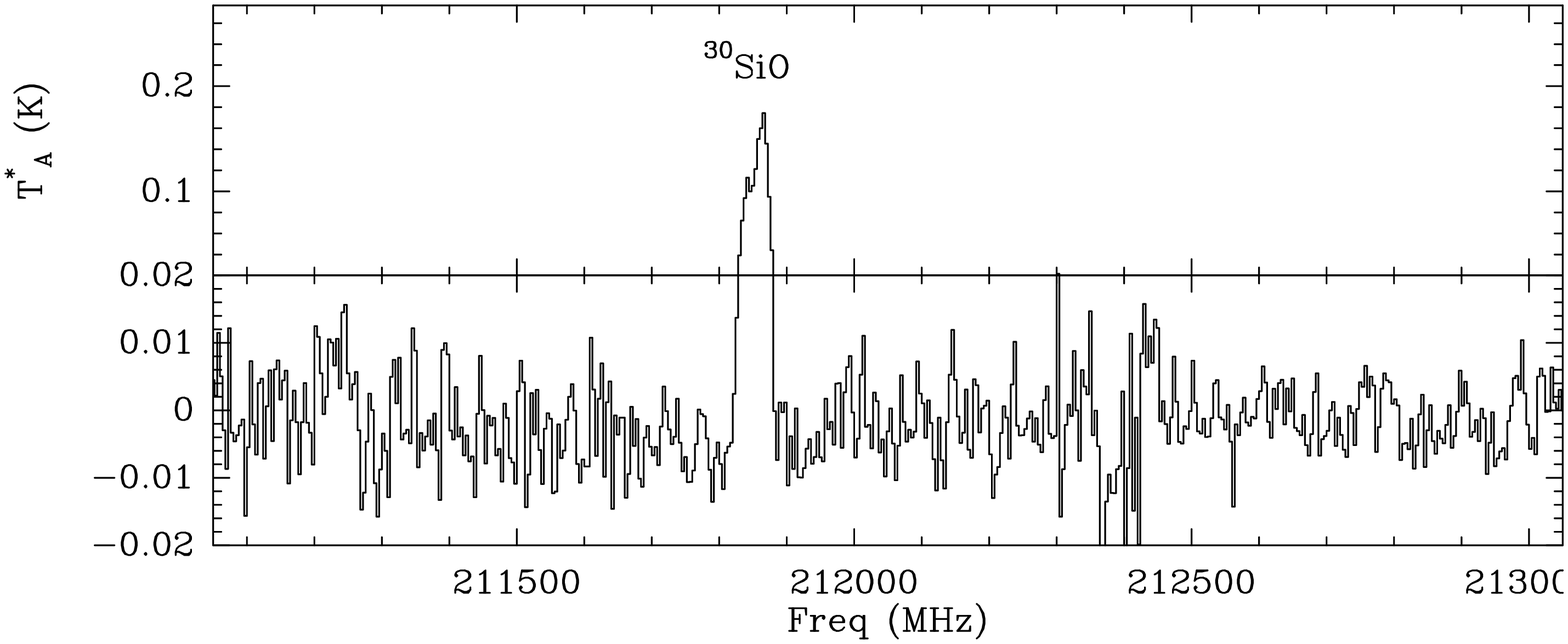} 
\includegraphics[angle=0,width=12cm]{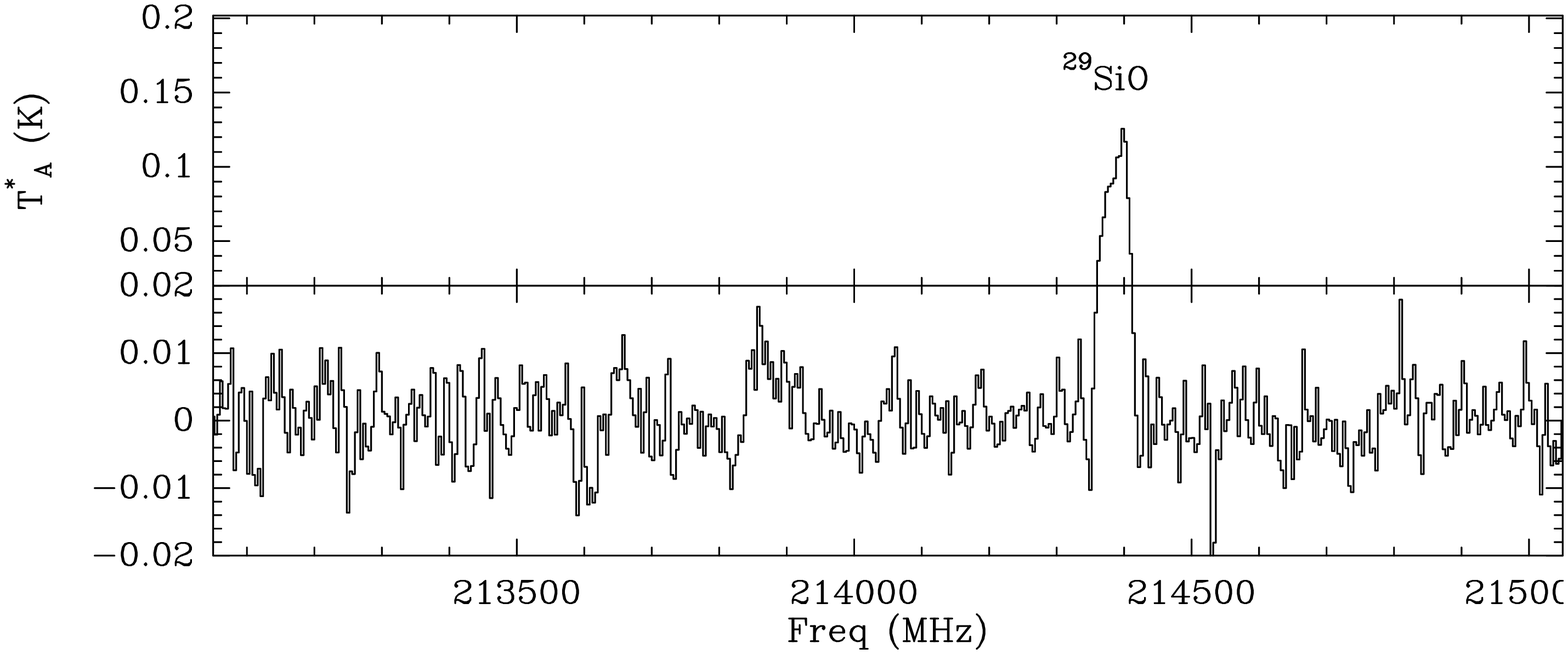} 
\caption{. (continued) } 
\label{Fig1mm}%
\end{figure*} 
\begin{figure*}[h!] 
\centering 
\ContinuedFloat 
\includegraphics[angle=0,width=12cm]{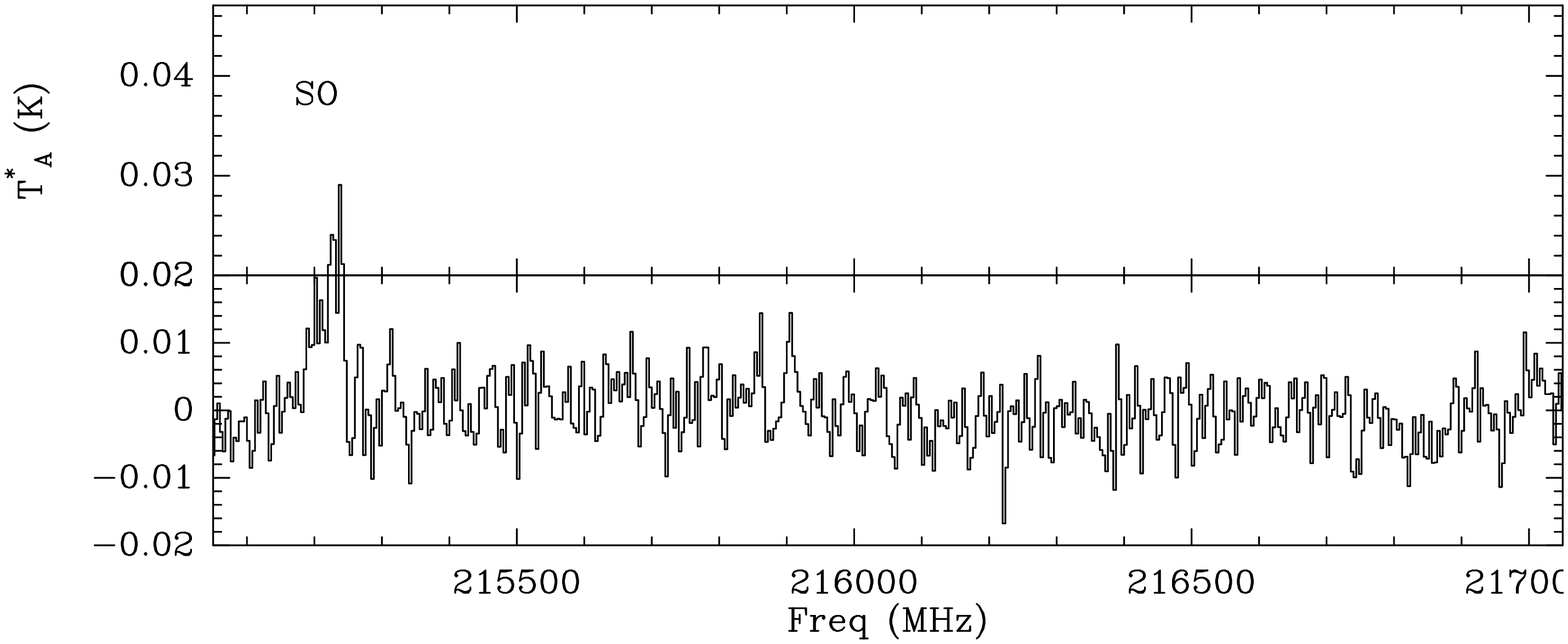} 
\includegraphics[angle=0,width=12cm]{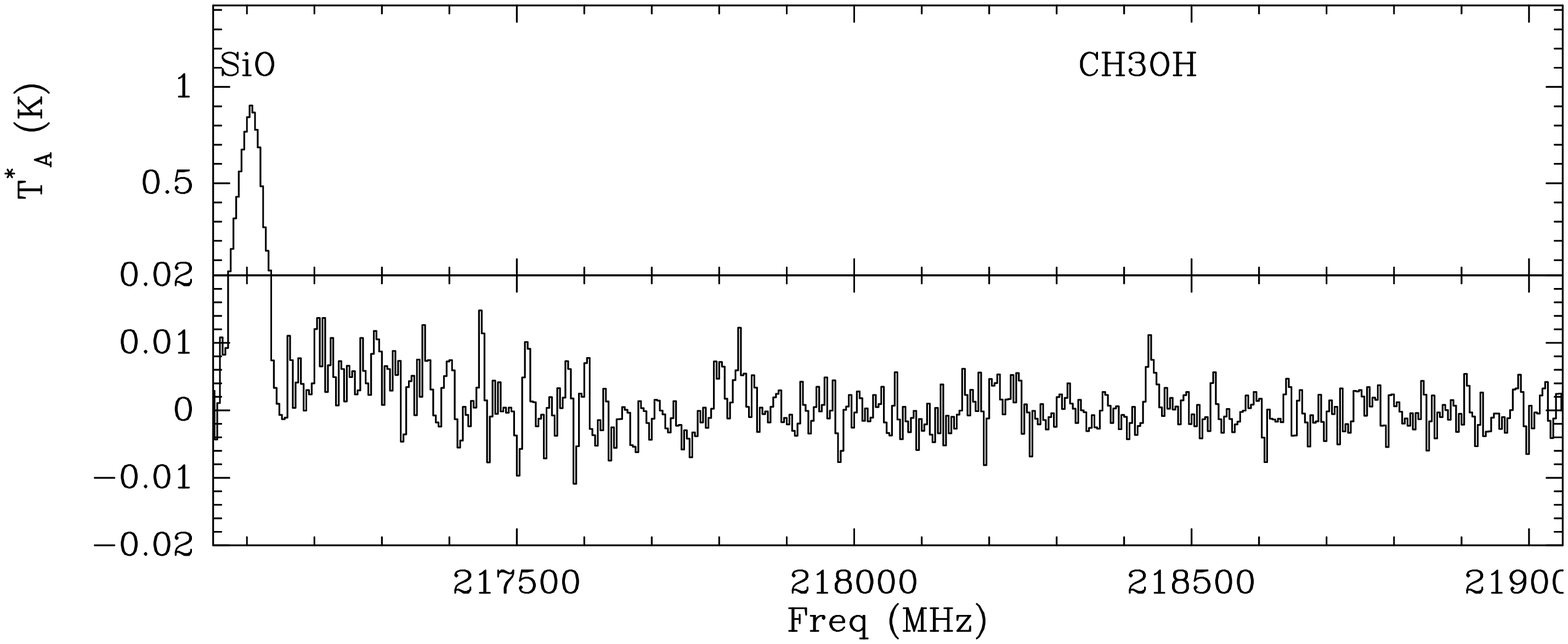} 
\includegraphics[angle=0,width=12cm]{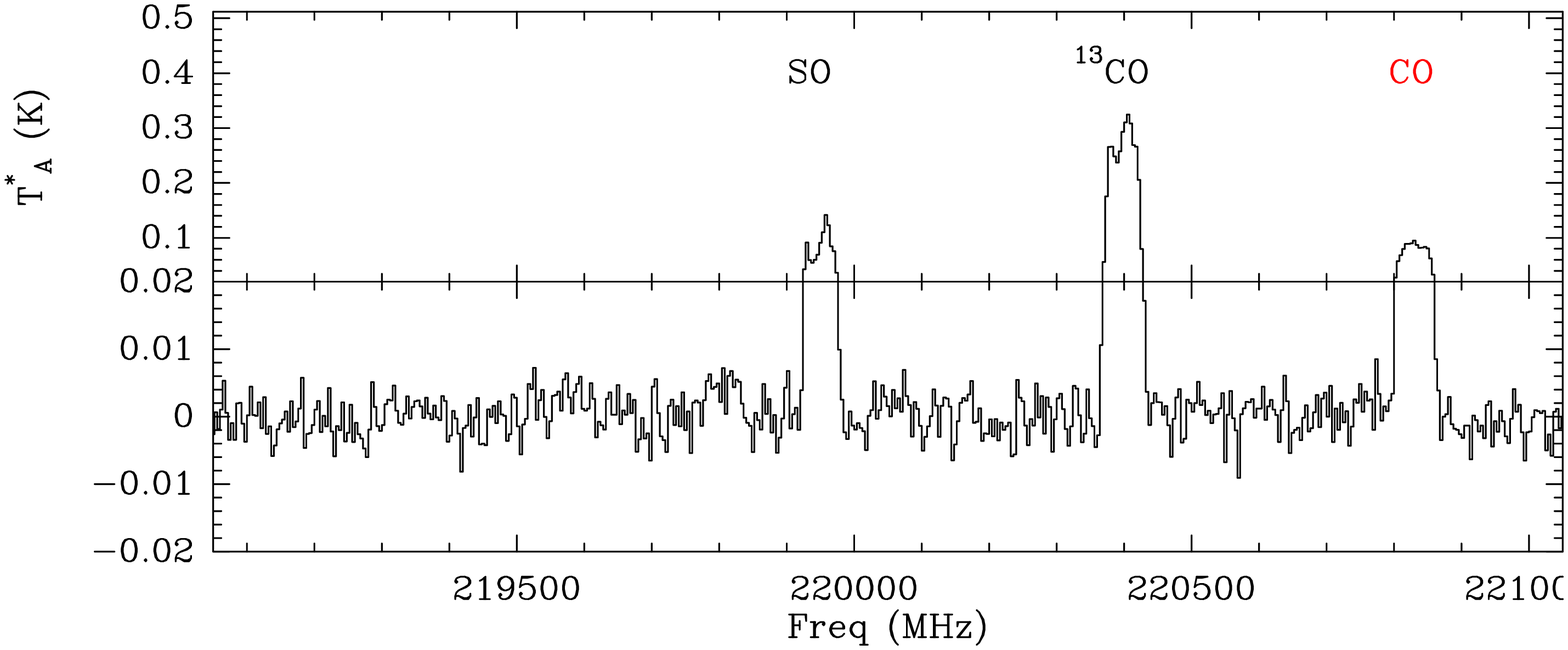} 
\includegraphics[angle=0,width=12cm]{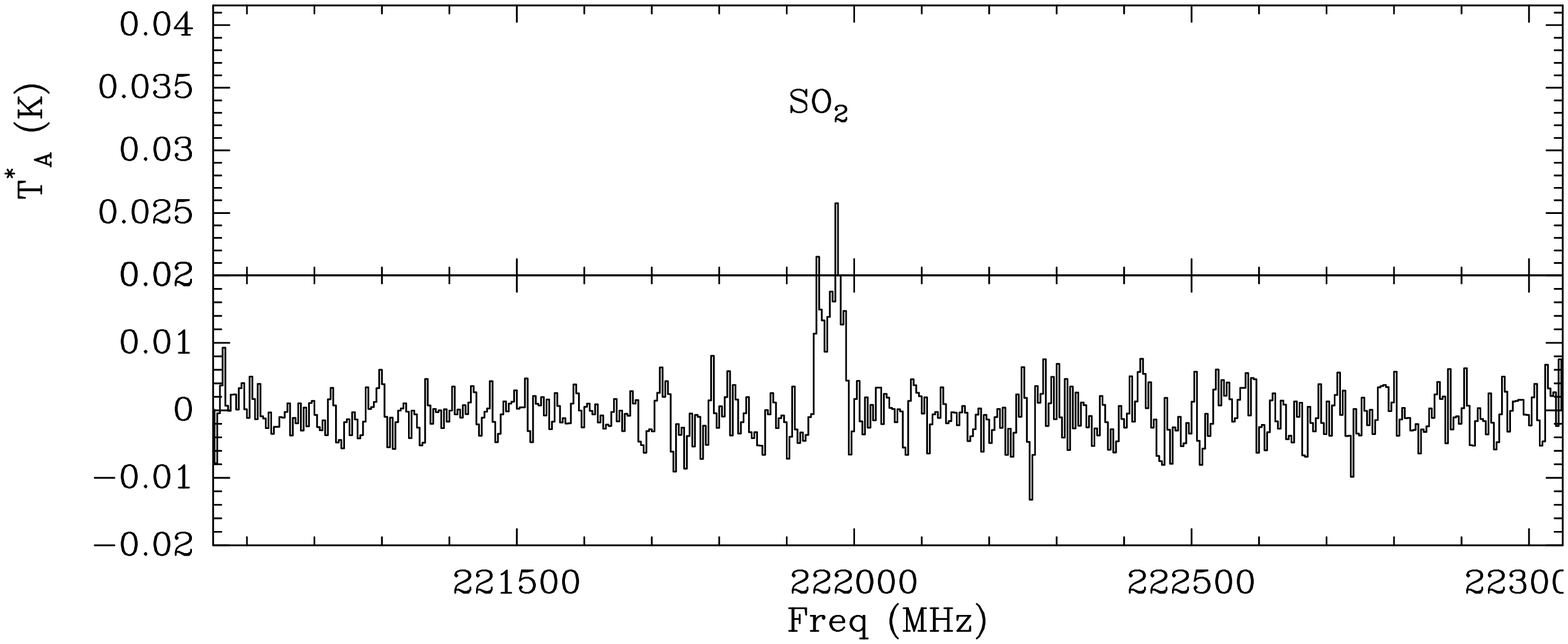} 
\caption{. (continued) } 
\label{Fig1mm}%
\end{figure*} 
\begin{figure*}[h!] 
\centering 
\ContinuedFloat 
\includegraphics[angle=0,width=12cm]{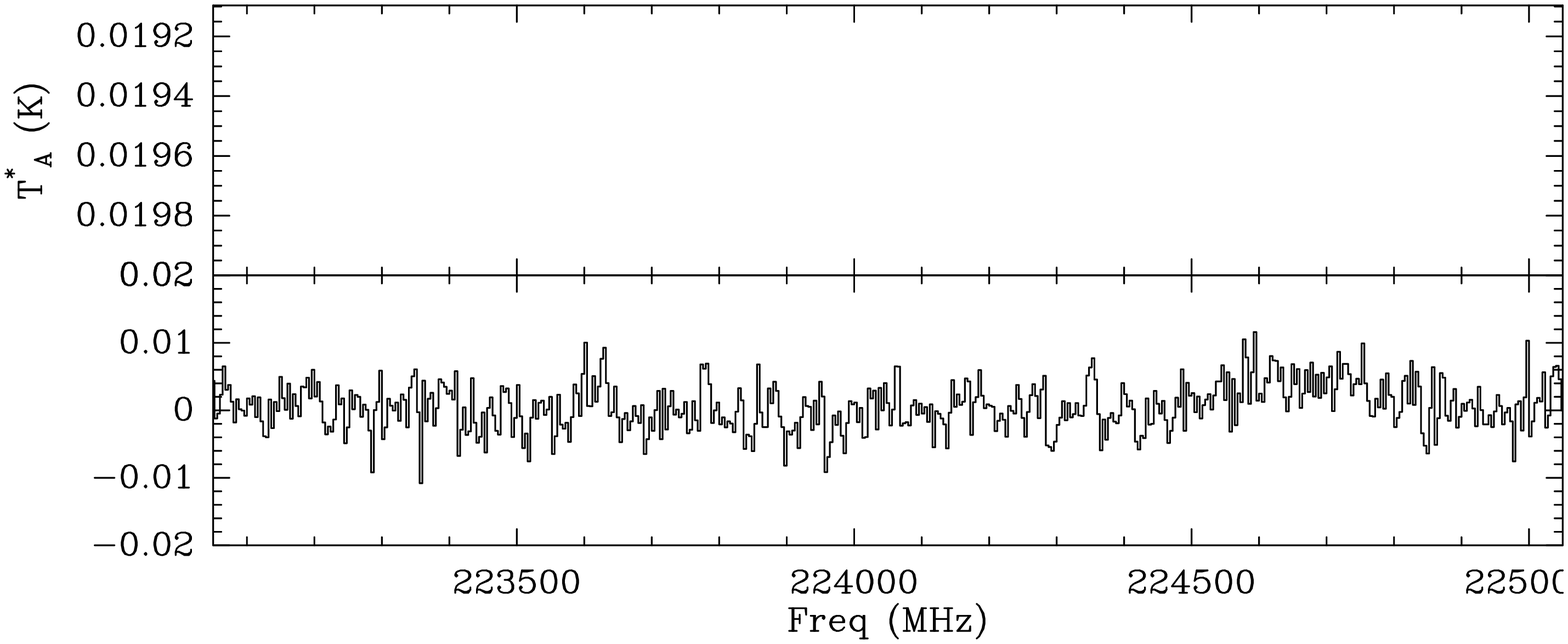} 
\includegraphics[angle=0,width=12cm]{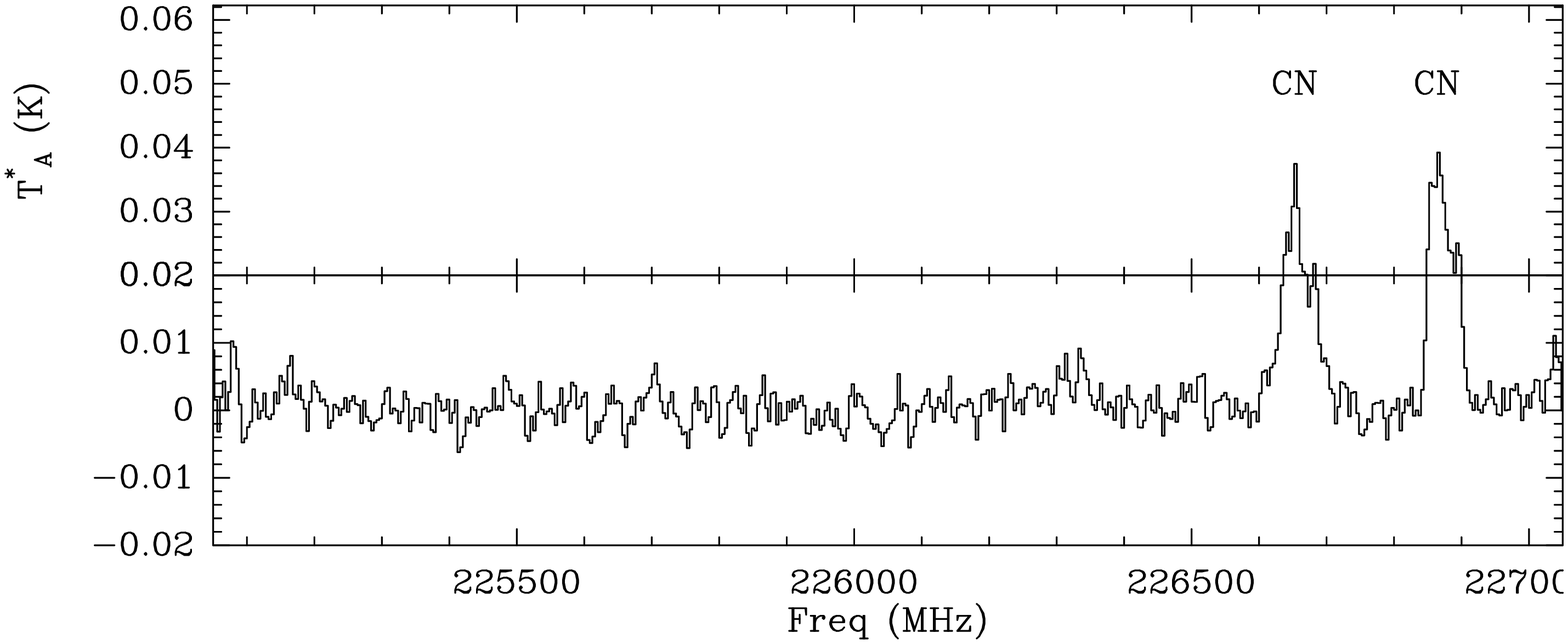} 
\includegraphics[angle=0,width=12cm]{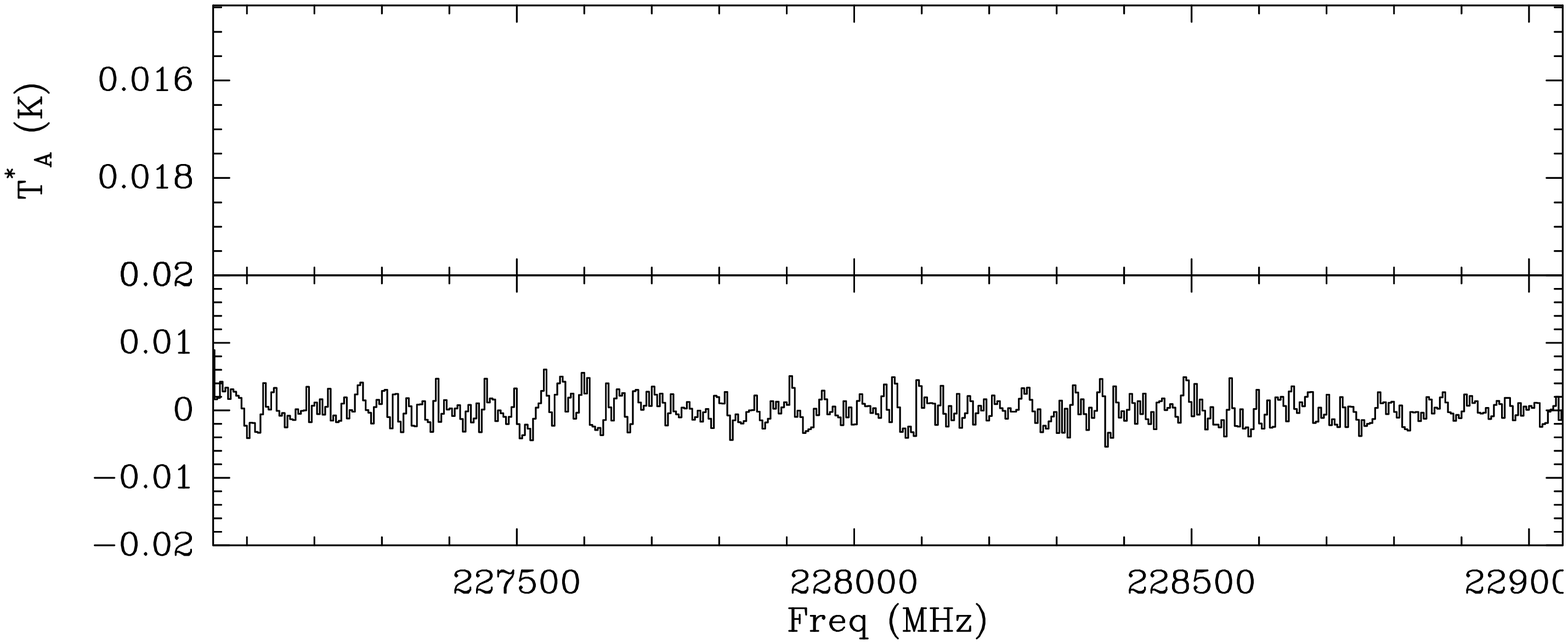} 
\includegraphics[angle=0,width=12cm]{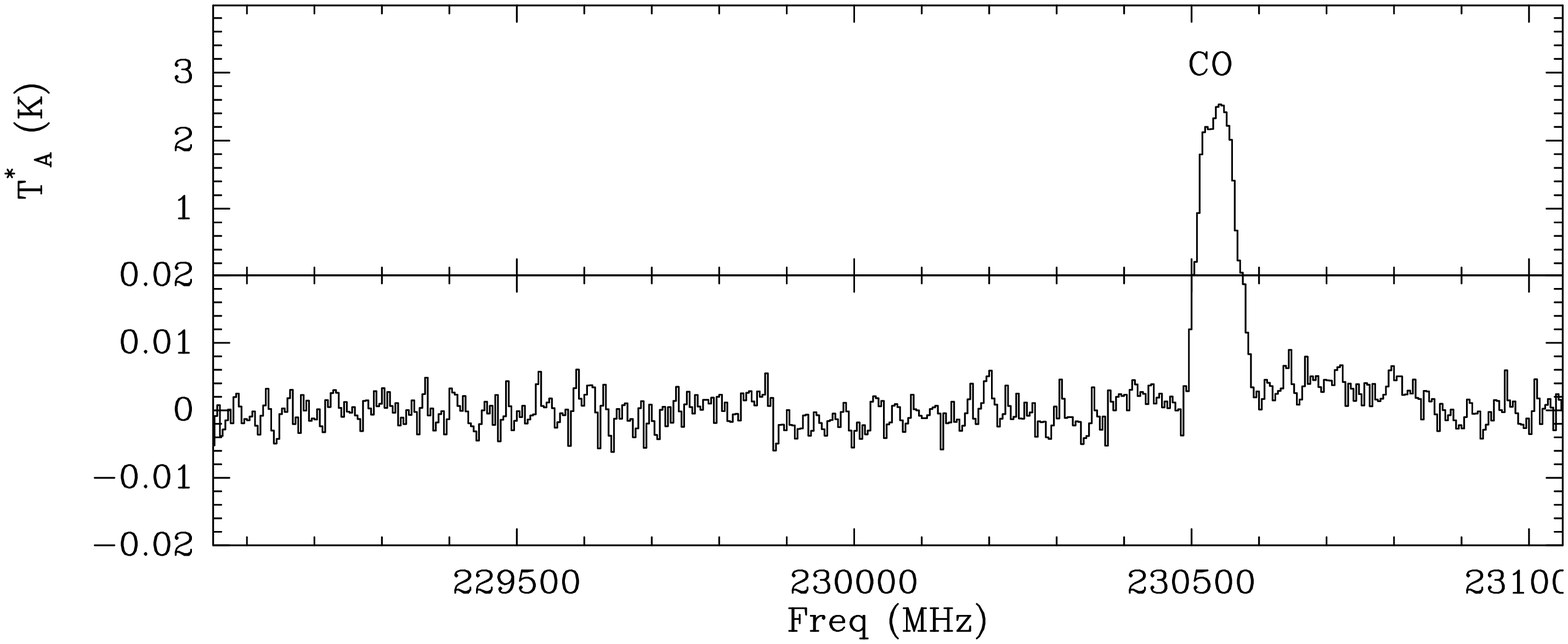} 
\caption{. (continued) } 
\label{Fig1mm}%
\end{figure*} 
\begin{figure*}[h!] 
\centering 
\ContinuedFloat 
\includegraphics[angle=0,width=12cm]{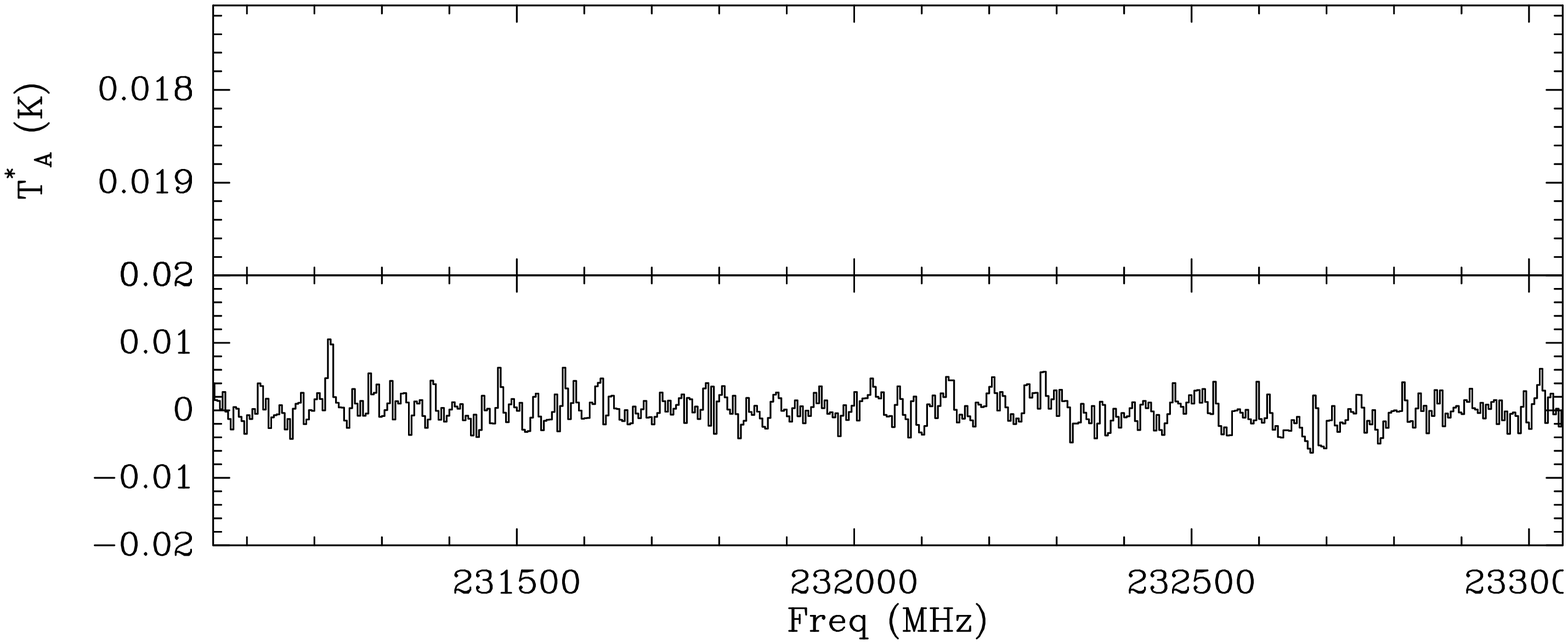} 
\includegraphics[angle=0,width=12cm]{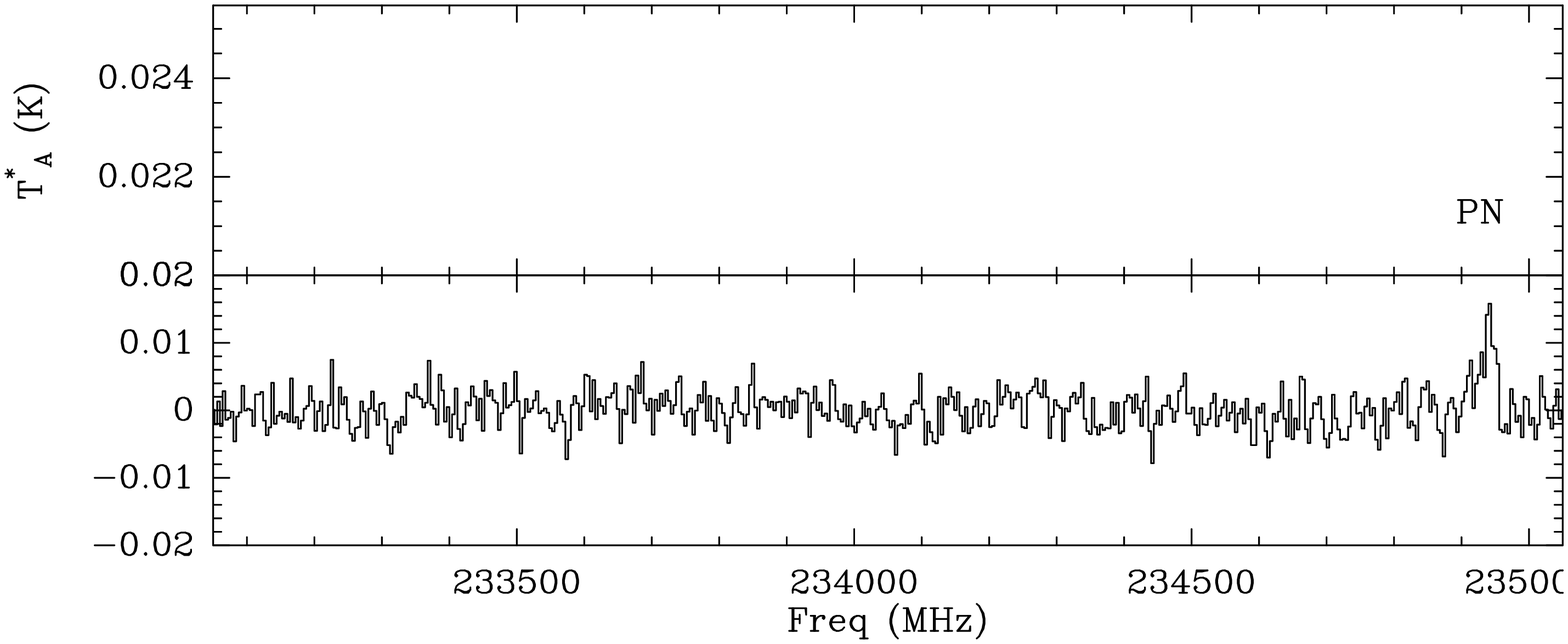} 
\includegraphics[angle=0,width=12cm]{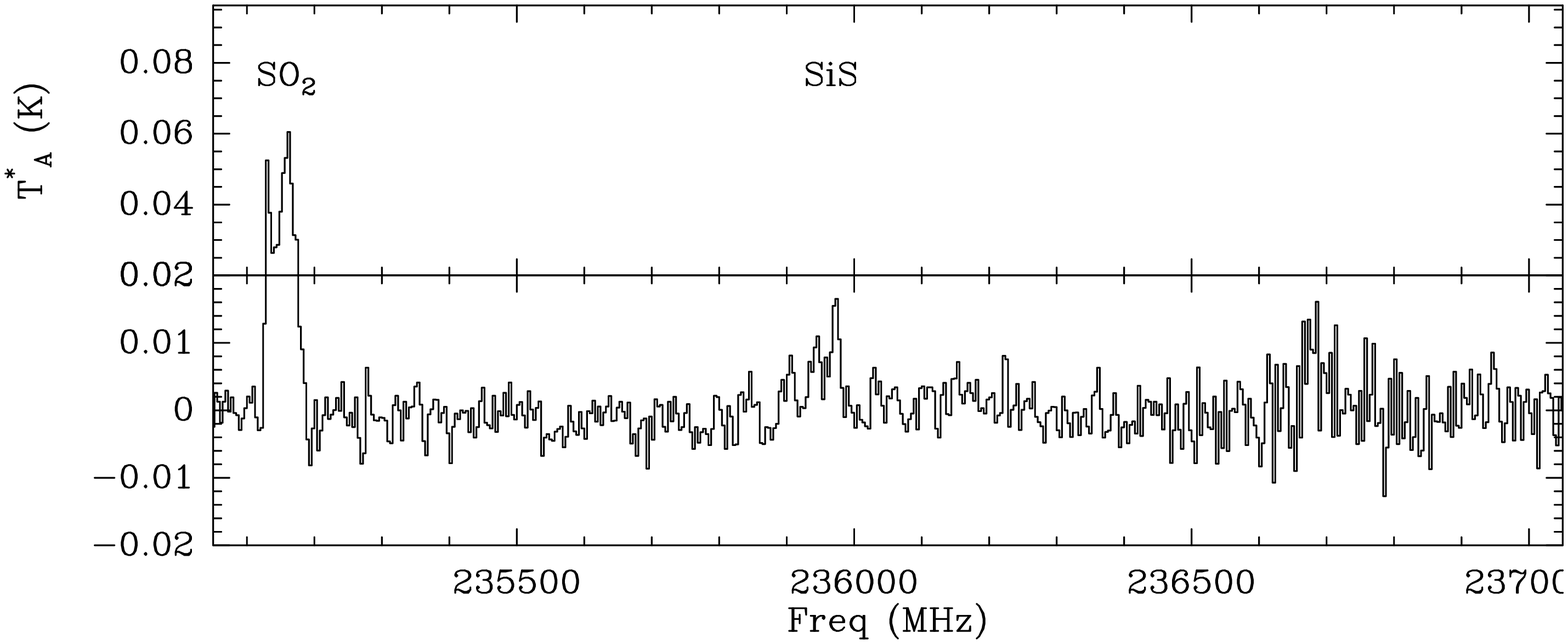} 
\includegraphics[angle=0,width=12cm]{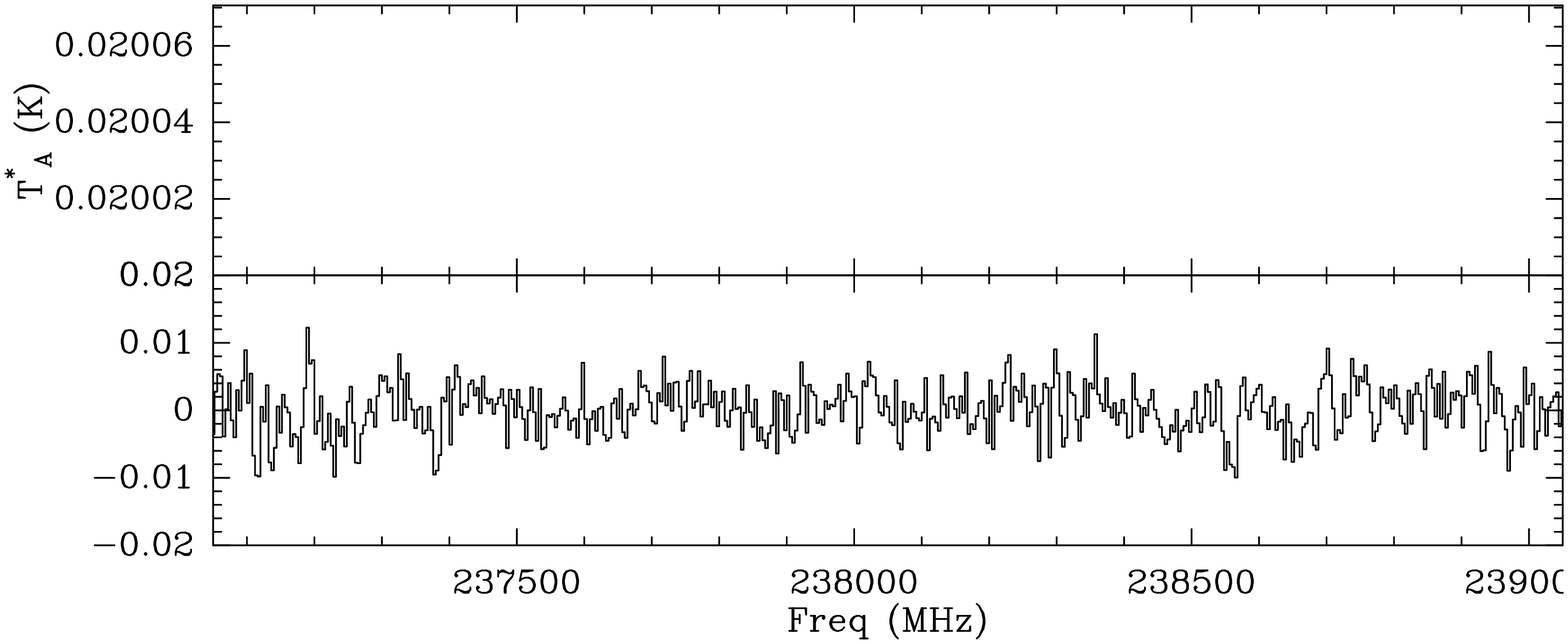} 
\caption{. (continued) } 
\label{Fig1mm}%
\end{figure*} 
\begin{figure*}[h!] 
\centering 
\ContinuedFloat 
\includegraphics[angle=0,width=12cm]{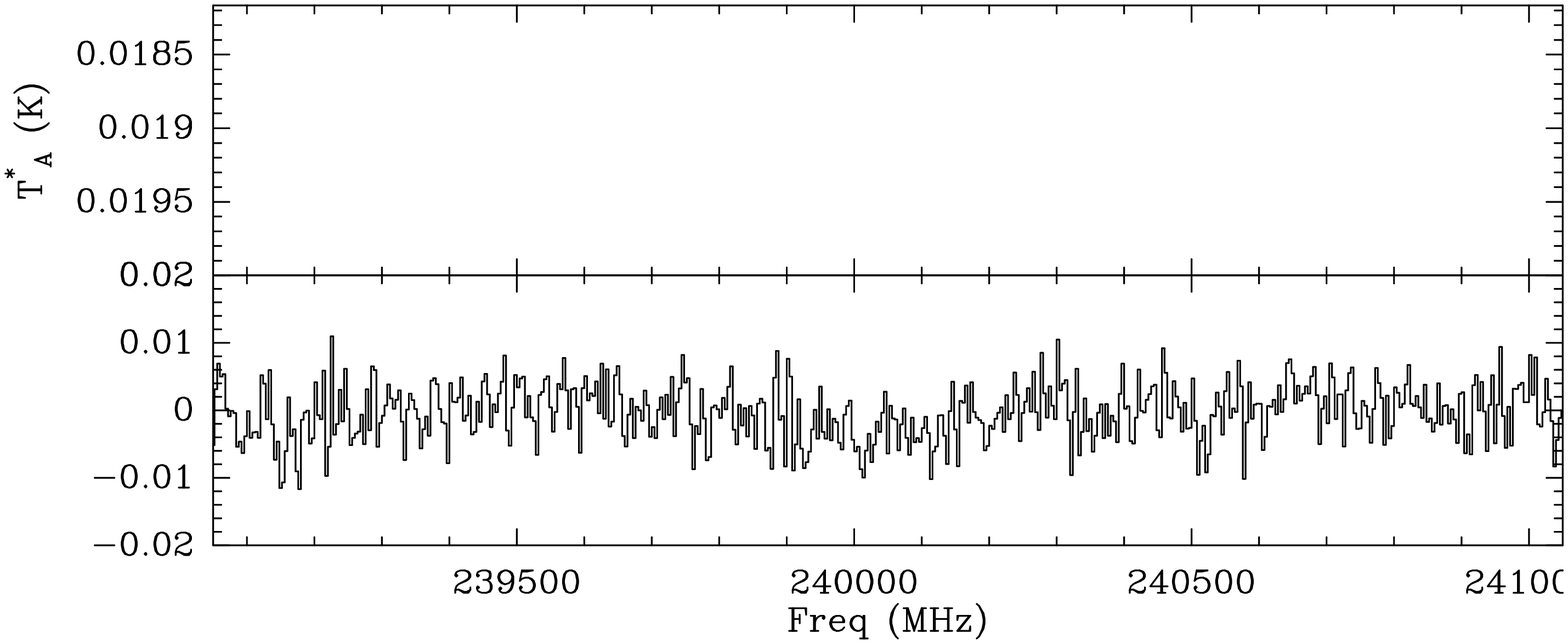} 
\includegraphics[angle=0,width=12cm]{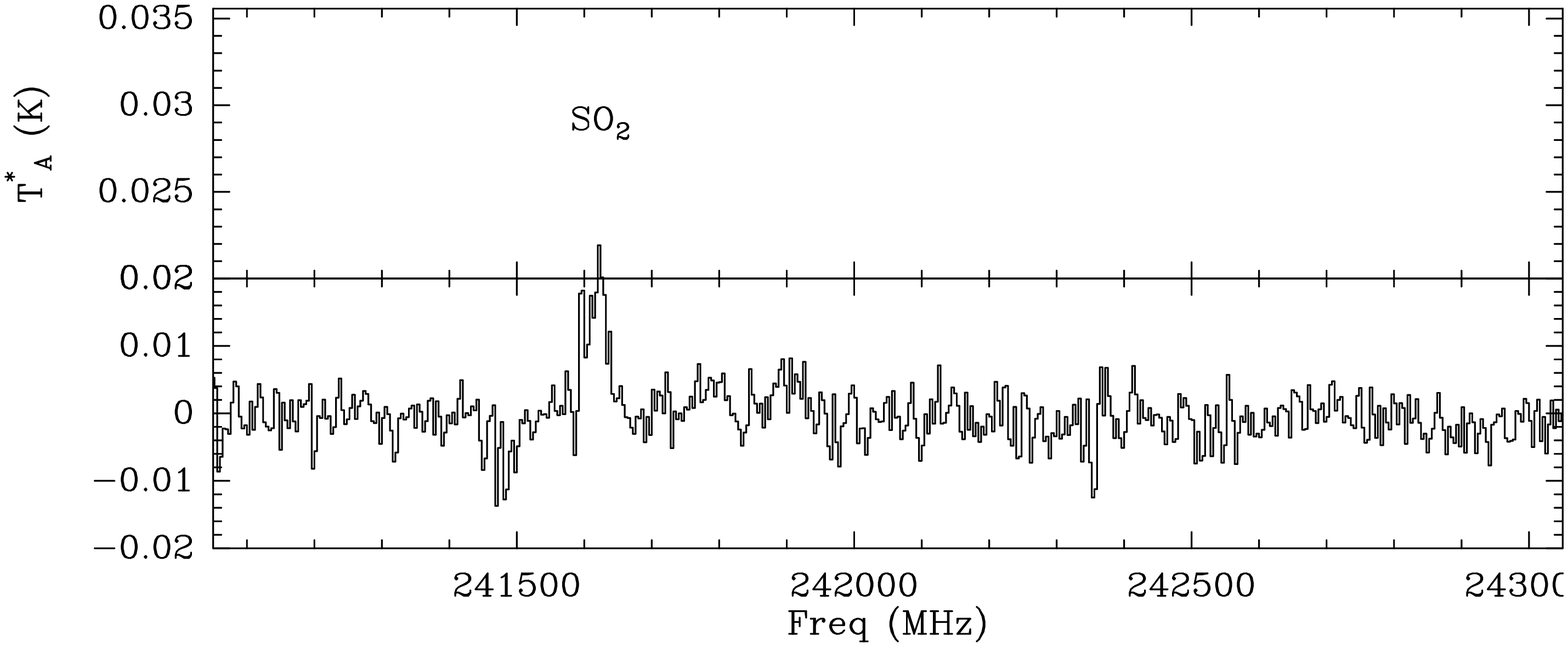} 
\includegraphics[angle=0,width=12cm]{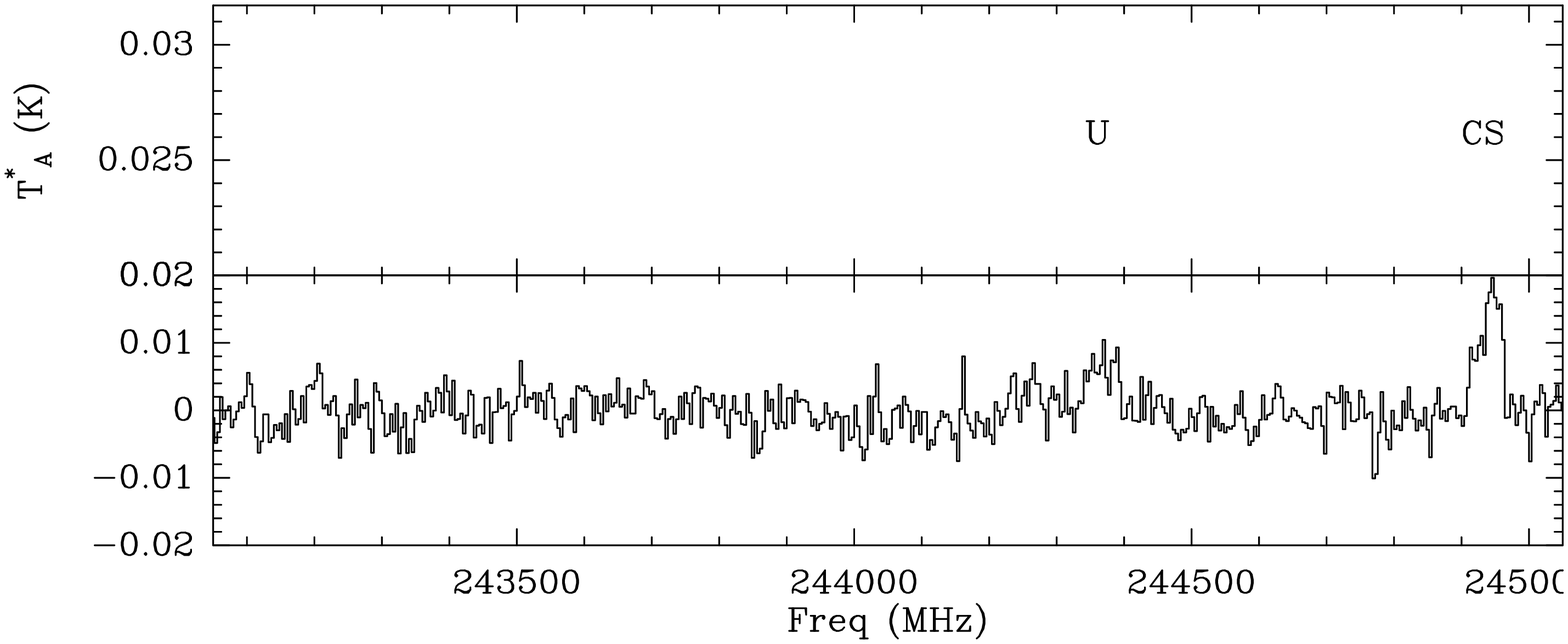} 
\includegraphics[angle=0,width=12cm]{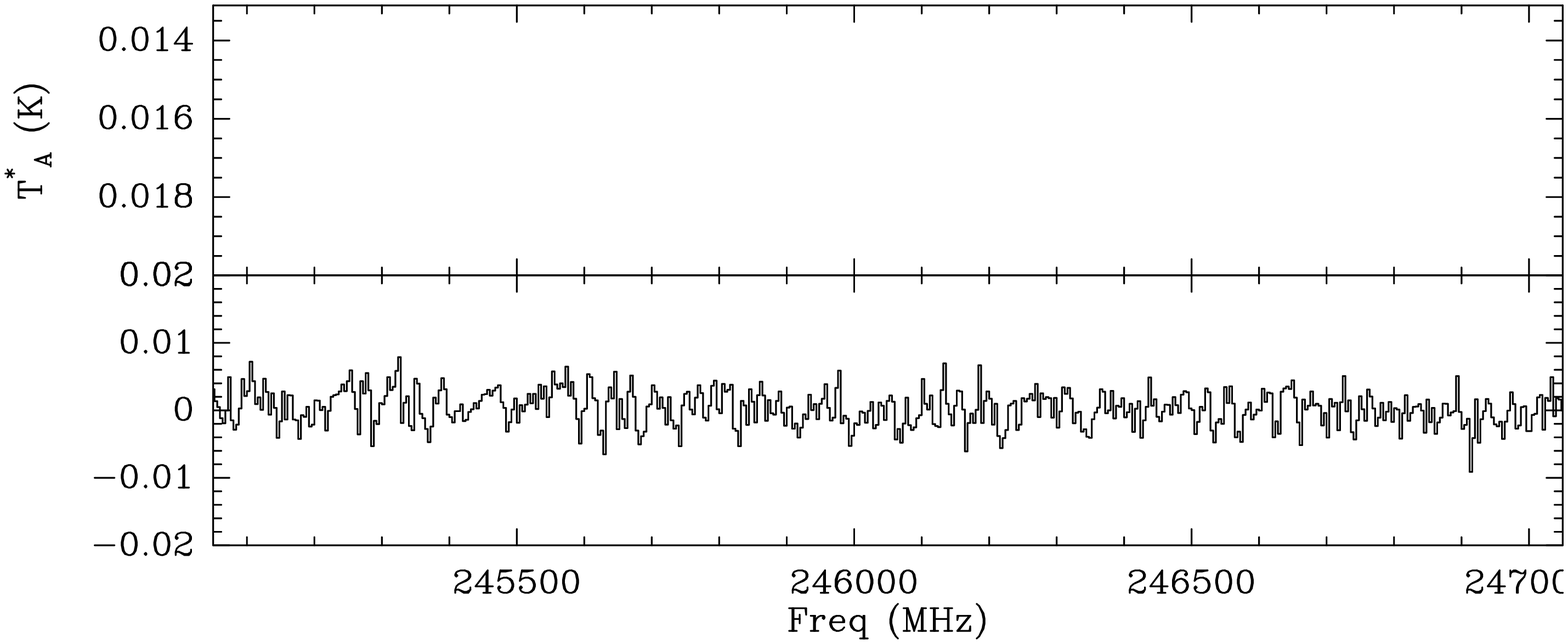} 
\caption{. (continued) } 
\label{Fig1mm}%
\end{figure*} 
\begin{figure*}[h!] 
\centering 
\ContinuedFloat 
\includegraphics[angle=0,width=12cm]{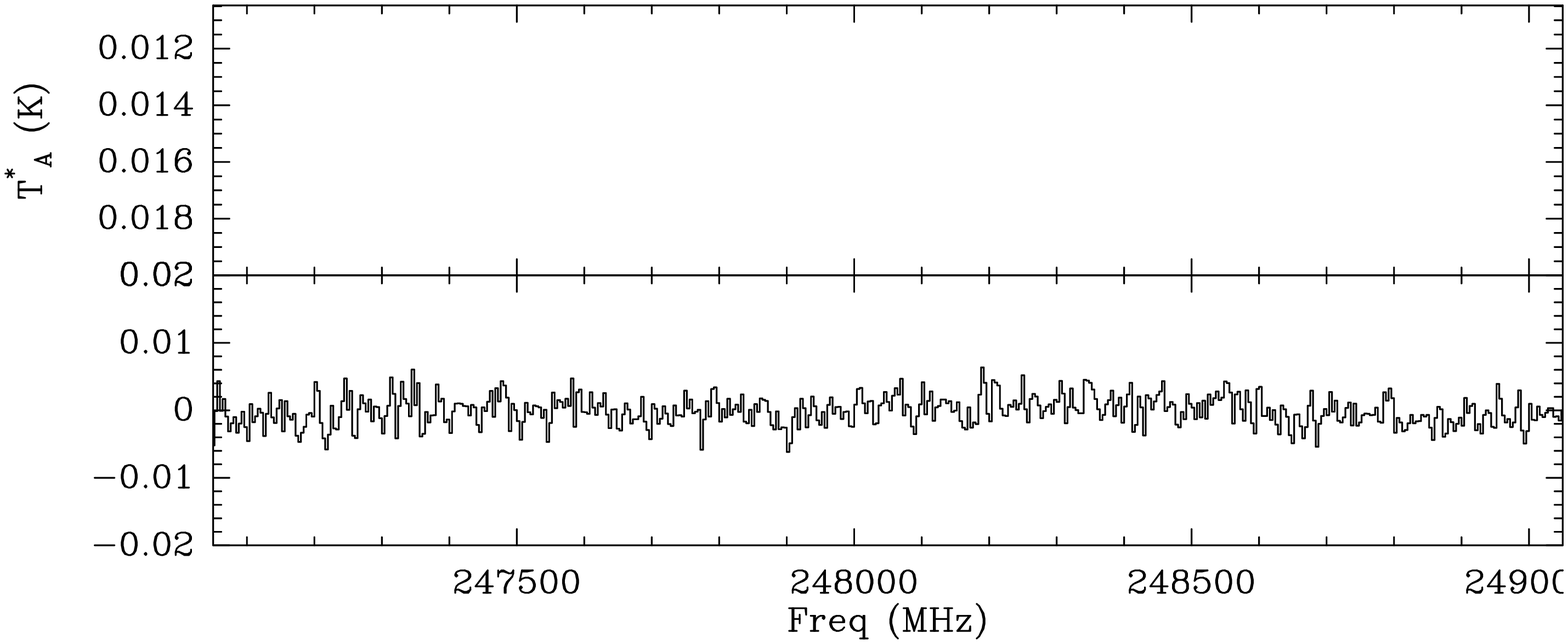} 
\includegraphics[angle=0,width=12cm]{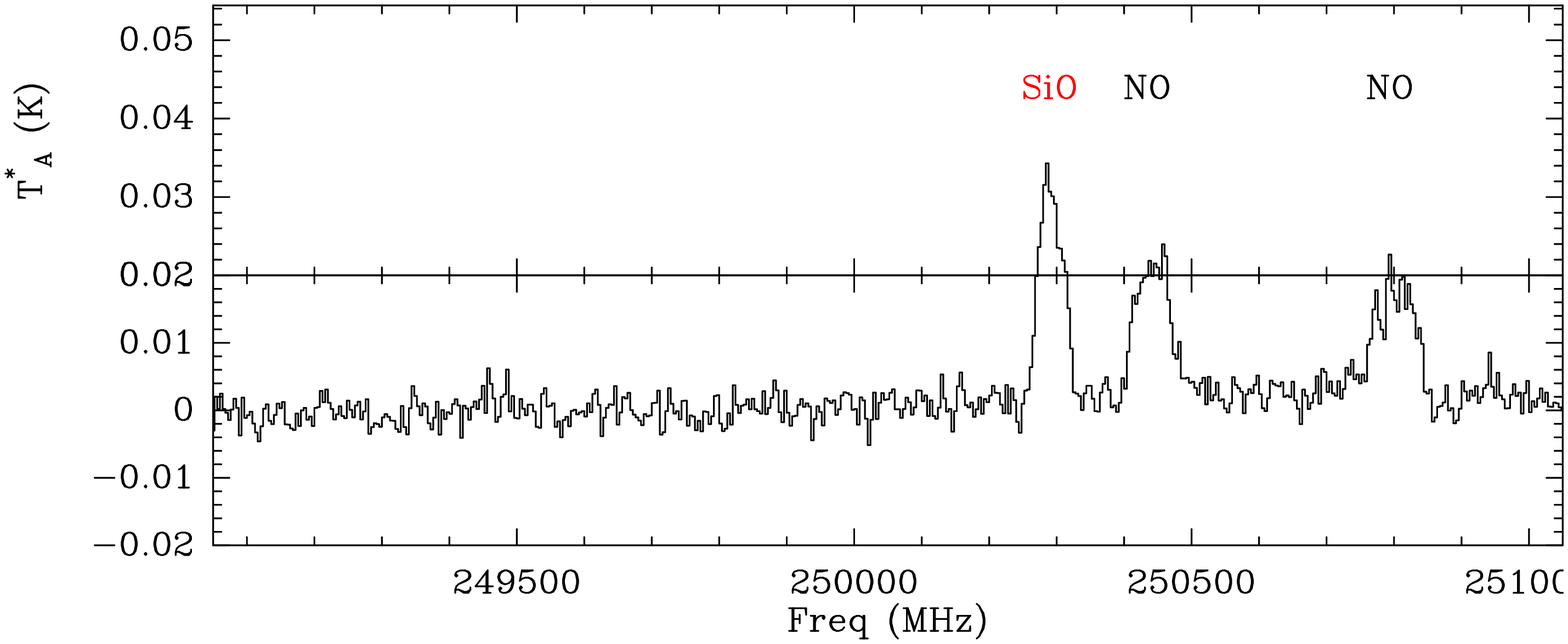} 
\includegraphics[angle=0,width=12cm]{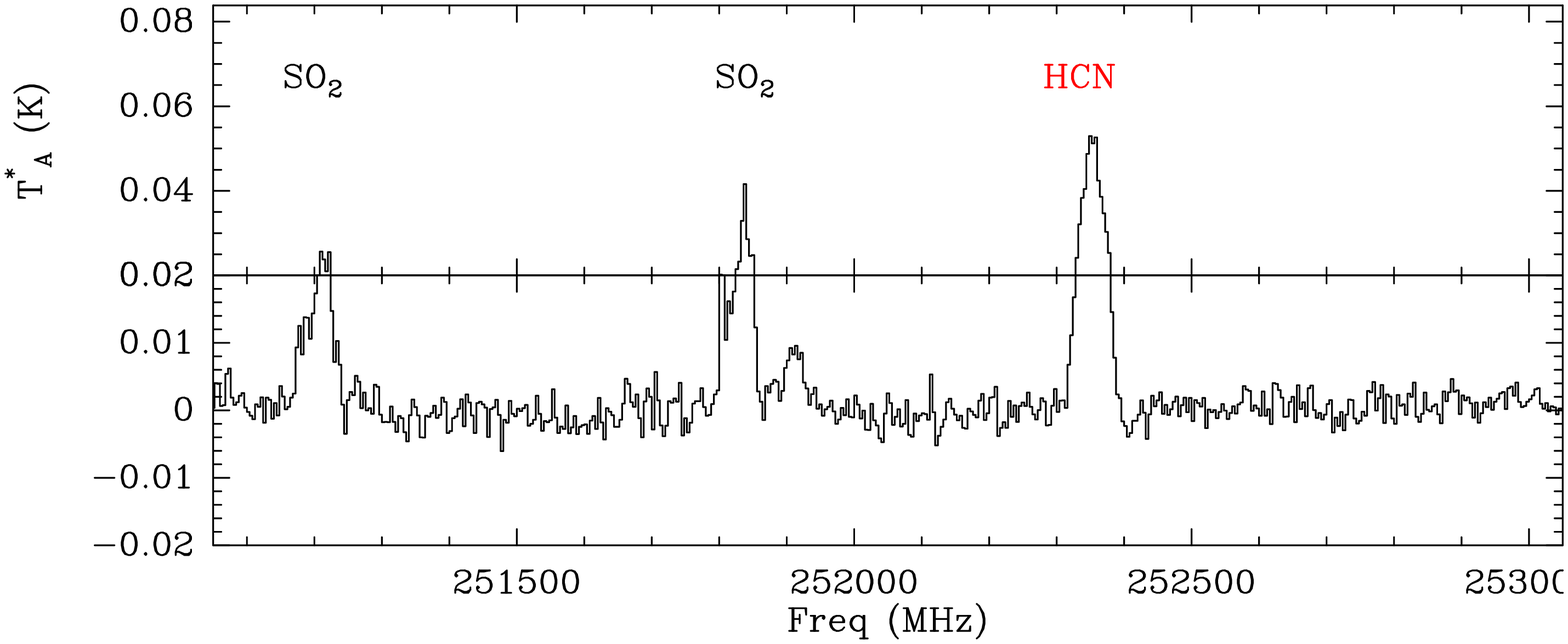} 
\includegraphics[angle=0,width=12cm]{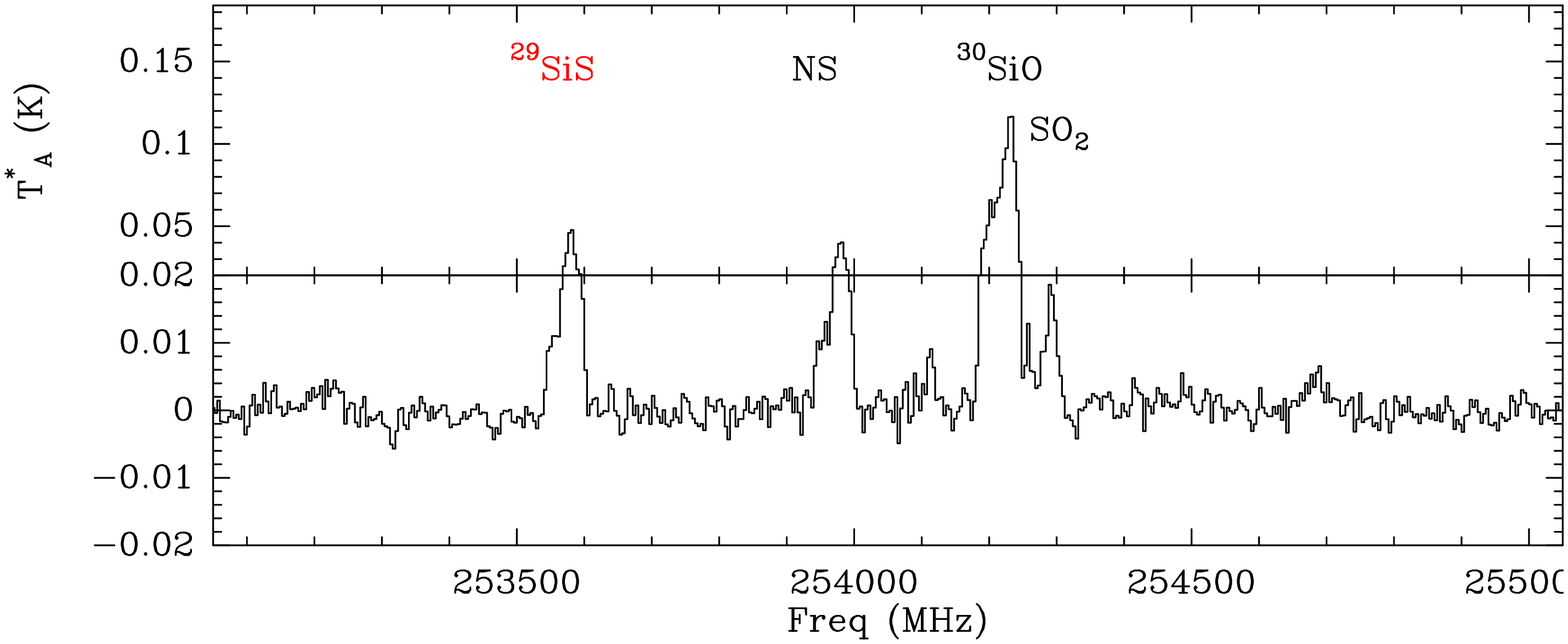} 
\caption{. (continued) } 
\label{Fig1mm}%
\end{figure*} 
\begin{figure*}[h!] 
\centering 
\ContinuedFloat 
\includegraphics[angle=0,width=12cm]{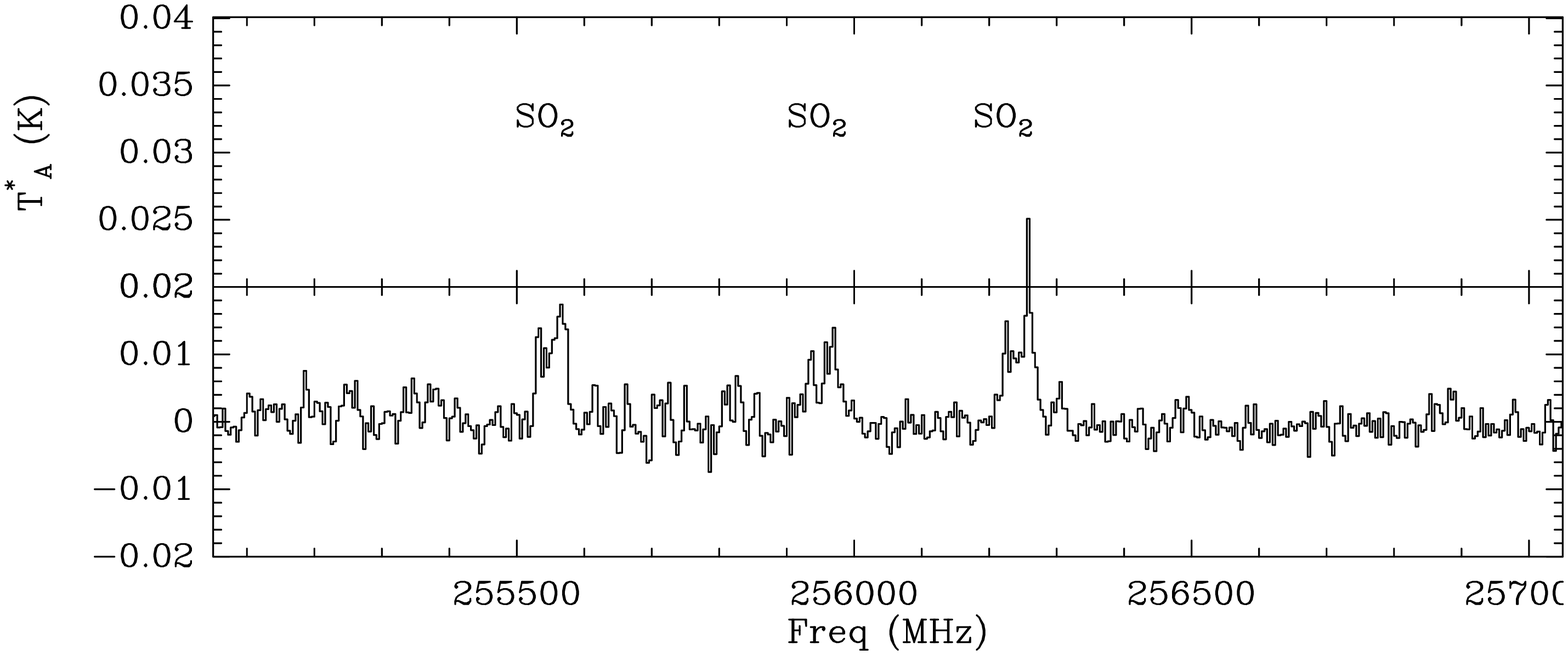} 
\includegraphics[angle=0,width=12cm]{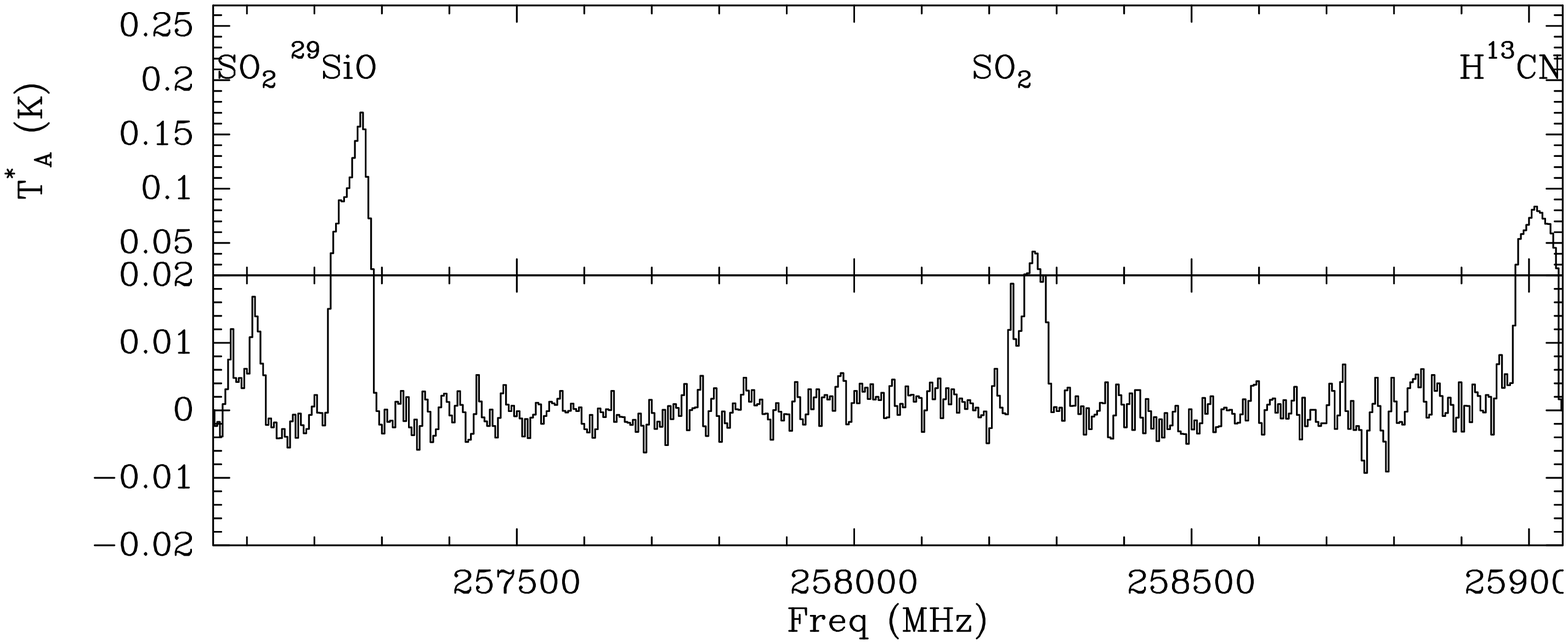} 
\includegraphics[angle=0,width=12cm]{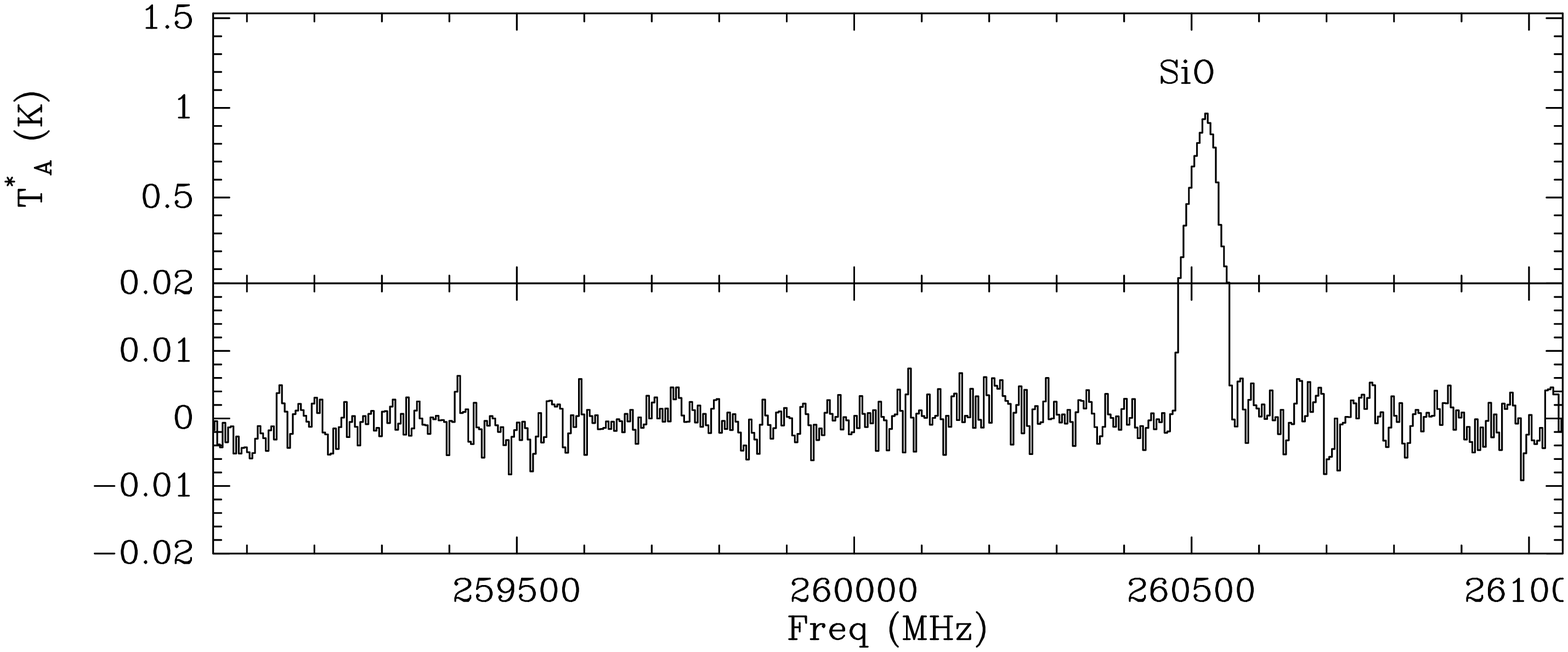} 
\includegraphics[angle=0,width=12cm]{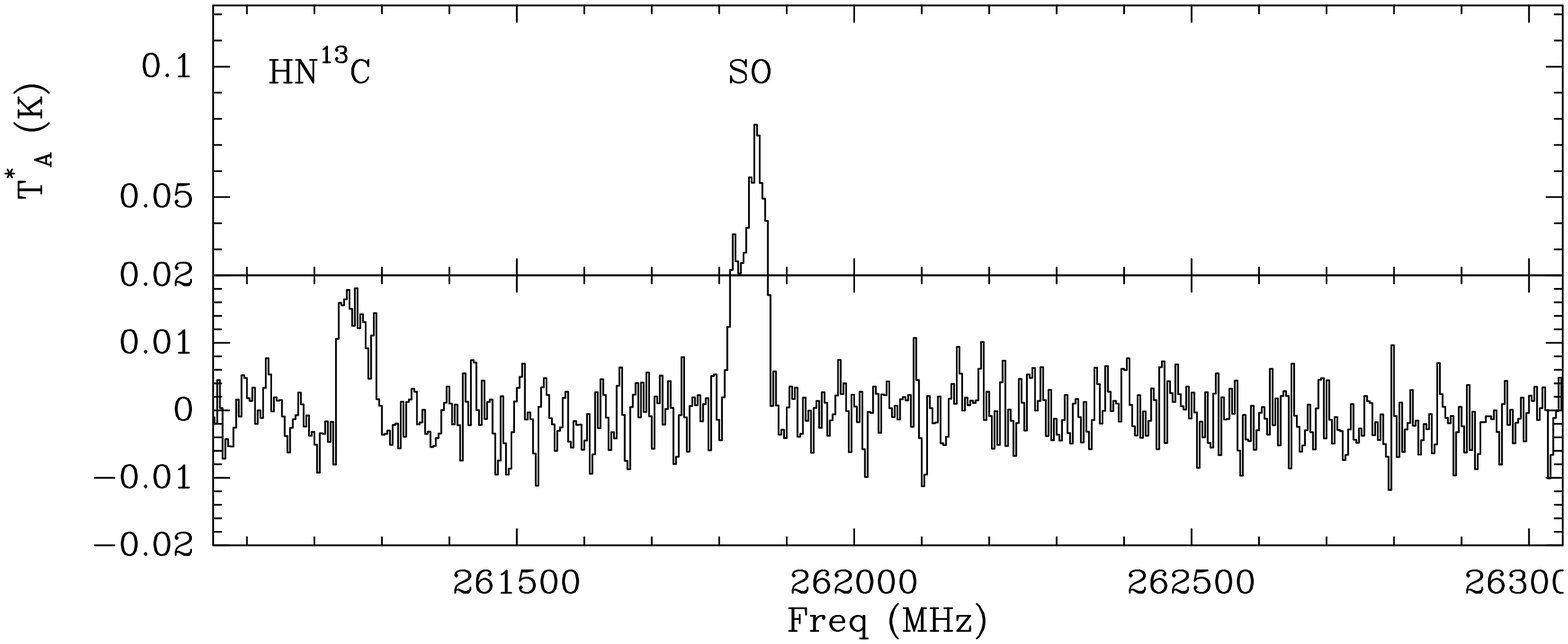} 
\caption{. (continued) } 
\label{Fig1mm}%
\end{figure*} 
\begin{figure*}[h!] 
\centering 
\ContinuedFloat 
\includegraphics[angle=0,width=12cm]{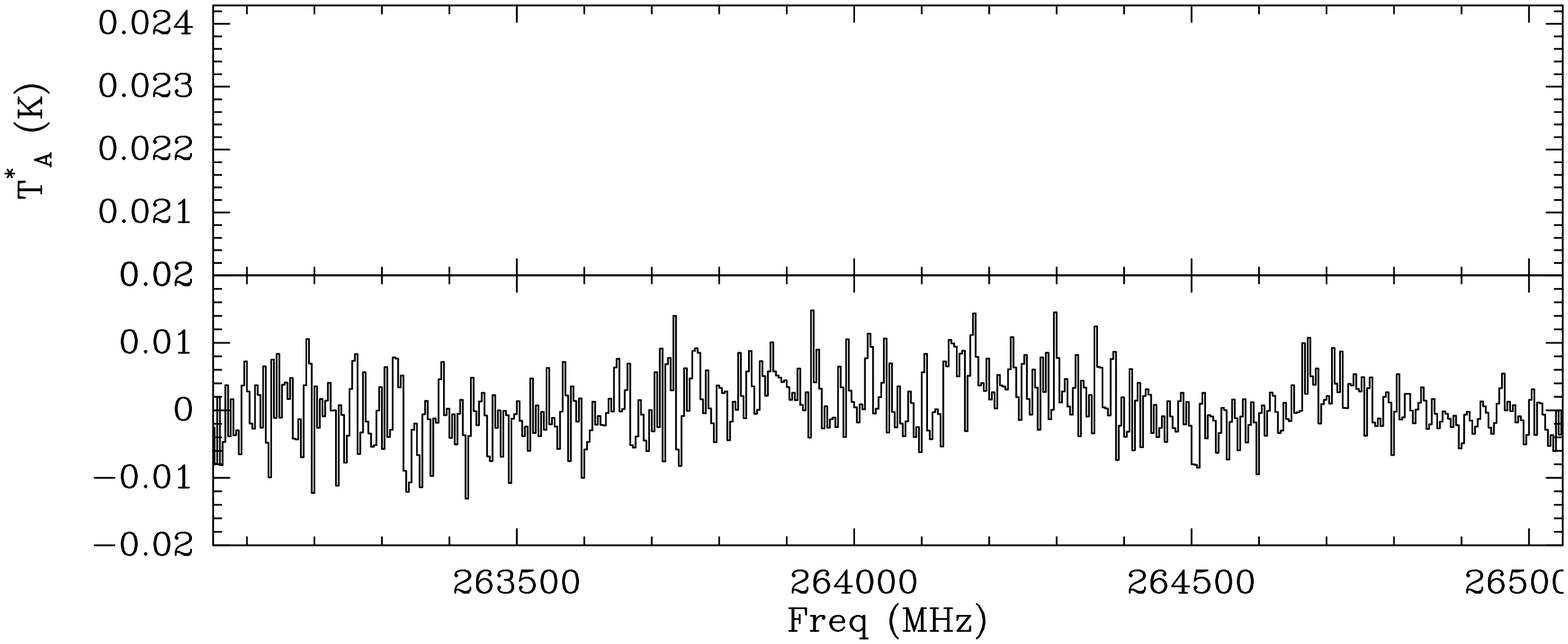} 
\includegraphics[angle=0,width=12cm]{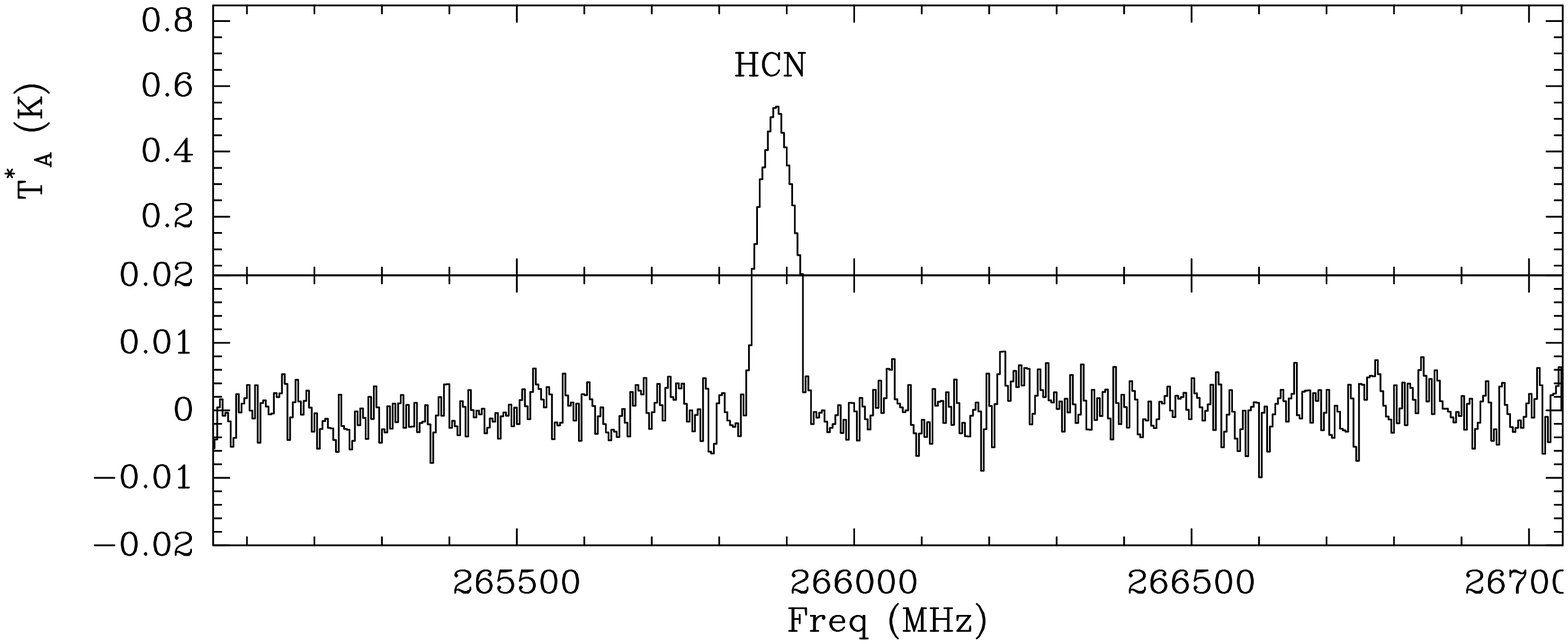} 
\includegraphics[angle=0,width=12cm]{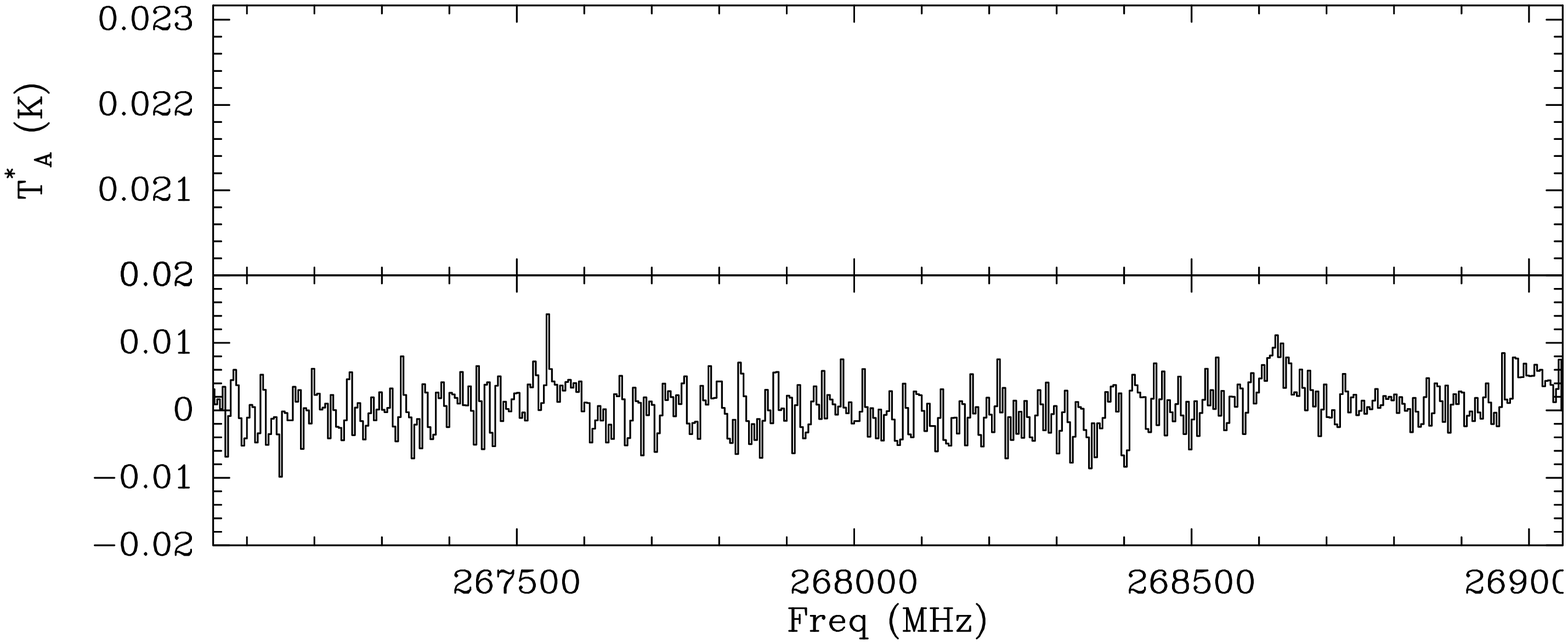} 
\includegraphics[angle=0,width=12cm]{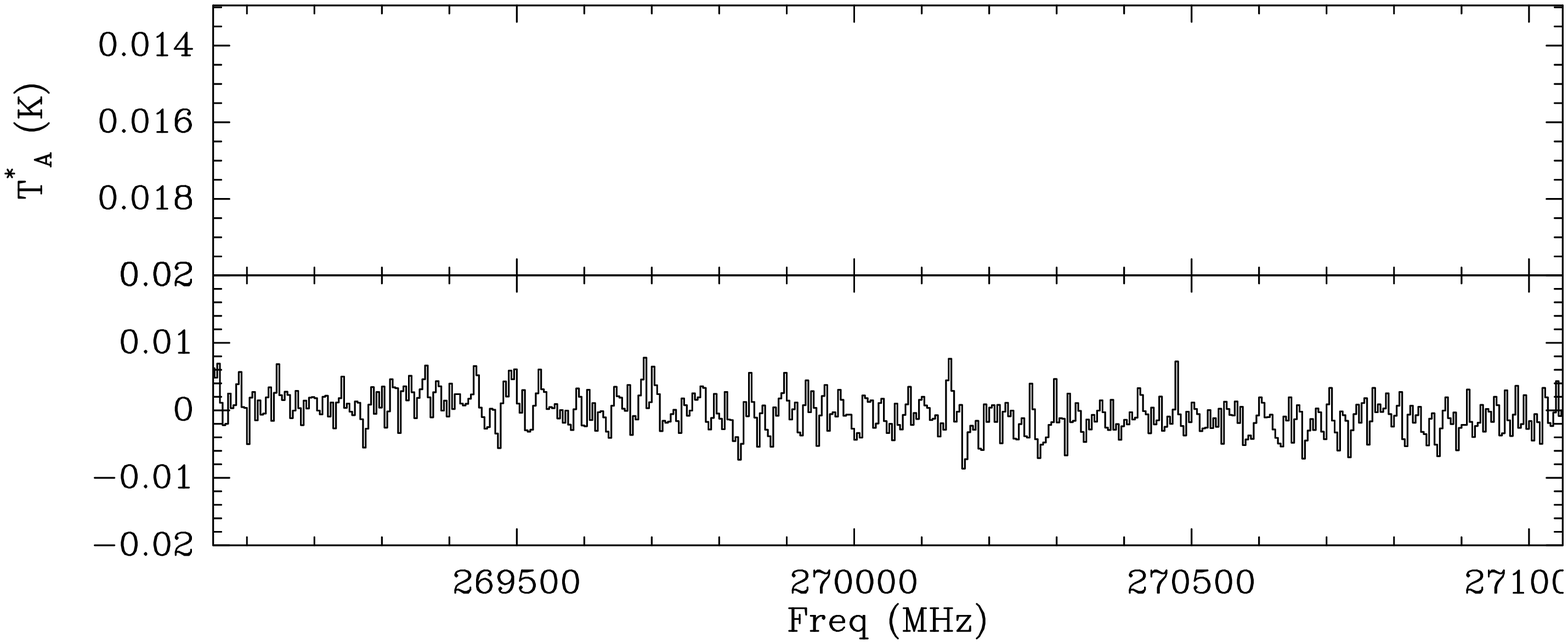} 
\caption{. (continued) } 
\label{Fig1mm}%
\end{figure*} 
\begin{figure*}[h!] 
\centering 
\ContinuedFloat 
\includegraphics[angle=0,width=12cm]{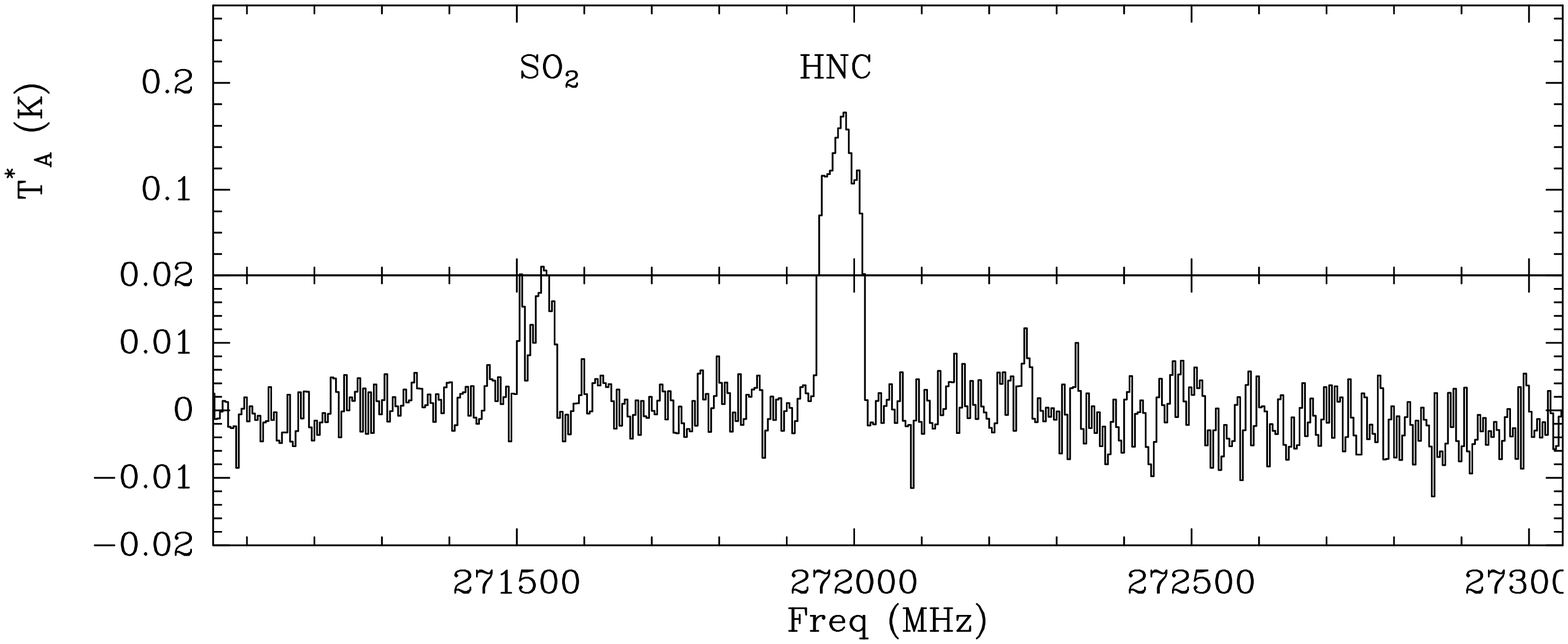} 
\includegraphics[angle=0,width=12cm]{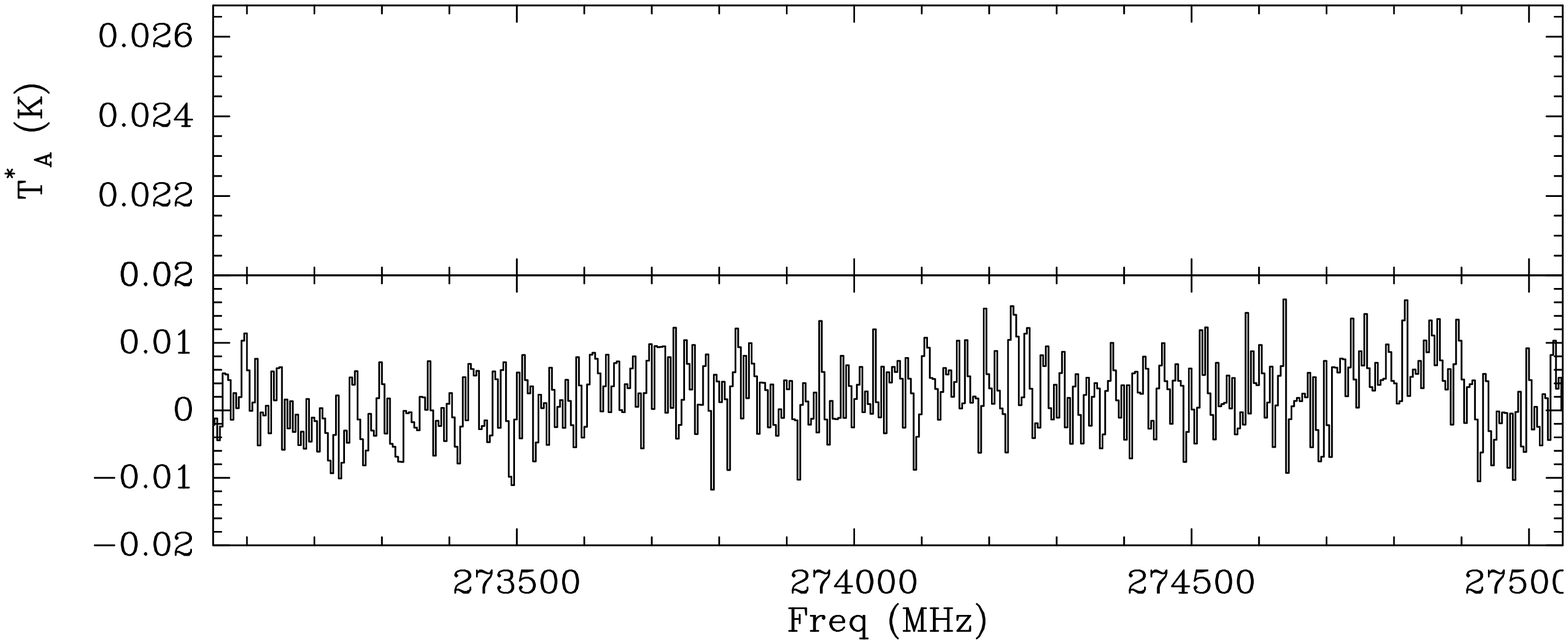} 
\includegraphics[angle=0,width=12cm]{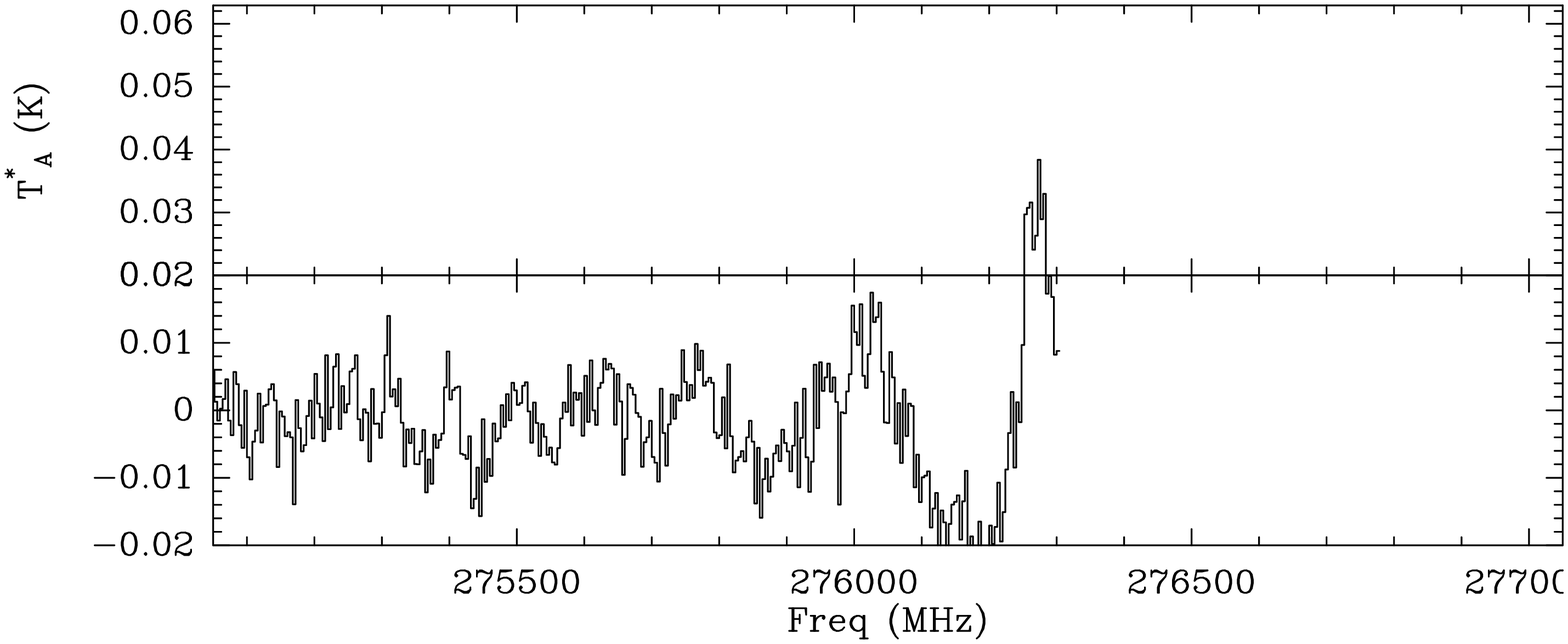} 
\caption{. (continued) } 
\label{Fig1mm}%
\end{figure*} 

\begin{figure*}[h!]
   \centering
   \includegraphics[angle=0,width=8cm]{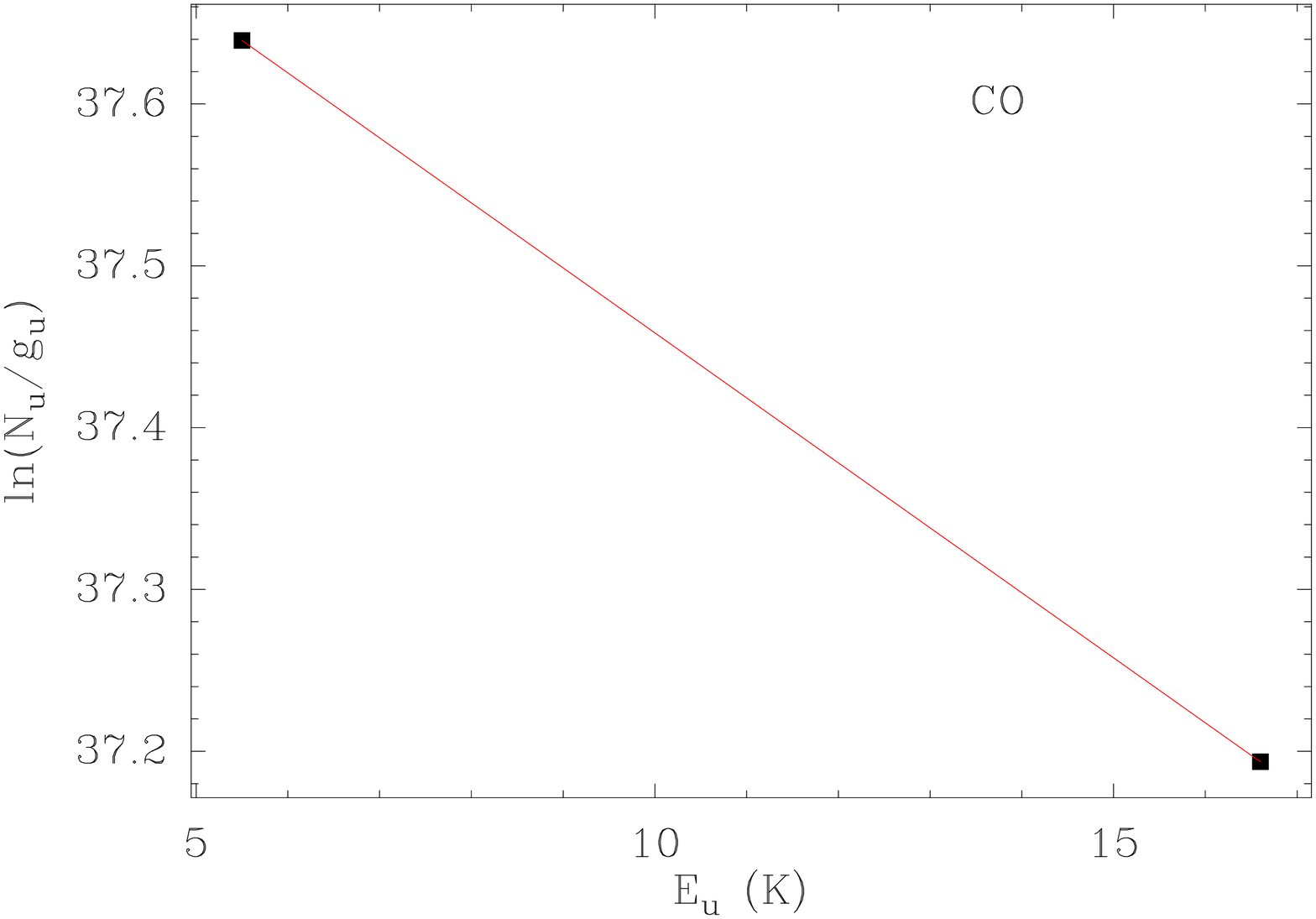}
   \includegraphics[angle=0,width=8cm]{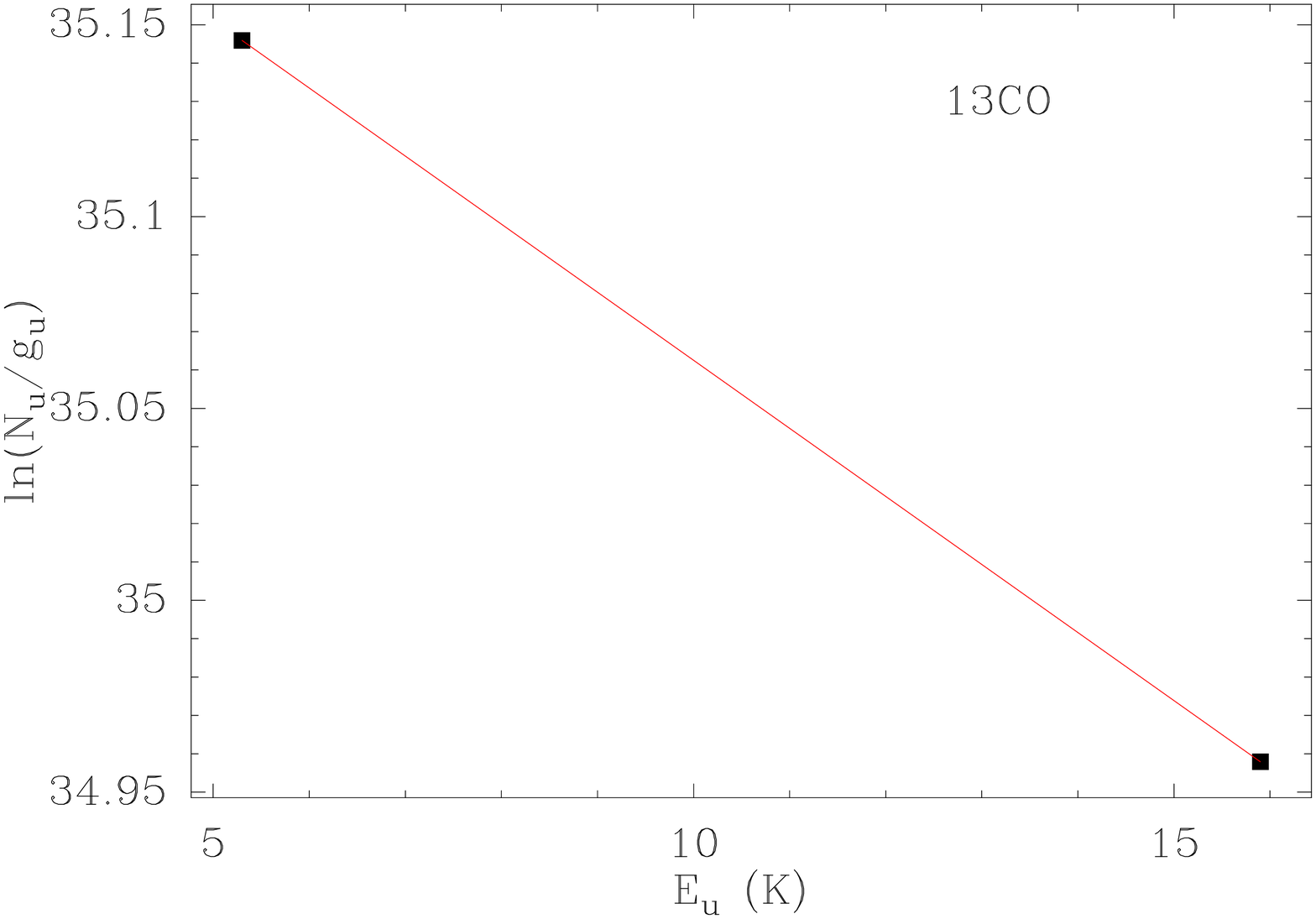}
   \includegraphics[angle=0,width=8cm]{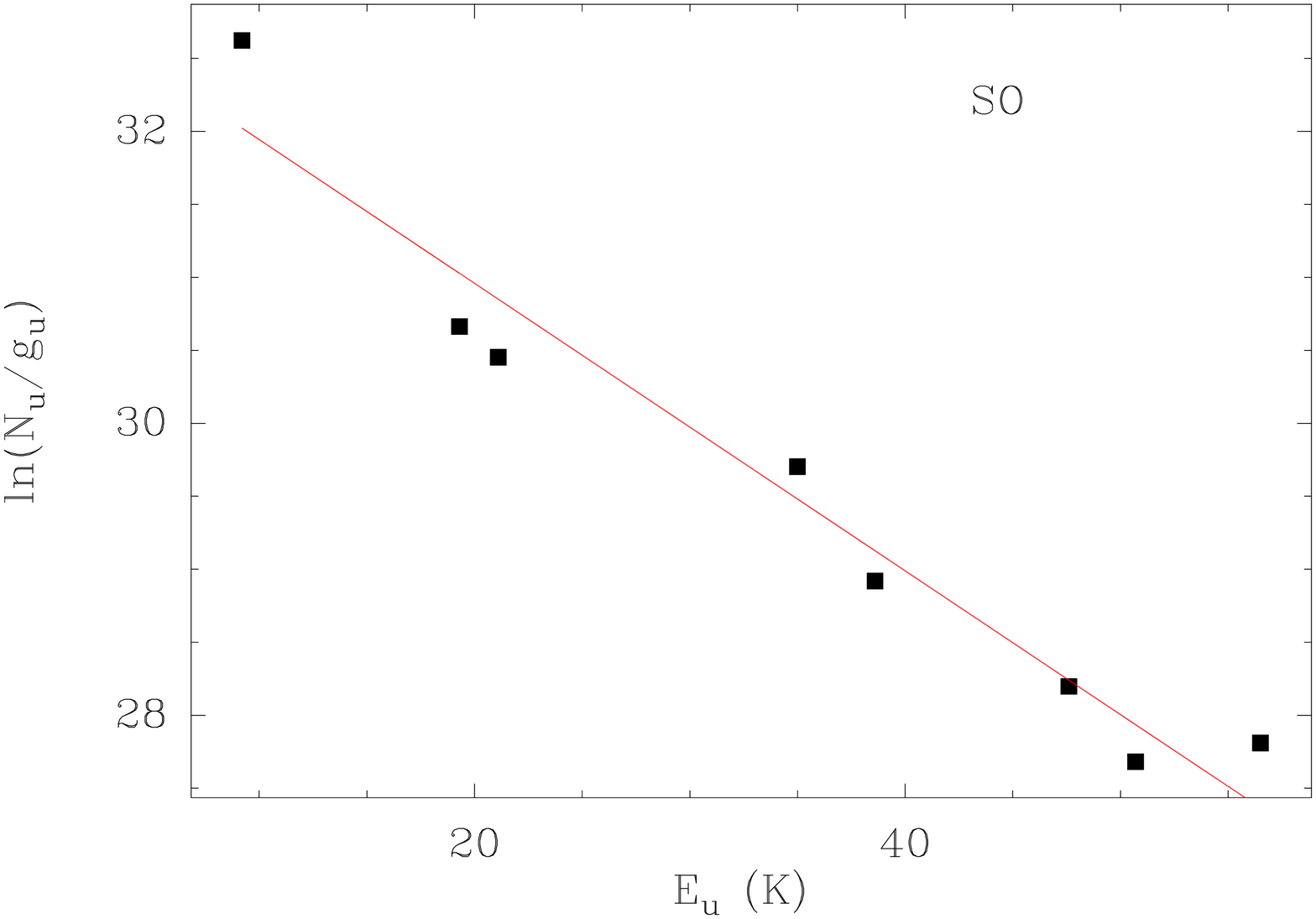}
   \includegraphics[angle=0,width=8cm]{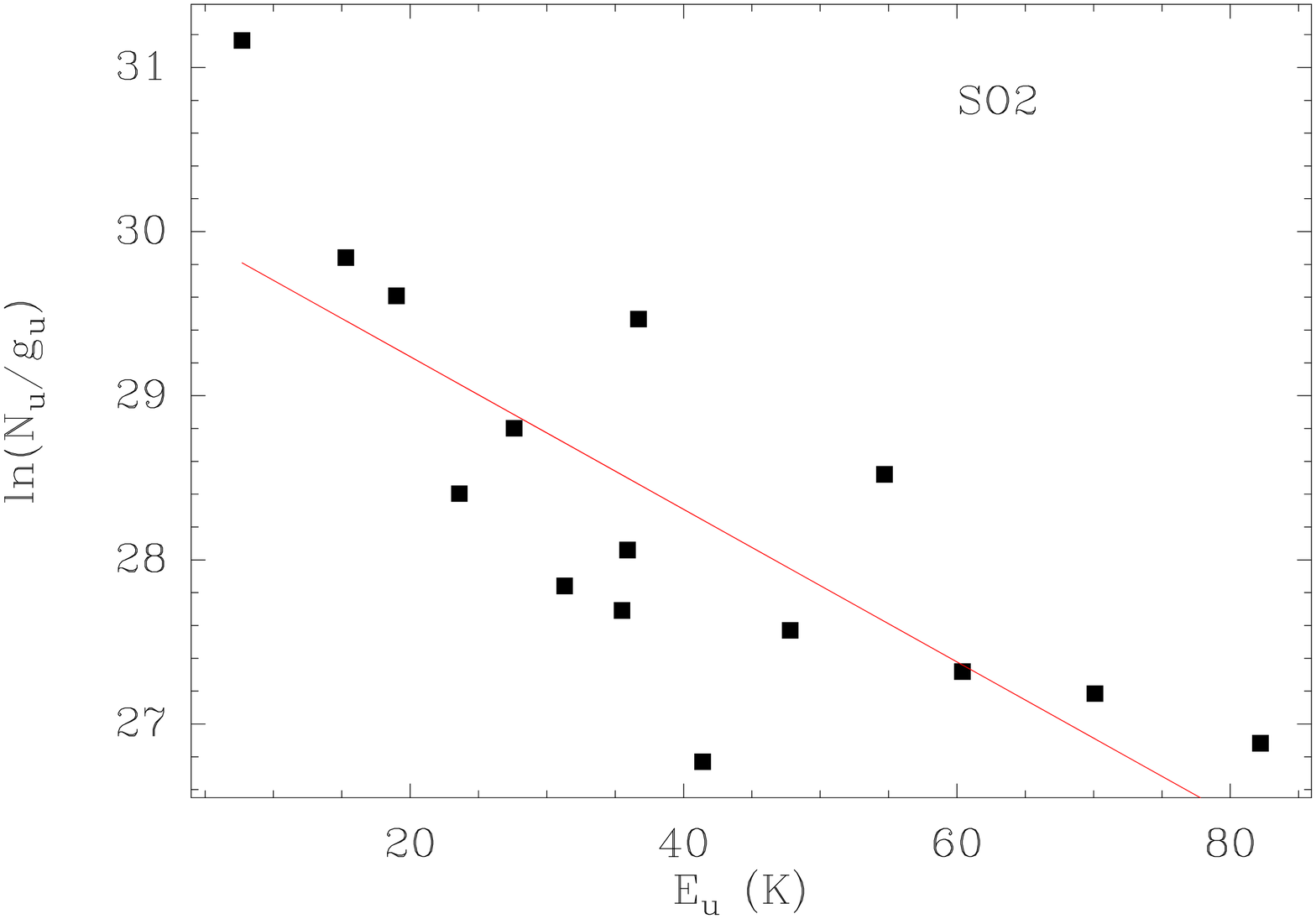}
   \includegraphics[angle=0,width=8cm]{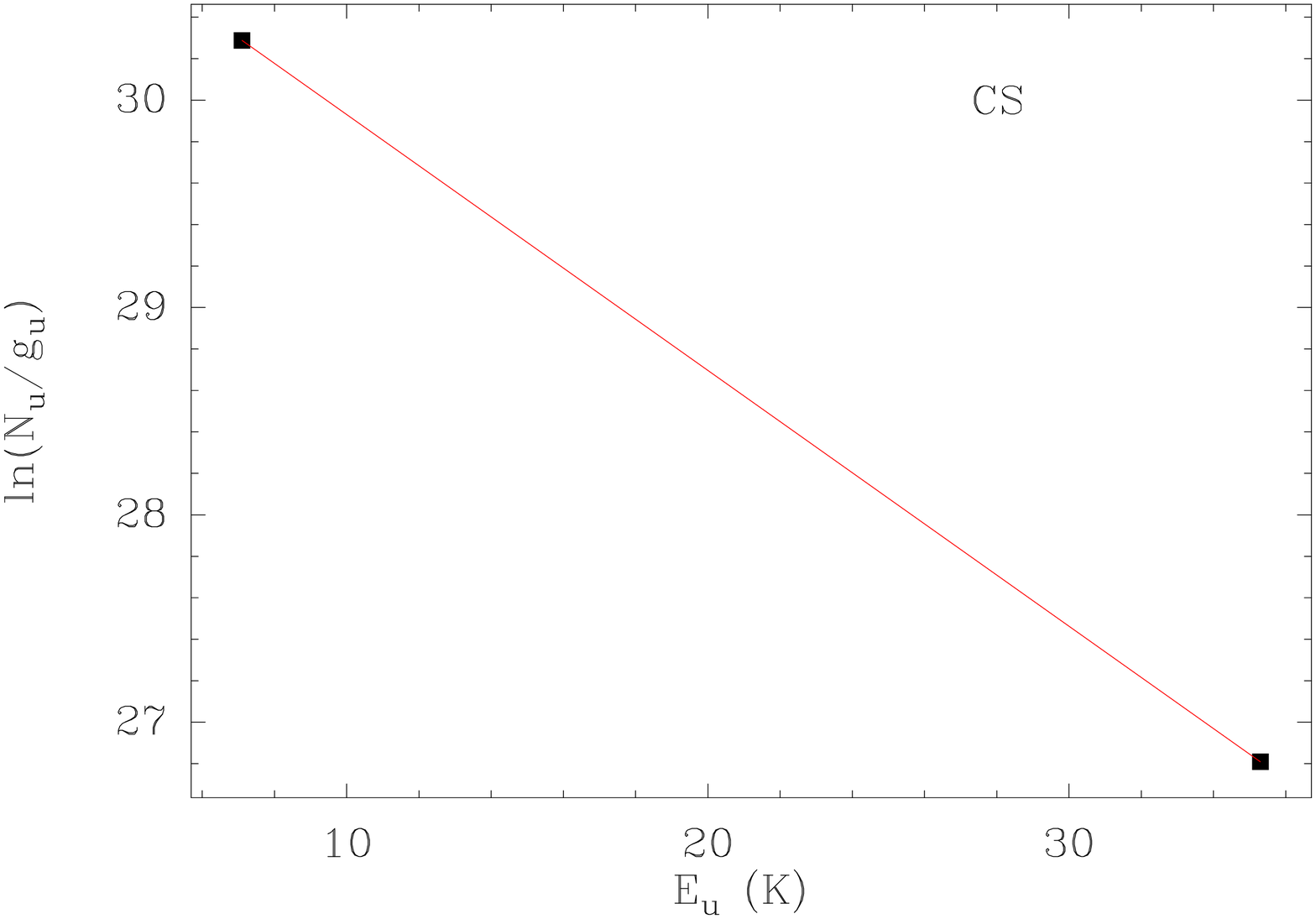}
   \includegraphics[angle=0,width=8cm]{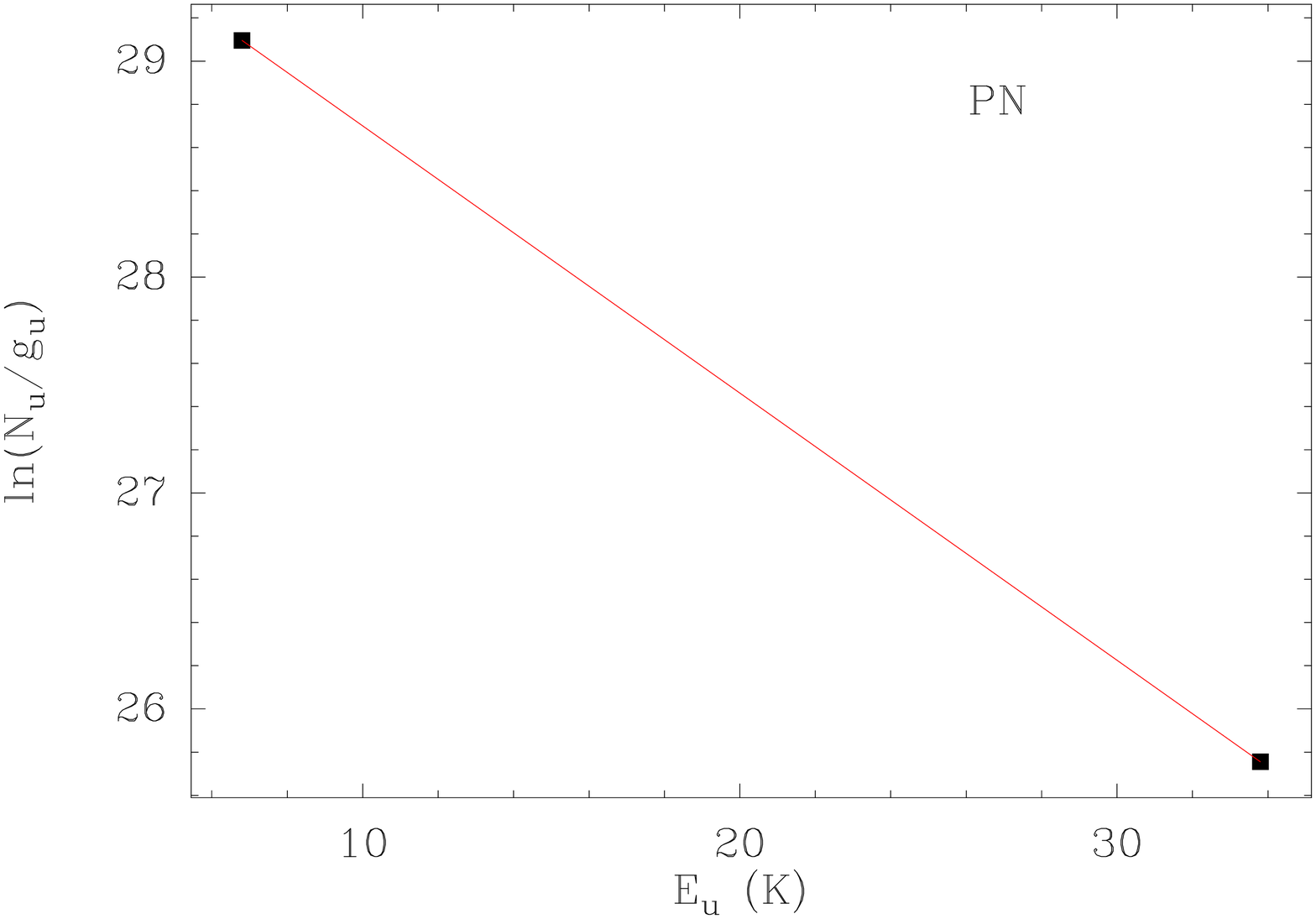}
   \includegraphics[angle=0,width=8cm]{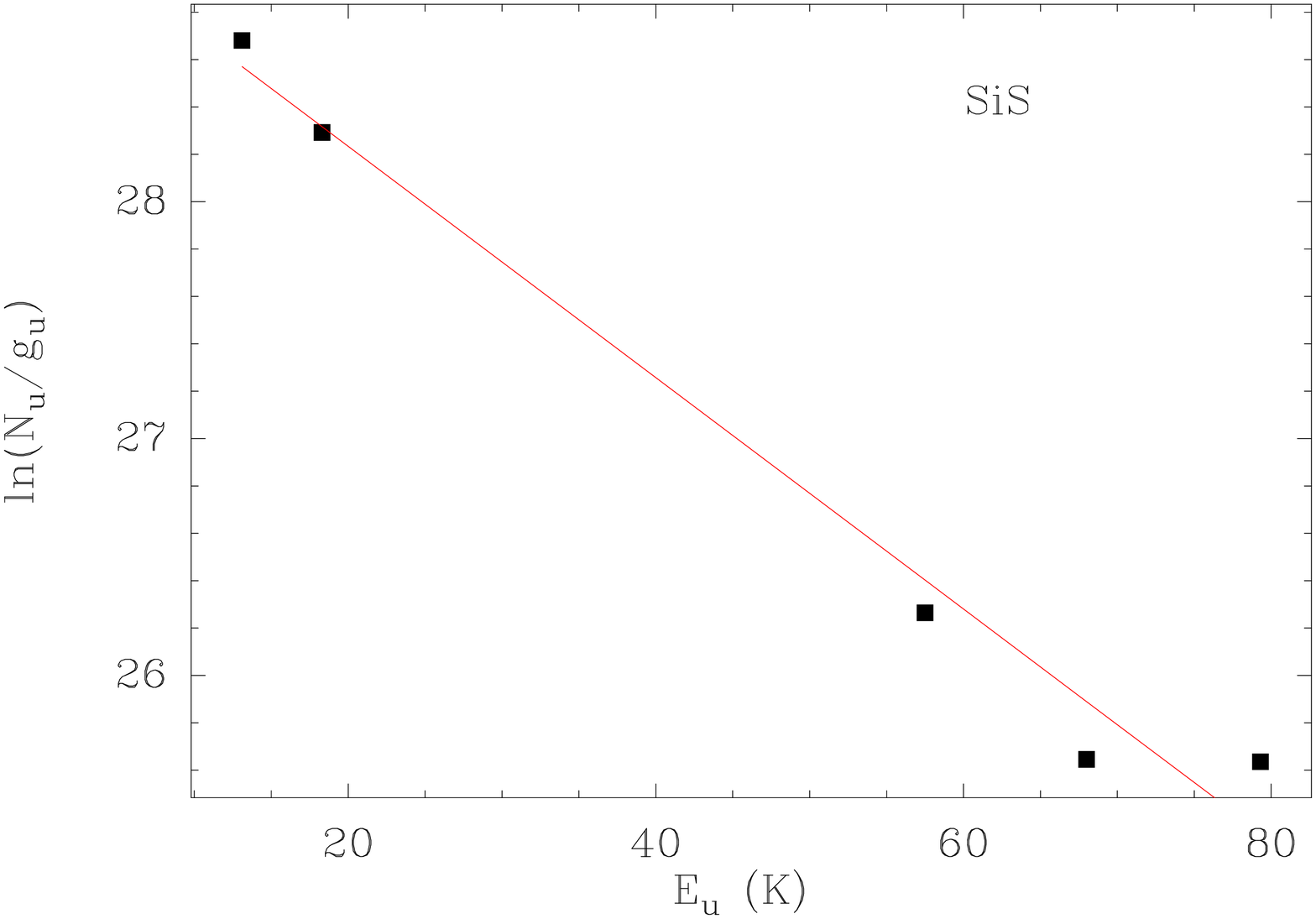}
   \includegraphics[angle=0,width=8cm]{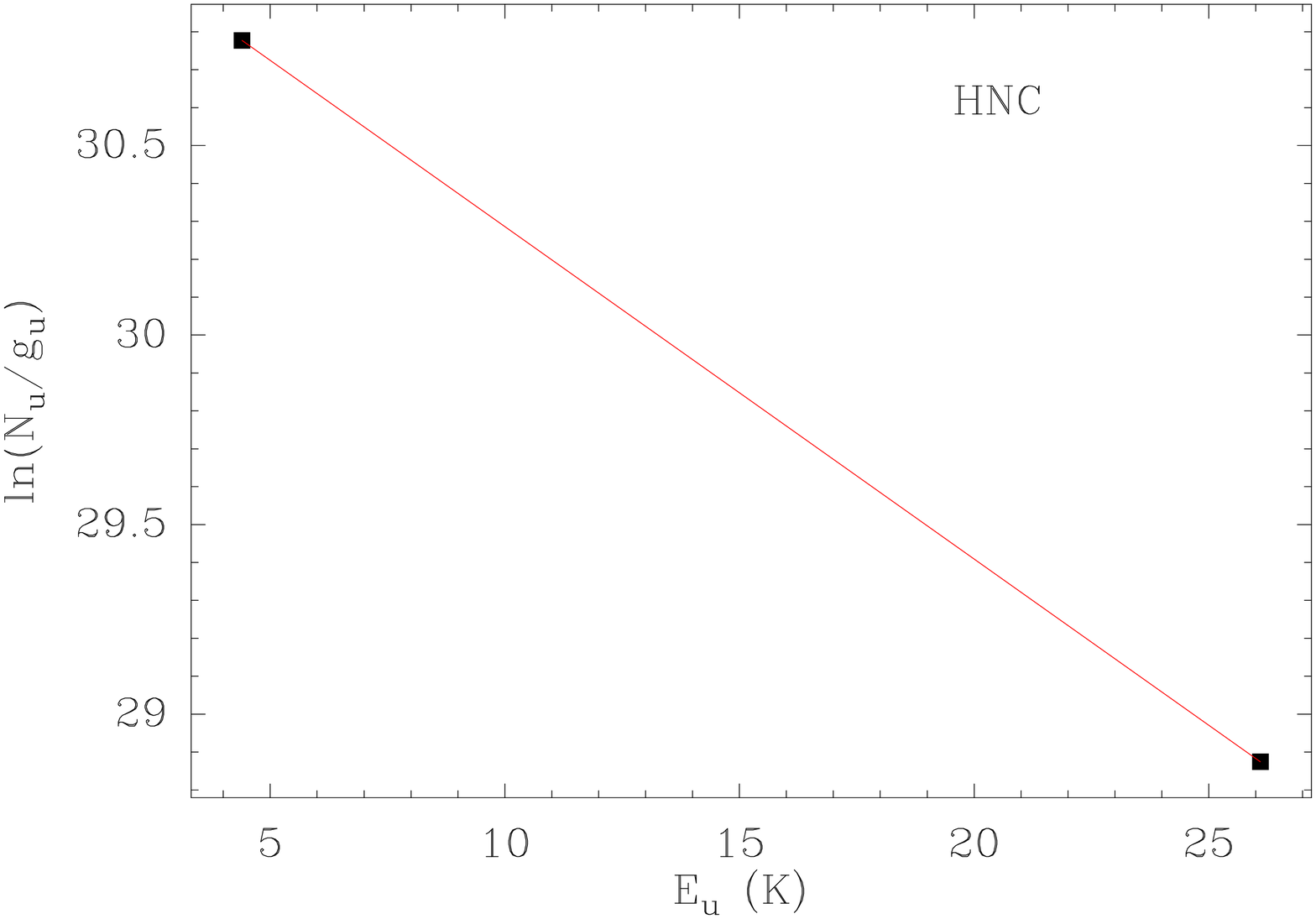}
   \caption[]{Rotational diagrams for CO, SO, SO$_2$, CS, PN, SiS, and HNC.}
              \label{rtd}%
\end{figure*}

\begin{figure*}[h!]
   \centering
   \includegraphics[angle=0,width=8cm]{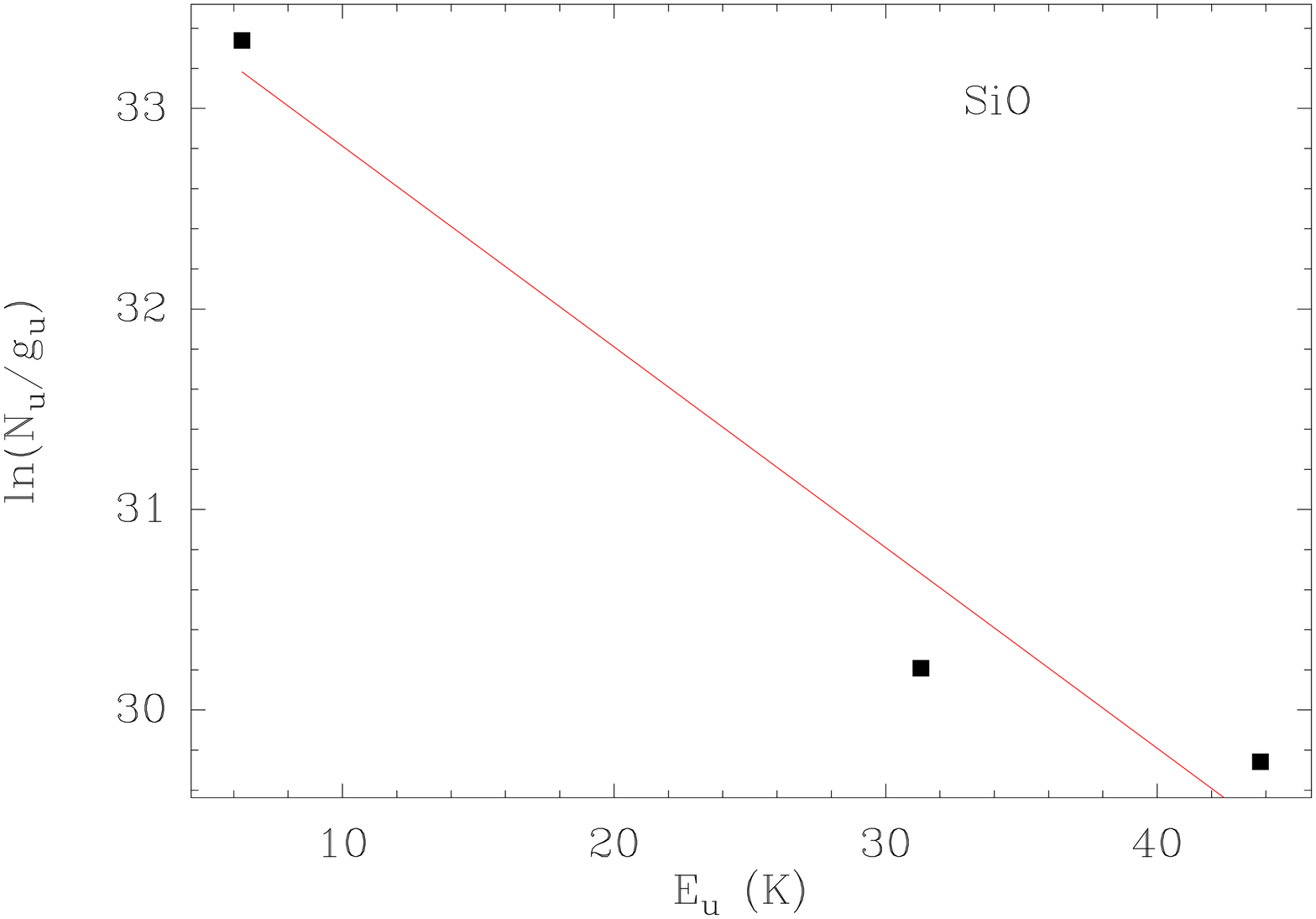}
   \includegraphics[angle=0,width=8cm]{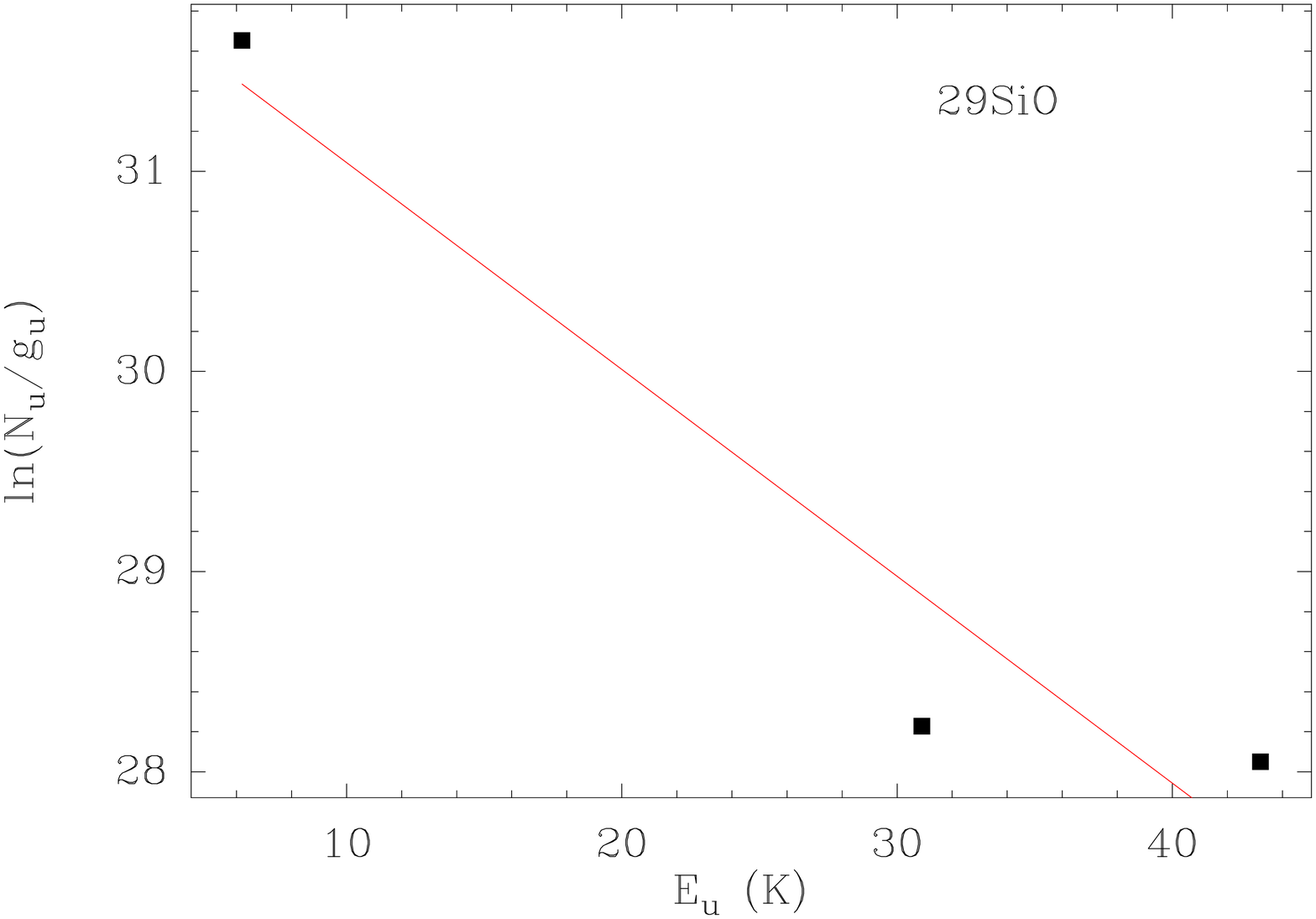}
   \includegraphics[angle=0,width=8cm]{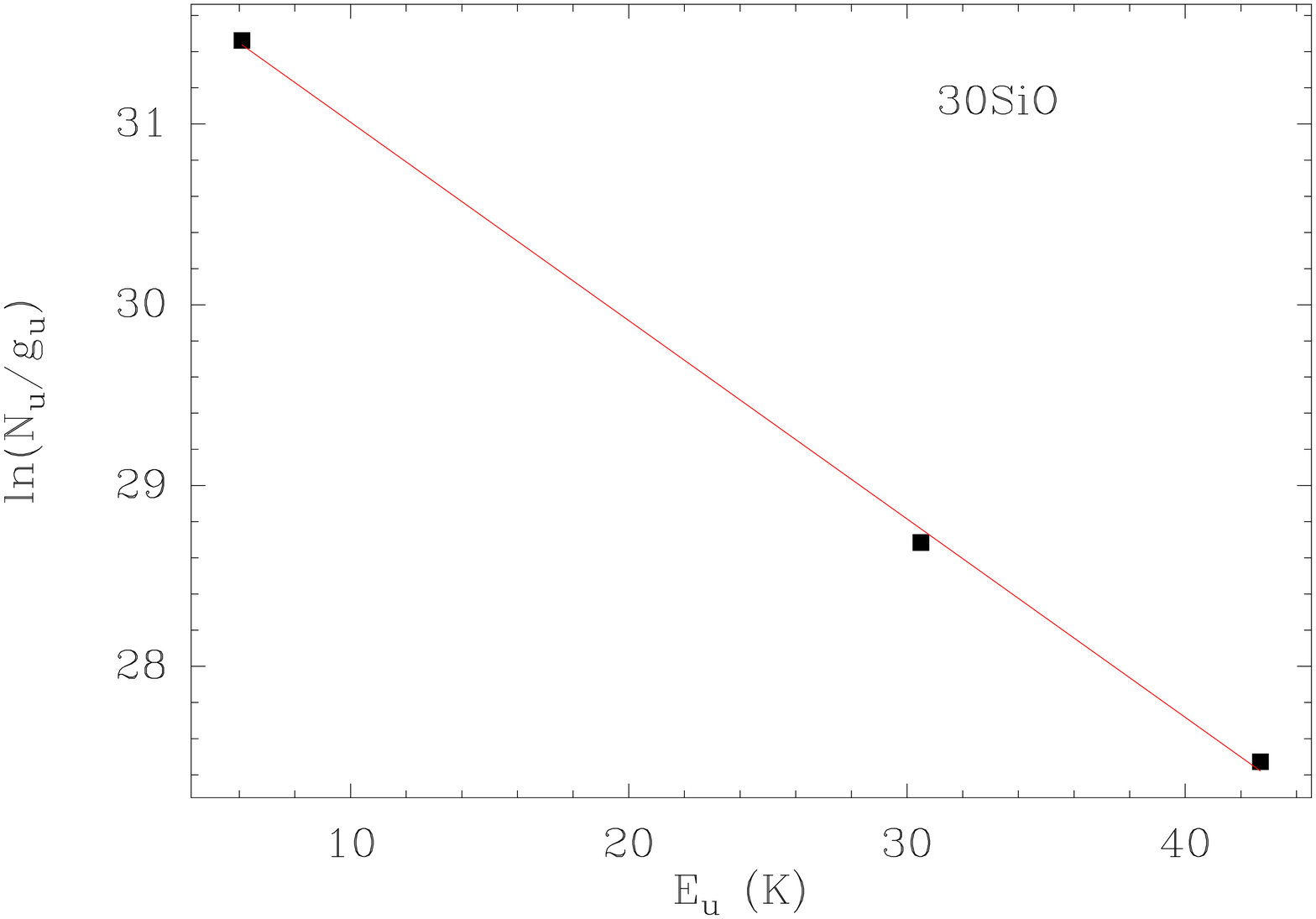}
   \includegraphics[angle=0,width=8cm]{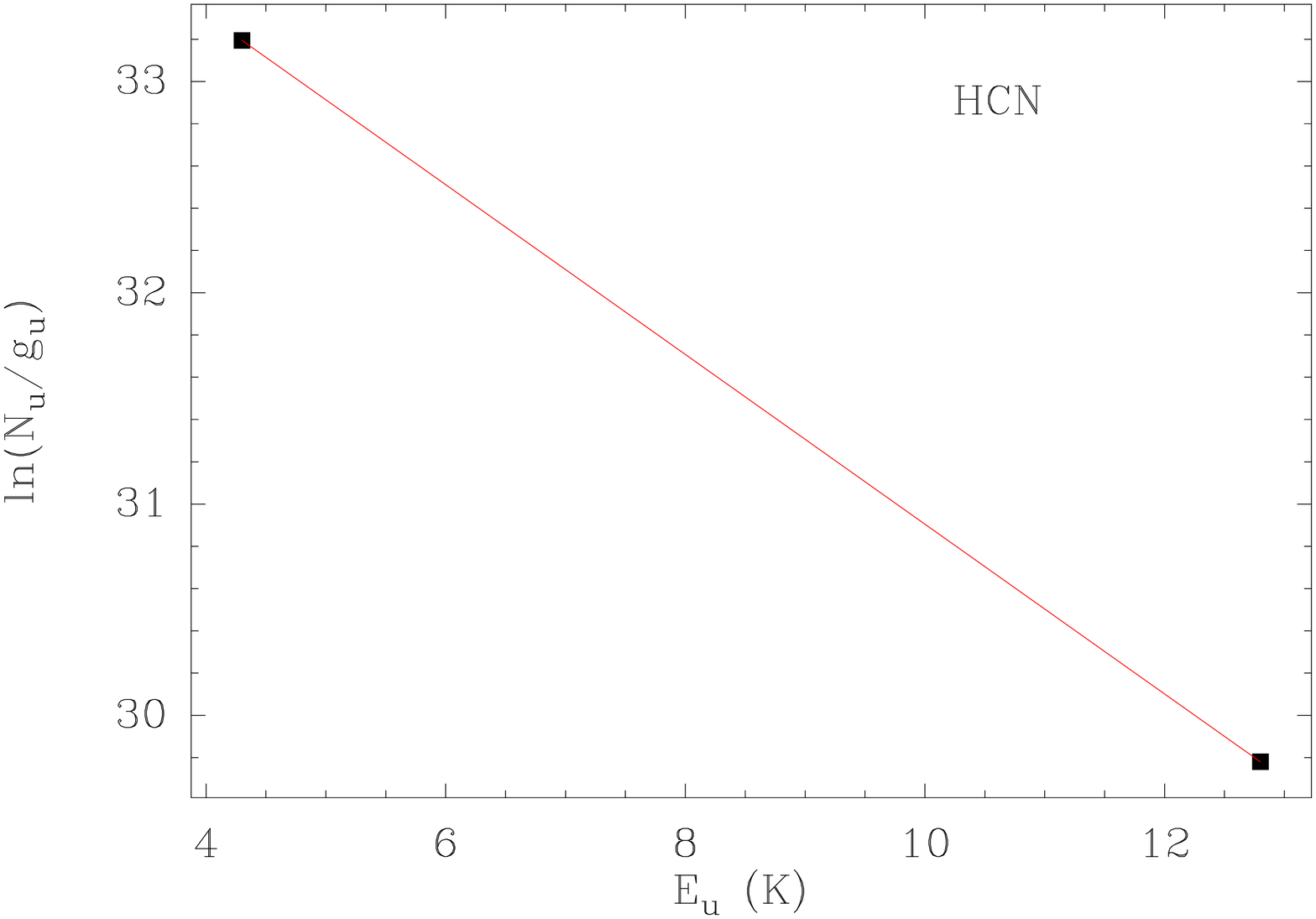}
   \includegraphics[angle=0,width=8cm]{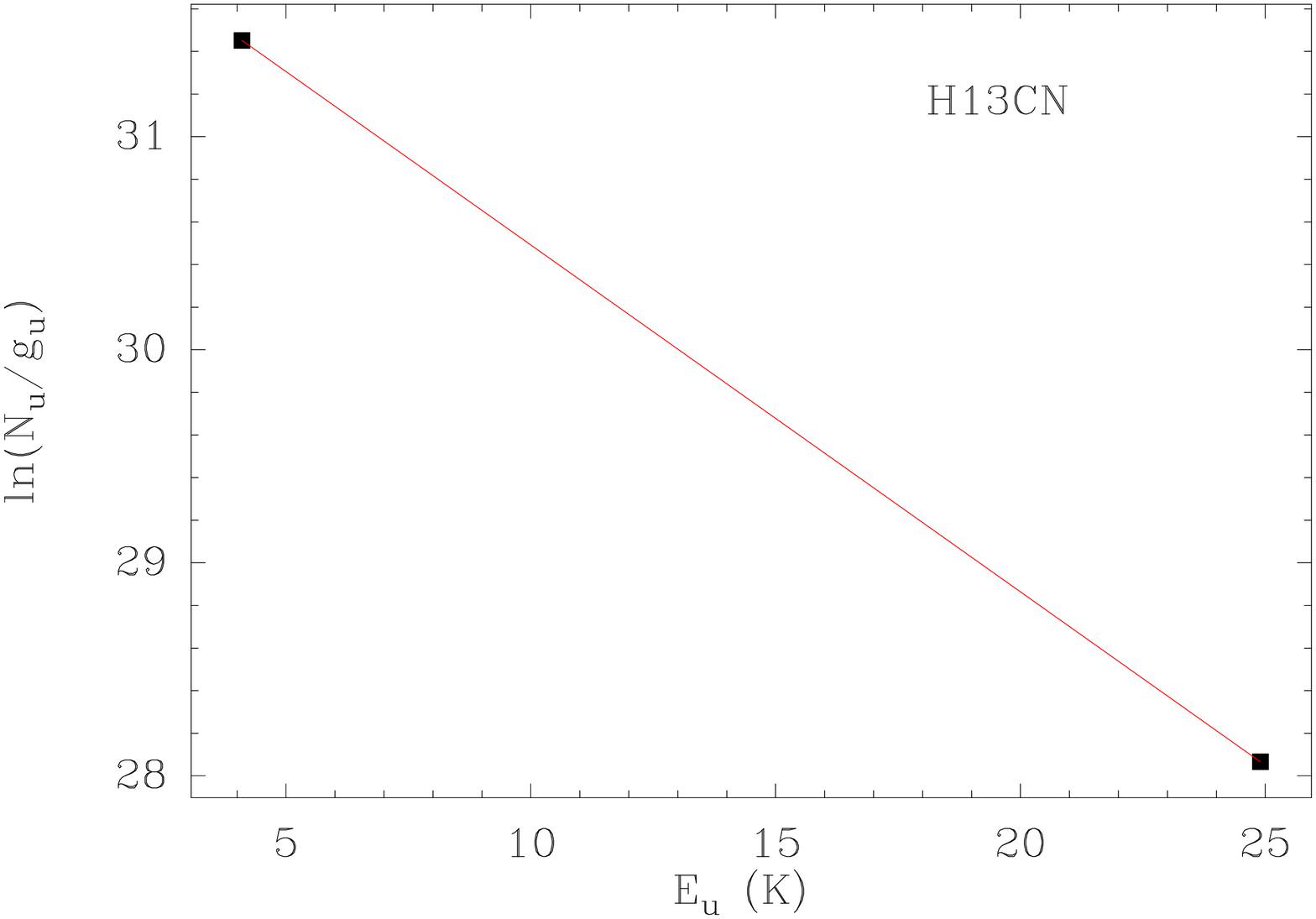}
   \caption[]{Rotational diagrams of SiO, HCN, and their isotopologues.}
              \label{rtd2}%
\end{figure*}

\end{document}